%% file: astro-ph.tex
\documentclass{article}
\usepackage{emulateapj,apjfonts,psfig}

\begin{document}

\lefthead{Evolution of E/S0s...}
\righthead{Im et al.}

\submitted{Accepted for Publication in the Astrophysical Journal}

\title{The DEEP Groth Strip Survey X: Number Density and Luminosity 
 Function of Field E/S0 Galaxies at $z < 1$}

\author{Myungshin Im\altaffilmark{1}, Luc Simard\altaffilmark{1,2},
 S. M. Faber\altaffilmark{1}, David C. Koo\altaffilmark{1},
 Karl Gebhardt\altaffilmark{1,6},  Christopher N. A. Willmer\altaffilmark{1,3},
 Andrew Phillips\altaffilmark{1},   Garth Illingworth\altaffilmark{1},
 Nicole P. Vogt\altaffilmark{4}, \& Vicki L. Sarajedini\altaffilmark{5}}

\altaffiltext{1}{UCO/Lick Observatory,
 Department of Astronomy \& Astrophysics, University of
 California, Santa Cruz, CA 95064}
\altaffiltext{2}{Steward Observatory, University of Arizona, 933 North
 Cherry Avenue, Tucson, AZ 85721}
\altaffiltext{3}{On leave from Observatorio Nacional, Rua General Jose Cristino
 77, 20921-030 Sao Cristovao, RJ, Brazil}
\altaffiltext{4}{Institute of Astronomy, University of Cambridge, 
 Madingley Road, Cambridge, CB3 0HA, United Kingdom}
\altaffiltext{5}{Department of Astronomy, Wesleyan University,
 Middletown, CT 06459} 
\altaffiltext{6}{Department of Astronomy, University of Texas, Austin}

\begin{abstract}
  We present the luminosity function and color-redshift relation of
 a magnitude-limited sample of 145 
 mostly red field E/S0 galaxies at $z \lesssim 1$ 
 from the DEEP Groth Strip Survey (GSS).
   Using nearby galaxy images as a training set, we develop a quantitative
 method  to classify E/S0 galaxies based 
 on smoothness, symmetry, and bulge-to-total light ratio.
  Using this method, we identify 145 E/S0s at $16.5 < I < 22$ within the 
 GSS, for which 44 spectroscopic redshifts ($z_{spec}$) are available.
 Most of the galaxies with spectroscopic redshifts  (86\%) 
 form a {\it red envelope} in the redshift-color diagram,
 consistent with predictions of spectral synthesis
 models in which the dominant stellar population is
 formed at redshifts $z \gtrsim 1.5$.  
  We use the tight correlation
 between $V-I$ and $z_{spec}$ for this red subset to estimate 
 redshifts of the remaining E/S0s to an accuracy of $\sim$10\%,
 with the exception of a small number (16\%) of blue interlopers 
 at low redshift that 
 are quantitatively classified as E/S0s but are not contained within
 the red envelope.
   Constructing a luminosity function of the full sample of 145 E/S0s,
 we find that  
 there is about 1.1--1.9 magnitude brightening in rest-frame $B$ band
 luminosity back to  $z \simeq 0.8$ from $z=0$, consistent with other studies.
  Together with the red colors, 
 this brightening is consistent with models in which the bulk of stars
 in red field E/S0s formed before $z_{for} \gtrsim 1.5$ and have been
 evolving rather quiescently with few large starbursts since then.
  Evolution in the number density of field E/S0 galaxies is harder to
 measure, and uncertainties in the raw counts and their ratio to
 local samples might amount to as much as a factor of two.  
  Within that uncertainty, 
 the number density of red E/S0s to $z \simeq 0.8$ seems 
 relatively static, being comparable to 
 or perhaps moderately less than that of local E/S0s depending
 on the assumed cosmology.
 A doubling of E/S0 number density
 since $z = 1$ can be ruled out
 with high confidence (97\%) if $\Omega_{m}=1$.
 Taken together, our results are consistent with the hypothesis
 that the majority of luminous field E/S0s were already in place 
 by $z \sim 1$, that the bulk of their stars were already
 fairly old, and that their number density has not 
 changed by large amounts since then.
\end{abstract}


\section{Introduction}

\input{intro.tex}

\section{Observational Data}

  Our goal is to identify moderate-to-high redshift red E/S0s 
 and measure their redshifts photometrically from HST images alone.
 We proceed in two steps.  First, we develop and tune our
 selection method for finding E/S0 galaxies
 by testing it on a subsample of GSS galaxies
 having spectroscopic redshifts.  This test allows us 
 to verify simultaneously that we are selecting red-envelope galaxies and that
 our photometric redshift scheme works for such galaxies.
 Note that the completeness of this spectroscopic training sample does not
 matter since the completeness
 of the final sample is limited only by magnitude.

  The Deep Extragalactic Evolutionary Probe (DEEP) is a study of 
 faint field galaxies using spectra obtained with
 the Keck 10-m telescope and high-resolution images 
 obtained by the Hubble Space Telescope (HST). 
  As a part of DEEP, we have extensively studied the 
 Groth Strip, which is comprised of 28 contiguous HST WFPC2 fields
 covering roughly 134 arcmin$^2$ on the sky (Groth et al. 1994;
 Koo et al. 1996). In this section, we summarize 
 the basic data from which our E/S0 galaxies are selected.

  We adopt the Vega magnitude system
 as in Holtzman et al. (1995) and  
 in previous DEEP papers.  The zero-point of the 
 magnitude system depends on the detailed shape 
 of the filter and CCD response function.  The $I$-band magnitude
 measured through the F814W filter is simply referred to as ``$I$''.
  Likewise, for the 
 $V$-band magnitude measured through the F606W filter we will
  use the symbol ``$V$''.  Readers are alerted that
 this $V$-band magnitude from F606W can be up to 1.0 magnitude brighter than 
 the conventional Johnson $V$ magnitude,
 depending on the color and redshift of the galaxy.
  The $I$-band magnitude with the F814W filter is roughly 
 equal to the Cousins $I$ magnitude, with
 $I_{F814W}-I_{Cousins} \simeq 0.04$--0.1 (Fukugita, Shimasaku \& 
 Ichikawa 1995).   Vega-calibrated magnitudes can be converted to 
 the AB-magnitude system using the following relation:
 $I_{AB}-I=0.44$ and $V_{AB}-V=0.11$ (Simard et al. 2000).

\subsection{Photometric data}

  Galaxies were imaged by WFPC2 in both the F814W filter
 and the F606W filter, with exposure times of 4400 s 
 and 2800 s respectively (one deep field has total exposure times of 25200 s in
 F814W and 24400 s in F606W).
  The limiting magnitude for the detection of objects is about $I=24$, and 
 about 3000 galaxies are detected to this 
 magnitude using FOCAS (Groth et al. 2000; Vogt et al. 2000).
  For the $I$ and $V$ magnitudes, we use the model-fit total magnitudes 
 throughout the paper as described below. 
  To obtain structural parameters and morphological parameters,
 the surface brightness of each object is fit by a 2-dimensional 
 bulge+disk model profile using the GIM2D software package of 
 Marleau \& Simard (1998) and Simard et al. (1999, 2000).
  Here, the disks are assumed to have an exponential profile, 
 and the bulge profiles are assumed to follow the $r^{\frac{1}{4}}$ law
 (de Vaucouleurs 1948).  This procedure is 
 similar to that of Ratnatunga, Griffiths, \& Ostrander (1999) 
 and Schade et al. (1995), 
 and returns model-fit structural parameters that include 
 total magnitudes ($I_{tot}$, $V_{tot}$),
 bulge-to-total light fraction ($B/T$), disk scale length ($r_{0}$), 
 bulge effective radius ($r_{e}$), half-light radius ($r_{hl}$), and position
 angles and ellipticities for both the bulge and disk components. 
 Sizes are defined along the major axis of the model fit.
 Instead of listing each magnitude  
 as $I_{tot}$ or $V_{tot}$, we omit the subscript ``$tot$''.
  For a more complete description of the basic photometric catalog
 of GSS galaxies, see Vogt et al. (2000) and Koo et al. (1996).
  For further description of the model-fit procedure
 and a complete list of derived structural parameters,
 see Simard et al. (2000).  

\subsection{Spectroscopic data}

   Spectroscopic data were taken during the 1995-1998 Keck observing runs
 using multislit masks on the Low Resolution Imaging Spectrograph 
 (Oke et al. 1995).
   Each mask contains about 30--40 objects. The blue part of the 
 spectrum is covered with a 900-line/mm grating, and the red part of 
 the spectrum is covered with a 600-line/mm grating, giving us 
 spectral coverage from roughly $5000$ $\AA$ to $8000$ $\AA$. 
  Spectral resolution is FWHM $=\sim 2.5~\AA$ and $\sim 4~\AA$
 for the blue and the red settings respectively,
 providing sufficient resolution to 
 measure internal kinematics of faint galaxies at moderate to high redshift
 down to a velocity width of  $\sigma \simeq$ 70 km~sec$^{-1}$.
  The high resolution also resolves the
 [O II] doublet emission at 3727 $\AA$, which enables 
 redshifts to be measured from this single line alone.
 It also improves removal of the night sky
 lines, thus increasing the likelihood of identifying
 lines in the red part of the spectrum.
   For spectroscopic observations,
 objects were mostly selected based on
 mean $(V_{1\farcs5}+I_{1\farcs5})/2$ magnitude
 (i.e., roughly an $R$ magnitude) 
 through a $1\farcs5$ aperture, and aperture
 color $(V-I)_{ap}$ through a $1\farcs0$ aperture. 
 A higher priority was given to the selection of red objects 
 with  $(V-I)_{ap} > 1.75$, which roughly 
 corresponds to a passively evolving stellar population beyond $z = 0.7$: 
 therefore, our sampling strategy provides more spectra of red 
 E/S0s at $z > 0.7$ than expected with a uniform
 sampling strategy.  The typical exposure time for each mask 
 was 1 hour in each wavelength setting, and 
 objects that did not yield redshifts on the first try
 were attempted on subsequent exposures.
 The total number of redshifts is about 590, and 
 the sample is 90\% complete 
 down to $I = 23$. 
  This limit is
 about 1.5 magnitudes fainter than that of 4-m class surveys 
 (e.g., CFRS) and comparable to other Keck telescope faint-galaxy surveys 
 (Cohen et al. 1996; Cowie et al. 1996).

  For further description of the spectroscopic sample, 
 see Phillips et al. (2000) and Koo et al. (1996).

\section{Selection of E/S0 Galaxies}

\subsection{Overview}

  The ultimate goal is to identify red, moderate-to-high redshift E/S0s 
 and to measure their redshifts photometrically from HST images alone.
  This section tests the adopted selection procedure 
 using simulated images of local galaxies and shows that the
 number of selected galaxies is reasonably stable versus redshift.  

  In detail, there are two selection criteria that are
 somewhat in conflict.  On the one hand, we seek to restrict
 the sample to objects whose redshifts can be estimated from $V-I$ alone.
 This requires that the selected objects {\it lie on the red envelope}
 of galaxies in color versus redshift,
 and thus that they be morphologically quite ``pure."  On the other
 hand, we seek a classification system that yields {\it stable numbers of
 galaxies} as the quality of data declines at high $z$.  This requires
 locating the data cuts such that equal numbers of objects are scattered
 out of the sample by errors as are scattered in.  However,
 the in-scattered objects
 are in general {\it not} red and will not yield good photometric redshifts.
 Being blue, they appear to have low reshift,
 and they inflate the number of E/S0s
 counted at low $z$. 

 The first step in developing the selection procedure
uses the spectroscopic redshift subsample to
verify that the selection procedure finds mainly red-envelope
galaxies, by plotting candidate E/S0s with redshifts in 
$V-I$ vs. $z$.  Having finalized the selection procedure, we again use the
spectroscopic redshift subsample to calibrate the relation between
spectroscopic redshifts and  photometric
redshifts determined from $V-I$ color (for the red subset of galaxies).
A final photometric sample is selected from the full database
(lacking redshifts) and numbers and densities are calculated in
Section 4.

 Traditionally, local E/S0s have been visually classified.
   However,  visual classification is subjective, and
  boundaries are vague between morphological types (e.g., E/S0s vs. S0/a's).
  Naim et al. (1995) compared independent visual classifications of
 local galaxies by various experts and found that 
 the average uncertainty over all types is $\delta T \sim 2$, where
 $T$ is the morphological type defined by the RC3.
 The same uncertainty applied to 
 galaxies with $T \leq 0$ (E/S0s).
  There are other difficulties as well; 
  morphological features can be lost 
 when spiral galaxies are viewed edge on, or when the number of
 resolution elements declines at high redshift and is insufficient to resolve 
 spiral arms or other morphological details (so-called pixel smoothing).

  To lessen these difficulties,  we establish a quantitative scheme
  that promotes a more objective and reproducible 
 morphological classification. Such a scheme also enables
 us to use simulations to test efficiently 
 the effects of  pixel smoothing and low signal-to-noise 
 ($S/N$) that afflict high-redshift data.

  The properties of E/S0s suggest that two parameters should suffice
 for the purpose of quantitative classification. 
  One parameter should describe how featureless and symmetric the appearance 
 of the galaxy is, since E/S0s are characterized as smooth, featureless,
 symmetric objects.
  Another parameter should describe how the light is distributed within
 the galaxy, since E/S0s have a centrally concentrated surface brightness
 profile well fit by the $r^{\frac{1}{4}}$ law.

  As the two morphological 
selection parameters, we have therefore chosen to use 
 bulge-to-total light ratio, $B/T$, to describe the concentration
 of light, and the residual parameter, $R$, to describe the smoothness and
 symmetry of the galaxy morphology. Both parameters are 
 byproducts of the 2-dimensional surface brightness profile 
 fitting technique described in
 Section 2.1.

\subsection{Bulge-to-total light ratio, $B/T$}

   The quantity $B/T$ is the fraction of the light contained in the bulge
 component.
   Suppose that $L_{B}$ and $L_{D}$ are the total light in the bulge and 
 disk respectively.
   Then, $B/T$ is given as

\begin{equation}
 B/T=  \frac{L_{B}}{L_{D}+L_{B}},
\end{equation}

\noindent
 The quantity $B/T$ is well correlated with the concentration parameter ($C$), 
 an alternative indicator of the concentration of light that has been used 
 in previous studies of galaxy classification (e.g., Abraham et al. 1996).

  Kent (1985) shows that almost all S0 galaxies in his
 magnitude-limited sample of nearby galaxies have $B/T > 0.3$.
  For E galaxies,  Scorza et al. (1998) find that some of the 28 E galaxies
 in their sample have  $B/T$ as small as  $\sim 0.5$.
  Thus, adopting a $B/T$ cut at around 0.4 is a reasonable choice 
 for selecting E/S0s.
  Also, $B/T$ is useful for excluding   
 dwarf spheroidal or other late-type galaxies that may have a smooth appearance but
 have SB profiles closer to an exponential law 
 (e.g., Im et al. 1995b and references therein).

  However, previous works have found that some
 late-type galaxies can have $B/T > 0.4$ and/or high $C$ values (e.g., Kent 1985).
  For example, about 20--30\% of galaxies in the Kent (1985) sample
 with $B/T > 0.4$ are later than E/S0s.
  Therefore, $B/T$ (or $C$) alone is not sufficient to 
 select E/S0 galaxies, since early-type spirals and late-type galaxies
 with $B/T > 0.4$ would significantly contaminate the sample.
 For this reason we add a second parameter based on galaxy symmetry and smoothness.

\subsection{Residual parameter, $R$}

   Our 2-dimensional 
 surface-brightness fits provide a smooth and symmetric
 best-fit model for each galaxy. Subtracting the model  image
 from the actual image yields a residual image. 
 These residual images contain valuable information on
 galaxy morphology (Naim, Ratnatunga, \& Griffiths 1997; Schade et al. 1995).
 Figure \ref{fig:resimage} presents representative images of E, Sbc, and Im
 galaxies taken from the catalog of nearby galaxies of Frei et al. (1996).
 Corresponding residual images and 1-D surface
 brightness profiles along the major axis are also shown. 
  The residual image of the elliptical galaxy (top) is smooth 
 in comparison to that 
 of the spiral galaxy (middle), which shows spiral structure. The model fit 
 to the Im galaxy (bottom) is poor, and many features  
 can be seen in its residual image.

  To quantify the residual images, we use the residual parameter, $R$,
 as defined by Schade et al. (1995, 1999). 
  The quantity $R$ is called the ``asymmetry parameter'' in
 these papers, but we adopt a different name, ``residual parameter'',
 since  $R$ measures how irregular the galaxy is, how prominent the spiral arms are, 
 and how discrepant the galaxy surface brightness distribution is
 from the simple bulge+disk model, in addition to mere asymmetry.

   $R$ is defined as

\begin{equation}
 R=R_{T} + R_{A}, 
\end{equation} 

\noindent
 with 

\begin{equation}
 R_{T}=\frac{\Sigma\, \frac{1}{2} |R_{ij} + R_{ij}^{180}|}
            {\Sigma\,I_{ij}}
       -\frac{\Sigma\, \frac{1}{2} |B_{ij} + B_{ij}^{180}|}
            {\Sigma\,I_{ij}} \label{eq:rt}, 
\end{equation}

\begin{equation}
 R_{A}=\frac{\Sigma \, \frac{1}{2} |R_{ij} - R_{ij}^{180}|}
            {\Sigma \,I_{ij}}
      -\frac{\Sigma \, \frac{1}{2} |B_{ij} - B_{ij}^{180}|}
            {\Sigma \,I_{ij}} \label{eq:ra}, 
\end{equation}

\noindent
 where $R_{ij}$ is the flux at pixel position $(i,j)$
 in the residual image, $R_{ij}^{180}$
 is the flux at $(i,j)$ in the residual image rotated by 180 degrees, 
 $B_{ij}$ and $B_{ij}^{180}$ are similar quantities measured for 
 background noise, and $I_{ij}$ is the flux at $(i,j)$ in the object image. 
 The sum is over all pixels, and thus each pixel pair appears twice.
  We call $R_{A}$ the ``asymmetric residual parameter'' since it is closely
 related to conventional asymmetry parameters, and we call
 $R_{T}$ the ``total residual parameter'' since it measures 
 the absolute strength of the residuals.
  The second terms in equation (\ref{eq:rt}) and (\ref{eq:ra}) are
 approximate corrections for the contribution from random background noise
 (see Appendix A).
  When noise dominates the image, the first terms in  $R_{T}$ 
 and $R_{A}$ may become quite large since we may be adding up the absolute 
 values of noise, and $R$ may be overestimated.
 The noise is not negligible 
 even when the $S/N$ of the image is high (see also Conselice et al. 2000).  
 Therefore, it is essential to include  
 a background noise correction  in equations (\ref{eq:rt})
 and (\ref{eq:ra}). 

  An approximate expression for the 
 error in $R$ coming from the background noise is derived in Appendix A.
 The result is 

\begin{equation}
 \delta R \simeq 1.7 (S/N)^{-1}.
\end{equation}

\noindent
 This error 
 does not contain contributions due to centroiding errors, which are present
 even at very high $S/N$.
 Conselice et al. (1999) have shown 
 that the rms error in their asymmetry parameter, $A$,
 has a minimum value of $\delta A \sim 0.02$ even with perfect $S/N$.  
  This directly translates to $\delta R \simeq 0.02$.
 Therefore, for the final error in $R$  we adopt the following relation:

\begin{equation}
 \delta R = \sqrt{(0.02)^{2} + (1.7 (S/N)^{-1})^{2}}. \label{eq:dr}
\end{equation}

  In Schade et al. (1995, 1999), $R_{T}$ and $R_{A}$ are calculated within 
 a 5 kpc radius from the center of each galaxy image.
  Calculating $R$ within a fixed physical radius may not
 always be preferred since this would sample different parts of galaxies 
 depending on how extended in physical size they are. 
  To lessen such a problem, we calculate  $R$ within 2 $r_{hl}$.
  This has the additional advantage that we do not need to 
 recalculate $R$ when different cosmological parameters are 
 assumed.  

  As an illustration of the advantage of $R$ over more conventional
 asymmetry parameters for selecting E/S0 galaxies, 
 consider a spiral galaxy with perfectly 
 symmetrical spiral arms.  Conventional asymmetry parameters, computed 
 by subtracting the 180 degree-rotated image from the original image
 (Abraham et al. 1996; Conselice et al. 2000; Wu 2000), 
 would measure the galaxy to be ``symmetric'' with no information 
 whatsoever about spiral arms.
 With $R$, 
 one is able to quantify the prominent features caused by spiral arms
 in the residual image.

\subsection{Test on a local galaxy sample}

   This section demonstrates that $R$ and $B/T$ can be used
 to select E/S0s without substantially contaminating the sample with
 other galaxy types. Figure \ref{fig:resimage}   
 shows $B/T$ and $R$ values below the residual image of each galaxy. 
  The general trend is such that 
 early-type galaxies have small $R$ and large $B/T$, while
 late-type galaxies have large $R$ and small $B/T$. 

  Figure \ref{fig:a-bt} plots $R$ vs. $B/T$ for 
 the subset of 80 galaxies from Frei et al. (1996)
 that have enough background area necessary to determine a
 proper sky background subtraction.
  The Frei et al. sample contains a wide variety of Hubble 
 types and is thus well suited for studying the relation between
 our quantitative morphological selection criteria and visual classifications.

  The red squares  in Figure \ref{fig:a-bt} represent galaxies
 with RC3 type less than or equal to --3 (E or E/S0), green triangles
 represent galaxies with T=--2 (S0), stars are for objects with $-2 < T \leq 0$
 (S0 or S0/S0a), and the black crosses represent other types ($T > 0$). 
  The box drawn in Figure \ref{fig:a-bt} corresponds to the border
 defined by  $R \leq 0.08$
 and $B/T \geq 0.4$. Objects inside or on
 the border of the box are classified as  
 QS-E/S0s (``QS'' denotes ``quantitatively selected''), and 
  we find that nearly all of the selected objects are E/S0s with $T \leq -0$
 (16 out of 17).
  Importantly, there are almost no spiral or peculiar galaxies
 ($T > 0$) in the box (only one object). 
 In fact, the residual parameter ($R$) cut by itself
 can provide a sample dominated by E/S0s.
   Thus, $R$ alone can be used to define {\it local}
 E/S0s, but $B/T$ will become important at high redshift when pixel smoothing
 degrades image quality (see Section 3.5). 

   Note that roughly one-third of the Frei et al. galaxies
 with $T \leq 0$ are missed with these selection criteria.
   The objects omitted are mostly of borderline type, with 
 $-2 < T \leq 0$ (S0, S0/S0a). Some of these could be included
 by loosening the selection criteria, at the expense of contaminating
 QS-E/S0s by non-E/S0s.  
  We have examined images of the missed objects with $T \leq 0$
 and found that several have a non-smooth or irregular 
 appearance, whereas the QS-E/S0s 
 selected using the above criteria
 appear to be truly regular systems.  Adoption of these stricter criteria
 is in keeping with our wish to minimize contamination by blue interlopers.
  On balance, our scheme of selecting E/S0s
 is probably somewhat more conservative than the morphological
 typing of the RC3, and we seem likely to miss about 20--30\% of E/S0s
 with RC type  $T \leq 0$.
  We make use of this fact below to correct the counted numbers of
 E/S0s in local surveys. 
  Effectively, our morphological cut roughly corresponds to $T \simeq -2$, 
 but we expect that more $-2 \lesssim T \lesssim 0$ objects will be
 chosen if we apply our method on noisier images.

\subsection{Tests on ``shrunken'' local galaxy images}

   We have shown in the previous section that our quantitative scheme
 is effective at selecting local E/S0s.
   However, one cannot blindly apply  the 
 above criteria to HST images of faint galaxies
 because of pixel smoothing.  
 As we look at more distant objects, each pixel of a given image 
  samples a larger physical area.
  Because of this  effective smoothing, our $R$ (as well as other asymmetry 
 parameters in the literature) is underestimated
 for galaxies with small apparent size.
  Note that $B/T$ is much less susceptible
 to pixel smoothing since the bulge+disk fit procedure
 incorporates the effects of pixel binning.
 However, it is not entirely free from 
 systematic errors arising from either pixel smoothing or reduced
 $S/N$ (see Simard et al. 2000).

   In order to see how important pixel smoothing and $S/N$ are,
  we block-average images of the same 80 Frei et al. galaxies 
 to simulate galaxy images with different half-light radii of 
 roughly 5, 3, 2, 1.5, and 1 pixel.  Note that the
 apparent sizes of distant galaxies in the GSS sample
 average $r_{hl,med}\sim 5$ pixels at the magnitude limit of
 $I \simeq 22$, and
 nearly all are larger than 3 pixels, so these tests are conservative
 (see Figure 6; also Simard et al. 2000).
  Background noise is added to the resultant image
 so that the $S/N$ of each simulated image is $S/N \simeq 40$--80,
 comparable to GSS galaxies with $I=21$--$22$ (see Figure \ref{fig:sn}).
  Sample postage stamp images 
 of simulated galaxies are available in Appendix B, along with 
 their morphological parameters.
   In Figure \ref{fig:rrcom}, we show the input values of $R$ (as derived from 
 the simulated Frei et al. images with $r_{hl}=5$ pix) vs. output $R$ for the simulated galaxies.
  Note that output $r_{hl}$ values from GIM2D are close to input values with 
 a random error of order of $\delta r_{hl}/r_{hl} \sim 0.08$,
 and a slight systematic bias ($\sim 10$\% or
 more) for underestimating sizes when input sizes are $\lesssim 2$ pixels.
  What we actually have for GSS galaxies are ``output'' values, and we find 
 that simulated Frei galaxies with input $r_{hl}
 \simeq 1.5$ pixels have output $r_{hl}$ of $r_{hl,output} \simeq 1.12$ pixels.
 Similarly, for  $r_{hl,input} \simeq 1$ pixels, we find $r_{hl,output} \simeq 0.72$
 pixels, and for $r_{hl,input} \simeq 2$ pixels, $r_{hl,output} \simeq 1.7$ pixels.

  Importantly, output $R$ does not change significantly from the input 
 value (20\% or less) when sizes of galaxies are sufficiently
 large ($r_{hl} \gtrsim 2$ pixels).
  When galaxy sizes become smaller than 1.5--2 pixels, the global
 shift of $R$ is significant (30\%).  

 Figure \ref{fig:btcom} is a similar comparison of output $B/T$ vs.  
 input $B/T$.  
  When apparent sizes are very small ($r_{hl} \lesssim 1.5$ pixels),
 $B/T$ values are again poorly determined. 
  Simard et al. (2000) perform more extensive tests 
 using artificially constructed galaxies with various 
 $B/T$ values and find a similar result.
  However, only a small fraction of GSS galaxies in
  our sample have $r_{hl} < 2$ pixels; hence the effect of 
 pixel smoothing on both the $R$ and $B/T$ cuts 
 should be small.  This is shown explicitly below.

  Figure \ref{fig:rall} shows the $R$-$B/T$ diagram for 
Frei et al. simulated galaxies with 
 output $r_{hl} \simeq$ 5, 3, 1.7, and 1.0 pixels. 
 The selection
 criteria shown on Fig. 5 ($R \leq 0.08$ and $B/T > 0.4$) select RC3 type 
 E/S0s fairly well, although some E/S0s, especially with $T > -2$,  
 are missed.  
 However, the number of QS-E/S0s increases from 15 for $r_{hl}=5$ pixels to
 19 for $r_{hl}=1$ pixel,   indicating that contamination
 from spiral galaxies becomes more important as galaxy size decreases.
 In order to compensate for this,
 we can try lowering the upper limit of the $R$ cut for smaller galaxies.
 Adopting $R=0.07$ for $2 < r_{hl} \leq 3$ pixels,
 $R=0.06$ for $1 < r_{hl} \leq 2$ pixels, and $R=0.05$ for $r_{hl} \leq 1$ pixel
 is found to  yield roughly stable numbers of E/S0s over the 
 whole range of galaxy sizes in the simulated 
 Frei et al. sample.
 However, we stress that the great majority ($ > 93$ \%)
 of GSS galaxies with $I < 22$ have
 $r_{hl} > 2.5$ pixels, as shown in Figure \ref{fig:smag}, next section;
 thus any such reduction in $R$ cut for small galaxies would not come
 into play for many objects.

\subsection{Final selection of Groth Strip QS-E/S0s}

   Using local galaxy images (Figure \ref{fig:resimage}),
  we have found that a constant boundary of 
  $R \leq 0.08$ and $B/T > 0.4$ selects E and S0 galaxies quite well
  without contaminating the sample with later galaxy types.  
   However,  when the object sizes
  are small ($r_{hl} \leq 2$ pixels), pixel smoothing starts to wash away
  detailed morphological 
  features, causing an underestimate of $R$ values and a consequent
  overinclusion of galaxies.
   Also, as implied in equation (\ref{eq:dr}), 
  errors in $R$ increase as $S/N$ becomes smaller 
  ($S/N \lesssim 50$).
These effects are dealt with in this section.

   Figures \ref{fig:smag} and \ref{fig:sn} 
 show  $r_{hl}$ vs. $I$  and $S/N$ vs. $I$ for galaxies in
 the GSS. Here, $S/N$ is defined as the $S/N$ within 
 one $r_{hl}$ radius 
of the object.  The number of galaxies with $r_{hl} < 3$ pixels
 is not large but is not completely negligible; therefore we need to 
 take into account the effect of pixel smoothing on $R$.
 Likewise, the median $S/N$  
 approaches $S/N \sim 30$ at our lower magnitude
 limit of $I=22$.   Using equation (5), we get an rms uncertainty $\delta R 
 \simeq 0.06$ at this brightness level, 
 which is significantly 
 larger than our rough estimate of the minimum scatter
 due to centroiding errors ($\delta R \sim 0.02$). 
 Thus the effect of $S/N$ decrease on $R$ needs to be considered as well. 
  After experimentation, we have adopted the final $R$ cuts shown
  in Table 1, which are a function of both $r_{hl}$ and magnitude.
  Tests below suggest that these $R$ cuts, which are
 rather stringent, may be dropping
 some E/S0s at the faintest and smallest levels.  
  However, we have retained them because they 
 efficiently  reduce the number of spurious
 blue interlopers while keeping the number of red E/S0s  
 fairly close to intact.
 The efficacy of the adopted  $R$ cuts is examined next.

  E/S0s are the reddest galaxies in the local universe.
  If our selection method is good at identifying E/S0s and if 
 these galaxies remain red at recent epochs
 (as will happen, for example, if the bulk of star formation occurs
 at $z > 1$),
 we expect our selected QS-E/S0s to populate a tight red envelope
 in the redshift-color diagram.
  Reassuringly, nearly all the objects selected this way 
  are indeed the reddest galaxies at each redshift, as shown in 
  Figure \ref{fig:i22sel}.
 This figure shows all 262 
 GSS galaxies with $z_{spec}$ at $16.5 < I < 22$.
 Circle size is proportional to 
 brightness, with the largest circles representing
 galaxies with $I < 20$, mid-sized circles objects with $20 < I < 21$,
 and smallest circles objects with $21 < I < 22$.
  Dashed lines indicate plausible ranges for the color of a passively evolving 
 stellar population formed at very early times. The upper dashed line represents a  
 0.1-Gyr burst model with 2.5 times solar metallicity, a Salpeter IMF (0.1 to 
 135 $M_{\odot}$), and $z_{for}=11$. The lower dashed line represents
 the same 0.1-Gyr model but with 0.4 times solar metallicity.
  To allow for color errors, we have added or subtracted
 0.15 mag in $V-I$ to and from the upper and lower dashed lines.
 
  Panel b) of Figure  \ref{fig:i22sel} shows the 84 galaxies 
 in the spectroscopic training sample that 
  satisfy the $R$ cut. 
  The great majority fall within
  the plausible color range of the passively evolving stellar population
  (51 out of 84 galaxies).   By the same token,
  only 33 out of 186 galaxies outside the red color boundaries 
 in panel a) are selected as low-$R$ galaxies.  
  Figure \ref{fig:i22sel}c likewise shows the 77 galaxies 
 that satisfy  $B/T > 0.4$. 
  Most of these again
  turn out to lie within the red color boundaries
 (50 out of 77 objects), and only a small fraction of blue 
 objects below the red boundary have high $B/T$ (27 out of
 186 galaxies).
 
  Finally, Figure \ref{fig:i22sel}d presents the 
 44 galaxies that satisfy both the $R$ and $B/T$ cuts.
  In addition to the previous dashed lines, we also plot 
 a solar-metallicity model with three different formation redshifts
 (11, 2, and 1.5).
  Now, only 6 of 44 selected objects (15\%) lie 
 outside the red color boundaries; these are the ``blue interlopers''
 referred to previously (filled circles).
  The remainder of the sample is found to follow the expected redshift-($V-I$)
 tracks of passively evolving stars.
  The nature of the blue interlopers is intriguing.
  Preliminary analysis of their spectra (Im et al., in preparation)
shows that most have strong,
narrow emission lines, suggesting that they are low-mass 
starbursts rather than  massive star-forming E/S0s (Im et al. 2000, in preparation);
they may be similar to the Compact Narrow Emission Line Galaxies
 (CNELGs; Koo et al. 1994; Guzman et al. 1997; Phillips et al. 1997).

  We now vary the  selection rules to see how the final sample depends on the
 precise criteria used.
  Figure \ref{fig:i22check} shows $V-I$ vs. redshift for samples
 selected based on $R$ and $B/T$ cuts that are slightly different from
 those adopted in Figure \ref{fig:i22sel}.
  The two figures on the left (panels a and c) show the results
 of loosening the cuts. The number of blue interlopers
 significantly increases, from 6 in Figure  \ref{fig:i22sel}d 
 to 11--12 here, while the number of 
 selected red QS-E/S0s increases by only 1--4.
  The two panels on the right (b and d) show the effect of stricter cuts.
  The number of blue interlopers 
 is not significantly reduced (4 here vs. 6 formerly), while
  the number of desirable red QS-E/S0s is decreased significantly,
 by 7--10 objects.
  Thus, the original cuts seem about optimal.

 Interestingly, there may be a bimodality in the colors of GSS
 galaxies such that the color distribution at a given redshift
 is double-humped.  A hint of this is seen in Figure \ref{fig:i22sel}
 and has been remarked on previously (Koo et al. 1996) (but
 note that no such feature is seen in CFRS data; Lilly et al. 1995).
 Such a hump would clearly help in the selection of red E/S0s; 
 setting criteria to cut in the valley would mean that object
 selection would be less sensitive to slight changes in the selection criteria.
 This approach can be tried in future if color-redshift surveys
 confirm the presence of the double humps.

   We next discuss likely systematic errors in the counted
  numbers of E/S0 galaxies at high redshift.
  Figure \ref{fig:eso}  shows images of all selected
  GSS QS-E/S0s with $z_{spec}$,
  ordered by redshift.  
    Blue interlopers (objects lying outside the red bands of 
  Figure \ref{fig:i22sel}) are separately presented at the end of the sequence.
  The pixel values of the galaxy images are roughly square-rooted 
 (more exactly, rescaled by the $\frac{1}{2.2}$th power of their
 values, as used in Frei et al. (1996)). 
  We find this scaling to be effective in bringing up
 faint details at low surface brightness, while making
 the bulge component look reasonably distinct when $B/T > 0.4$.    
  However, for eyes accustomed to looking at the linearly scaled
 images of most astronomical atlases, the scaling used here
 might make $B/T \sim 0.5$ objects appear rather 
 disk-dominated. 
  To avoid this confusion, we add similarly scaled images
 of local E/S0 galaxies from Frei et al. (1996) in the two bottom 
 rows of Figure \ref{fig:eso}.   The visual comparison
 between the local E/S0s and the Groth Strip QS-E/S0s confirms that the latter
 truly resemble the appearance of local E/S0s.
 Thus, aside from the blue interlopers (15\%), which will all appear at
 low $z$s from their photometric redshifts, 
the present classification scheme
 admits at most few additional 
spurious spirals and peculiar galaxies and is thus
 not likely to {\it over}estimate the number of distant E/S0s
 by even a small percentage.

  For the reverse comparison,  Figure \ref{fig:redim} shows
 images of 64 galaxies with $z_{spec}$ that lie within or close
 to the red color boundaries of Figure  \ref{fig:i22sel}
 but that do {\it not} meet
 the $R$ or $B/T$ cuts.
 Such objects could be real E/S0s that are improperly being lost.
  Comparison of this figure with Figure \ref{fig:eso}
 shows that red non-selected E/S0s 
 actually have a much higher frequency of 
 non-smooth morphological features
 (e.g., spiral arms, asymmetric nuclei), which are not apparent in 
  selected QS-E/S0s.
  Many also turn out to be 
 edge-on galaxies with low $R$ but also with low $B/T$.
  However, a significant number of the non-selected galaxies are
 indistinguishable visually from  the QS-E/S0s of Figure \ref{fig:eso}.
 We find about 10 such objects in Figure  \ref{fig:redim},
 of which 9 lie beyond $z = 0.6$.
 If these are truly E/S0s, they should be added to the 24
 QS-E/S0s in that redshift range from Figure \ref{fig:eso},
 which would mean that our numbers of high-redshift E/S0s are
 $\sim 35$\% too low.  These objects might overlap at least in part
 with the borderline S0-S0/a's missed in the test of local objects using the
 Frei et al. catalog  in Figure  \ref{fig:a-bt}.

  Clearly, resolution and $S/N$ effects can work both for or against selecting 
 E/S0s,  but Figure \ref{fig:redim} suggests 
 that, in our method, 
 they seem to work {\it against } 
 picking E/S0 galaxies at $z > 0.6$
 but do not seem to affect the low-redshift E/S0 selection very much.  
  Only  a 
 few of $z < 0.6$ red non-E/S0 galaxies in Figure \ref{fig:redim} would 
 resemble QS-E/S0s at $z > 0.6$ if their $V$-band images were reprocessed to
 appear  like $z > 0.6$ galaxies.  The great majority ($\sim 50$ \%) 
 of red non-E/S0 galaxies at $z < 0.6$ are edge-on galaxies with
 negligible bulge component ($B/T \lesssim 0.2$), while only 
 $\sim 25$\% of red, non-E/S0 galaxies at $z > 0.6$ are in such category.
 This supports the above idea, and also suggests that 
 red galaxies at lower redshifts can be more easily contaminated 
 by dust-extinguished edge-on disks than red galaxies at higher redshifts. 
  Again, the point is to establish that our counts are not likely to 
 {\it under}estimate the number of distant E/S0 galaxies by more than
 $\sim 35$\%.

   Finally, we note that our measured values of $R$ and $B/T$ are derived from
 the observed $I$-band image, which, for our sample at $z \sim 0.8$,
 corresponds to a rest-frame $B$ band. Since our local galaxy comparison
 sample is observed in the $B$ band,
 the morphological K-correction should be minimal
 when $0.6 < z < 1.2$.  For galaxies at $z < 0.6$, however, this could bias 
 object selection because $R$ and $B/T$ will be estimated at 
 redder rest-frame wavelengths. Since bulges are redder than disks
 and localized star formation is less prominent at redder 
 rest-frame wavelengths, the expectation is that $B/T$ would 
 be overestimated and $R$ would be underestimated when measured in rest-frame
 $V$ rather than $B$, and that consequently more objects would be selected  
 as E/S0s at $z < 0.5$.
  To test this, we have compared $B/T$ and $R$ values measured 
 in $V$ vs. $I$ and find there is no strong difference as long
 as both $V$ and $I$ sample light above rest-frame 
 4000 $\AA$. Re-selection of the sample at $z < 0.5$ 
 using $V$-band images rather than $I$-band shows further that 
 a $V$-band selected sample would be 
almost identical to the $I$-band E/S0 sample.
  We have attempted to estimate the rest-frame $B$-band $B/T$ by applying 
 K-corrections from Gronwall \& Koo (1996) and Gebhardt et al. (2000) 
 separately to bulge and disk components, and confirm  
 the above claim that rest-frame $B$-band $B/T$ is nearly identical
 whether estimated from observed $V$ or $I$.
  Therefore, we believe that morphological $K$-correction is not an important
 issue here.

\subsection{Selection of Groth Strip QS-E/S0s without spectroscopic redshifts}

  Our spectroscopic observations do not cover the entire Groth Strip,
 so we can more than triple the sample size by estimating 
 redshifts photometrically for galaxies 
 in regions where spectroscopic data are not available.
  For the GSS as a whole, the only photometric information we can use are 
 $I$, $V$, and $(V-I)$. Due to the wide range of 
 color space spanned by various types of galaxies,  it is not
 feasible to estimate  
 redshifts for all types of galaxies in the GSS using
 this limited photometric information.
   Nevertheless, it is possible to get reliable redshifts
 with $(V-I)$ only if we focus on a sample of E/S0s
 {\it preselected by morphology}, 
 since the previous analysis has shown that, for them,
 $(V-I)$ color and redshift are very well correlated (cf. Figure  \ref{fig:eso}).
   The correlation is virtually perfect at $I < 21$, where the $R$ and $B/T$ cuts
 select E/S0s with a tight $z_{spec}$--$(V-I)$ relation.   
   However,  at $I > 21$, blue interlopers make 
 photometric estimate of redshifts more challenging.
   In order to exploit the tight $z_{spec}$--$(V-I)$ relation 
 for accurate photometric redshifts,  we 
 exclude blue interlopers from consideration
 and use the remaining QS-E/S0s to fit $z_{spec}$ with 
 $V-I$ and $I$ polynomials.  We obtain the following relation:

\begin{eqnarray}
z_{phot} & = & A_{1}+A_{2} \,I+A_{3} \, I^{2}+A_{4} \, I^{3}+A_{5}\,I \times (V-I)+ \nonumber \\
         &   & +A_{6} \, (V-I)+A_{7}\,(V-I)^{2}+A_{8}\,(V-I)^{3},
\label{eq:zphot}
\end{eqnarray}

\noindent
 where the coefficients are $A_{1}=-2.7872 \times 10^{-2}$,
 $A_{2}=-1.7700 \times 10^{-1}$, $A_{3}=1.6784 \times 10^{-2}$, 
 $A_{4}=4.3770 \times 10^{-4}$, $A_{5}=2.7295 \times 10^{-2}$, 
 $A_{6}=2.0530 \times 10^{-3}$, $A_{7}=2.1158 \times 10^{-2}$, 
 and $A_{8}=1.6606 \times 10^{-3}$. The quantity 
 $z_{phot}$  estimated with equation (\ref{eq:zphot}) appears
 to underestimate systematically 
 the true redshift by a small amount at $z \gtrsim 0.8$.
 For that reason we make the following small correction 
 when $z_{phot} > 0.8$:

\begin{equation}
 z_{corrected} = 1.333 (z - 0.8) + 0.8.
\end{equation}

\noindent
Including terms in $I$ as well as $V-I$ in equation (\ref{eq:zphot})
reduces the residuals by about 10\%.

   Figure \ref{fig:zspeczphot} compares 
 $z_{spec}$ vs. $z_{phot}$ for the GSS QS-E/S0s having spectroscopic redshifts
(blue interlopers excluded).
   The RMS of $z_{phot}$ vs. $z_{spec}$ is about 10\% but increases at the 
 highest redshifts, as in  Figure \ref{fig:dz}.  
   The lines there indicate the rms of 
 $\Delta z \equiv z_{spec} - z_{phot}$ vs. $z_{spec}$, 
 and we adopt this as the error of $z_{phot}$.
   This error envelope will be used later in the estimate of 
 luminosity function parameters and in tests with Monte-Carlo simulations for 
 checking Malmquist-like bias.
   The error increases beyond $z > 0.8$ due to the fact that
 the main $z_{phot}$ indicator---the $4000~\AA$ break in the continuum of 
 the spectral energy distribution---passes through the F814W passband. 
   However, the combination of the following two facts
 makes $V-I$ and $I$ together useful for estimating redshifts 
 to reasonable accuracy ($<$ 15\%) even at $z > 0.8$.
   First, the observed $z$ vs. $V-I$ relation is not completely
 flat beyond $z > 0.8$,
 contrary to the predictions of the passive evolution models plotted 
 in Figure \ref{fig:i22sel}.
  The color-magnitude relation is at least partly responsible
 for this---the magnitude and redshift limits we adopt make
 only the intrinsically brightest, and thus the 
 reddest, objects detectable.  This effect acts to increase
 the average color vs.~ redshift, even when the color of any given
 galaxy would remain flat.  
   Second is the familiar fact that, at fixed intrinsic $V-I$, 
dimmer-appearing  galaxies are farther away.  Thus apparent
magnitude is by itself an indicator of redshift, independent
of color. 
  The fit at $z > 0.8$ is based on  
 more than 15 E/S0s with known $z_{spec}$ in this range;
 therefore, our $z_{phot}$ can be considered reliable within 
 the estimated errors even at $0.8 < z < 1.2$. 

   At low redshift ($z < 0.1$), there is a second concern that 
 small errors in photometric redshift (e.g., $\delta z =0.05$) 
 lead to large errors in 
 absolute magnitude.   The errors shown 
 in Figure \ref{fig:dz} imply
 that redshift errors remain fractionally small (10\%) 
 even at very low redshift.  
  However, in practice the
 errors are poorly known below  $z \lesssim 0.3$ since 
 there are only two galaxies in this redshift
 range. To check for a potential bias due to the effect of
 low-redshift errors, we repeat
 the  LF analysis below, increasing the lower redshift cut to $z=0.2$,
 and show that this has little effect.

  A cautionary remark must be made regarding 
 the photometric redshifts
 of blue interlopers. Since the number of blue interlopers
 is small ($<$15\% of QS-E/S0s), we do not try to exclude them 
 from the sample using additional color cuts. However,
 redshifts for the blue interlopers are underestimated using
 equation (\ref{eq:zphot}).  Fortunately, with these
 redshifts, blue interlopers tend to
 be the {\it faintest} QS-E/S0s at a given $V-I$ color,
 and thus they influence only the
 faintest part of the luminosity function at low redshift. 
   Figure \ref{fig:ccut}, which plots $I$ vs. $V-I$ for various samples,
 sheds further insight into the number of blue interlopers.
   The squares refer to the $z_{spec}$ sample. 
  Thick squares show the red QS-E/S0s,  while thin squares indicate 
 blue interlopers as defined previously in Figure \ref{fig:i22sel}d.
  The lines represent the color-magnitude relation for  
 a passively evolving elliptical with $M_{B}=-18.3$ and $M_{B}=-20.14$
 ($L^{*}$), assuming $q_{0}=0.1$, $h=0.7$,
 solar metallicity, Salpeter IMF, and $z_{for} = 5$. The majority of 
 red QS-E/S0s lie above this line, while all but one blue interloper
 in the $z_{spec}$ sample lie below the line. 
   The circled points denote the additional galaxies in the
 $z_{phot}$ sample.  As no independent
  redshifts are available for them, we do not have firm knowledge
  of which ones are blue interlopers.  
  However, the circles lying below
  the line are candidate blue interlopers
  according to $z_{phot}$; there are 14 of these, among 101 objects,
  similar to the 6 interlopers out of 44 objects in
  the $z_{spec}$ sample.  
  Redshifts of both kinds 
  of interlopers are likely to be severely underestimated,
 and we find that all of them have $z_{phot} < 0.42$ and $M_{B} > -18.2$. 
  Thus, the blue interlopers probably overestimate the faint end of
 the LF at low redshift.

  Figure \ref{fig:esop} shows images of 98 out of these 
101 QS-E/S0s in the range
 $16.5 < I < 22$  in the $z_{phot}$ sample, including blue interloper candidates.
  All red QS-E/S0s are presented, 
 and 11 out of the 14 candidate blue interlopers are shown at the end of the figure.
  We can again inspect the images of these objects as a sanity check
 for spurious late-type galaxies 
 and find that contamination by late-types and peculiars is 
 very small; by eye, only 2 out 
 of 98 galaxies look mis-selected.  Thus, aside from blue interlopers,
 the likely {\it over}estimate of distant E/S0 
 galaxies is again very small, even in the $z_{phot}$ sample.  
 (The opposite test of looking for objects missed 
 among ``red'' galaxies, which 
 we performed for the $z_{spec}$
 sample, is impossible here because it 
 requires a spectroscopic redshift to define a ``red'' galaxy.)
  The ellipticity distribution of the GSS QS-E/S0s is presented in the next section, 
 which further shows that they are similar to local E/S0s.  
  With the addition of the $z_{zphot}$ sample, we have 
 a final sample of 145 QS-E/S0s at $16.5 < I < 22.0$. Information on 
 these QS-E/S0s is listed in Table 2. 

  Figures \ref{fig:zi_zphot} and \ref{fig:zab_zphot} show
 redshift vs.~ $I_{tot}$ 
 and redshift vs.~ $M_{B}$ (K-corrected only) for all GSS galaxies. 
  QS-E/S0s are plotted as squares, and blue interlopers are plotted with 
 triangles. Thick symbols represent the spectroscopic sample, and
 thin symbols represent the photometric redshift sample.  Small crosses in
 Figure \ref{fig:zi_zphot} are the remaining galaxies (non-E/S0s) 
 with spectroscopic redshifts. 
  Also plotted in Figure \ref{fig:zi_zphot} are lines of three 
 different values of constant
 $M_{B}$, with (solid line) and without (dashed line) luminosity evolution.
 Note that $M_{B}=-20.14$ represents the $L^{*}$ of local E/S0s
 according to Marzke et al. (1998).
  The parameters for the open universe are adopted, 
 and for luminosity evolution
 we assume $E(z)=1.7 \times z$,a
s derived from our LF analysis for the open universe
 (see next section).  In Figure \ref{fig:zab_zphot}, we plot only lines for
 $M_{B} = -20.14$.

  A striking feature in Figures \ref{fig:zi_zphot} and 
 \ref{fig:zab_zphot} is that there seem 
 to be too many $L > L^{*}$ E/S0s beyond 
 $z > 0.6$ if the no-evolution line is used as a reference. This
 overabundance is not observed when we 
 count the number of $L > L^{*}$ 
 E/S0s with respect to the evoliving-luminosity line,
 and this can be considered a qualitative indicator of luminosity evolution. 
  In the analysis of the LF below, 
 we will quantify the amount of luminosity and
 number density evolution in detail. 
 A second important feature is the apparent 
 lack of $L > L^{*}$ E/S0s at $z < 0.2$, but this can be attributed to
 the bright magnitude limit 
 we adopted ($I=16.5$, see Figure \ref{fig:zi_zphot}). 
  In the LF analysis below, we adopt a default 
 redshift range of $0.05 < z < 1.2$, 
 despite the fact that few $L > L^{*}$ galaxies are seen at $z < 0.2$.
  The techniques we use are adaptive enough to adjust for this, but
 just to check,  
 we try increasing the lower redshift cutoff 
 and confirm that there is little
 effect.  A more serious deficiency of galaxies might also exist 
  at $z > 1.0$, but one that cannot be explained simply by magnitude limits.
  We will again vary the upper redshift cutoff and 
 find that our results are slightly more sensitive to
 this upper cut.

\subsection{Ellipticity distribution}

   The ellipticity distribution of elliptical galaxies is known to be quite
 different from that of spiral and   S0 galaxies,
 and thus can be used as yet another independent check on the
 Hubble types of the selected sample.
   Most ellipticals look  round or football-shaped with 
 modest ellipticities, or equivalently, large axis ratios.
   On the other hand, the ellipticity distribution of
 S0 galaxies is peaked at $\epsilon \sim 0.5$, while that of spirals 
 is nearly flat at almost all ellipticities 
(Sandage, Freeman, \& Stokes 1970; van den Bergh 1990; Fasano \& Vio 1991; Franx, 
 Illingworth, \& de Zeeuw 1991; Lambas, Maddox, \& Loveday 1992;
 J{\o}rgensen \& Franx 1994; Andreon et al. 1996; Dressler et al. 1997). 

   Figure \ref{fig:abtotal},
 presents the ellipticity distribution of the 145 GSS QS-E/S0s with
 $16.5 < I < 22$ as the thick histogram ($z_{phot}$ sample included).
   Ellipticities are measured at the $2~r_{e}$ isophote;
   they typically increase slightly with radius, but the increase
 is not large beyond $r \gtrsim 2~r_{e}$.
   Also plotted are the ellipticity distributions of local Es (dashed line) and S0s 
 (dotted line) in nearby clusters, taken from Dressler (1980).
   The thin line shows the combined nearby E and S0 ellipticity
 distributions for a model with a relative S0 fraction of 40\%.
 As a reference, we also plot an ellipticity 
 distribution of spirals with a dot-dashed line (Lambas et al. 1992).
   
   Since the ellipticity distributions of Es and S0s are distinctive, 
 it is worthwhile trying to separate the two types of galaxies in
 the GSS sample.   Since we do not have a reliable 
 scheme to distinguish Es from S0s quantitatively, 
 we instead divide the sample above and below $M_{B} < -20$.
   According to the LFs of local 
 field E/S0s (Marzke et al. 1994) and
 cluster E/S0s (e.g., Dressler et al. 1980),
 Es are more abundant than S0s at $M_{B} \lesssim -20$,
 while S0s are more abundant than E's at $M_{B} \gtrsim -20$.

   Figures \ref{fig:abbright} and \ref{fig:abfaint} show 
 the ellipticity distribution of Groth Strip QS-E/S0s divided this way.
  To estimate the absolute magnitude of each object, we assume an open universe
 with $\Omega_{m}=0.2$ and $h=0.7$; passive luminosity evolution (as 
 derived from the luminosity function in the next 
 section) and the $K$-correction are both taken into account.
   The resultant ellipticity distribution of luminous
 Groth Strip QS-E/S0s is well fitted with a model distribution 
 dominated by  Es, amounting to $\sim80$\% of the sample,
 while the ellipticity distribution of the faint sample resembles  
 a model distribution dominated by S0s, amounting to
 $\sim70$\%.
   Thus, the combined ellipticity distribution of local Es and S0s reproduces
 that of the Groth Strip QS-E/S0s fairly well,  and neither the ellipticity
 distribution of local Es alone nor that of S0s alone is 
 a good fit. Moreover, neither bright nor
  faint Groth Strip QS-E/S0s are consistent with
 the ellipticity distribution of local {\it spirals};
 late-type galaxies do not therefore appear to 
 contaminate our sample significantly.

\subsection{Comparison with previous studies}
 
  Studies of early-type galaxies by Brinchmann et al. (1998) and 
  Schade et al. (1999) have included objects from a part of the GSS.
  Using these overlaps, we compare our selection criteria with theirs.
  Brinchmann et al. (1998) use the AC-system, which originates from 
 Abraham et al. (1996). The AC-system classifies galaxies based upon
 their asymmetry (A) and concentration parameter (C). 
  According to this scheme,  early-type galaxies are selected 
 largely based on the concentration parameter, $C$, which correlates well with
 $B/T$.   As expected from Figure 5, we find that the AC method tends to include
 some later-type galaxies with large $C$ (or $B/T$).  Specifically, 
 of the 14 AC-classified 
 early-type galaxies in the Groth Strip, 9 are classified as
 QS-E/S0s by us, while 5 of them are classified as non-QS-E/S0s (for example,
 074\_2237 and 073\_3539 in Figure  \ref{fig:redim}). 
 In contrast, none of our Groth Strip QS-E/S0s 
 are classified as non-E/S0s by Brinchmann et al. (1998). 
  Thus, we conclude that our method
 is somewhat more conservative than the AC method
 in picking up only morphologically featureless E/S0s.

  Schade et al. (1999) compiled a list of elliptical galaxies 
 at $z < 1$. Their criteria are that the galaxy should have a 
 $r^{\frac{1}{4}}$ law-dominated profile and $R < 0.1$, slightly
 looser than the $R \lesssim 0.08$ criterion adopted here.  Eight ellipticals
 from Schade et al. (1999) are found in the our sample.
  Of these, 5 are classified by us 
 as QS-E/S0s, while one more is 
 a blue interloper. This shows again that our object selection criteria
 are more conservative than those of Schade et al. (1999),
 as expected from the tighter $R$ cut.
  The Schade et al. sample also has a 
 much larger scatter in $(V-I)$  vs. $z$ than ours, probably  
 due in part to the looser $R$ cut
 and possibly also to larger errors in their 
 ground-based photometry.

\subsection{Selection errors and biases for the distant sample}

  This section summarizes previous discussions of 
the selection effects and adds some new tests 
to produce an overall estimate of 
count uncertainties.  Here we consider internal errors in our own
counts only---errors in matching to local E/S0 counts are considered
in Section 3.

 Here is a summary of the tests:
 1) Varying $R$ and $B/T$ thresholds within plausible limits
 modulates the absolute number of counted galaxies by $\pm$25\%.
 2) Varying the thresholds as a function of galaxy size 
 affects the counts of small galaxies.  The finally adopted, 
 conservative thresholds lose about 20\% of Frei et al. simulated
 E/S0s at the smallest radii ($\lesssim$2 px), which affects
 low-luminosity and distant galaxies the most.
  3) Inspecting the $z_{spec}$ sample visually for all conceivable
 {\it interlopers} reveals no new objects other than the known blue
 interlopers (6 out of 20 objects below $z < 0.6$), and 
 the $z_{phot}$ sample gives essentially identical results.  Thus,
 the counts beyond $z = 0.6$ are unlikely to be biased too high
 by inclusion of late-type or peculiar objects, while the nearby counts
 below $z = 0.6$
 may be roughly 30\% too high due to the inclusion of blue interlopers
 in low-luminosity bins.
 4) Inspecting the $z_{spec}$ sample visually for all 
conceivable {\it omitted} E/S0s reveals only one or two new galaxies
that might be added to the 14 existing galaxies
at low redshift ($z < 0.6$),
but turns up an additional 9 objects that might plausibly be added
to the 24 existing objects at $z > 0.6$.  
Thus, the counts beyond $z = 0.6$ may be biased too low
by about 30\% due to omission of valid E/S0 galaxies, while the nearby counts
do not appear to be missing such candidates.

 The overall conclusion from these tests is that the counts are uncertain
at the 30\% level; they may be biased a little too high 
by this amount at faint absolute magnitudes 
for nearby redshifts ($z < 0.6$), and a little too low by a similar amount
at all absolute magnitudes for distant redshifts ($z > 0.6$).

\section{Volume density and luminosity evolution of Groth Strip QS-E/S0's}

  In this section, we construct the luminosity function (LF)  
  of distant Groth Strip QS-E/S0s and derive constraints on their
 number density and luminosity evolution. 
 The entire sample including 
 $z_{phot}$ is used unless otherwise noted.

\subsection{Luminosity function}

 To derive the evolution in luminosity and number density from the LF,
 we adopt two different approaches. One is to derive the LF parameters
 of the sample divided into two different redshift intervals
 ($0.05 < z < 0.6$, and $0.6 < z < 1.2$); evolution is then 
 measured  by comparing low- and high-redshift LFs
 to one another and to the local LF.  A second
 approach 
 introduces parameters for evolution and fits the (evolving) LF 
 for the whole sample simultaneously. For the first method (Method 1),
 we use the $1/V_{max}$ technique (Schmidt 1968;
 Felten 1976; Huchra \& Sargent 1973; Lilly et al. 1995a)
 to estimate the LF in magnitude bins at low and high redshift, and then
 apply the method of Sandage, Tammann, and Yahil (STY, 1979) to estimate 
 the LF parameters (with normalization provided by yet a third method).
 Method 1 is identical to 
 the approach adopted by Im et al. (1996).
  For the second method (Method 2), we follow an approach similar to that
 of Lin et al. (1999),
 in which luminosity  and  number density evolution
 are each parameterized versus redshift, and these parameters are then
 solved for together with other LF parameters using the whole
 sample simultaneously.

\subsection{Method 1}

\subsubsection{$1/V_{max}$ method}
   In the $1/V_{max}$ method, each galaxy in the sample is assigned 
 a $1/V_{max}$ value, where $V_{max}$ is the maximum volume within which 
 the galaxy would be observable under all relevant observational 
 constraints including magnitude and redshift limits.
  The quantity $V_{max}$ is calculated as

\begin{equation}
 V_{max}=\int^{min(z_{2},z_{m2})}_{max(z_{1},z_{m1})} (dV/dz)\,dz, 
 \label{eq:vmax}
\end{equation}

\noindent
  where $z_{1}$ and $z_{2}$ are the lower and upper limits
 of the redshift interval for which the LF is being
 calculated, $m_{1}$ and $m_{2}$ are the 
 apparent magnitude limits of the survey,
 $z_{m1}$ and $z_{m2}$ are redshifts where the galaxy would be located if
 it had apparent magnitude $m_{1}$ and $m_{2}$ respectively,
 and $dV/dz$ is the
 comoving volume element per unit redshift interval.

  An absolute magnitude of each galaxy in the F814W passband ($M_{I}$) 
 is calculated as 

\begin{equation}
  M_{I}=m_{I} - 5\,{\rm log}_{10}(d_{L}(z)) - 25 - K_{I}(z), 
  \label{eq:mi}
\end{equation}

\noindent
  where $d_{L}(z)$ is the luminosity distance in Mpc
 and $m_{I}$ is the apparent  magnitude of the galaxy. 
  For the $K$-correction, we use the present-day model SED  which 
 was used in the color cut in Section 3.6 (i.e., the BC96 model with
 0.1 Gyr burst, $z_{for}=5$, solar metallicity, and Salpeter IMF).
  This $K$-correction 
 is very similar to the $K$-correction used for Es in Gronwall \& Koo
 (1995), the standard set of $K$-corrections used in previous 
 DEEP publications. The difference between the two is roughly $0.08$ mag 
 at $z > 0.5$ and $\sim 0.3$ mag at $z \sim 0.2$, with the adopted 
 K-correction underestimating the luminosity in both cases.
  To obtain the rest-frame $M_{B}$ magnitude,
 we add 2.17 mag to $M_{I}$ since the $(B-I)$ color of the model E/S0 
 SED at $z=0$ is 2.17. 

  Our sample of QS-E/S0s is magnitude-limited at 
 $16.5 < I < 22$, as described in the previous section.
  Since the number of E/S0s with $z \gtrsim 1$ is 
 small with a faint apparent magnitude limit of $I = 22$
 (See Figure \ref{fig:zi_zphot}), and  
 since the accuracy of $z_{phot}$ rapidly drops due to $V-I$ 
 color degeneracy beyond $z \sim 1$,  
  we restrict the redshift interval to $0.05 < z < 1.2$.  
  With these selection criteria, 
 the total number of QS-E/S0s is 145;
 the number with $z_{spec}$ is 44.
  When there is luminosity evolution, the real $K_{I}$-correction should 
 include the luminosity dimming term, $E(z)$.
   This would change $z_{m2}$ and $z_{m1}$, thus affecting 
 $V_{max}$ as derived from equation (\ref{eq:vmax}).
   Intrinsically bright galaxy samples are nearly volume-limited
 (i.e., $min(z_{2},z_{m2})=z_{2}$ and $max(z_{1},z_{m1})=z_{1}$),
 except at very low redshift ($z \lesssim 0.2$).
  However, the volume at $z < 0.2$ is small compared to the 
 remaining volume (e.g., $0.2 < z < 0.6$), and the evolutionary 
 correction itself is small at low redshift;  
  thus, the evolutionary correction does not significantly affect the 
 bright end of the LF ($< 0.05$ dex in density). 
   As a result, $1/V_{max}$ tends to become bigger
 with negative $E(z)$.  However, the level of 
 change in $1/V_{max}$ is only $\sim$0.1 dex.

  Galaxies are then divided into different absolute magnitude bins,
 and the LF value for the $j$th bin is calculated 
 as the sum of all $1/V_{max}$ values of galaxies belonging to 
 that bin, i.e.,

\begin{equation}
  \phi(M)_j\,dM = \Sigma~\frac{1}{V_{max}}.
\end{equation}

\subsubsection{STY method}

  To estimate the parameters of the LF,
 we use the STY method (Sandage, Tammann \& Yahil 1979;
 Loveday et al. 1992; Marzke et al. 1994, 1998;
 Efstathiou, Ellis, \& Peterson 1988; Willmer 1997),
 assuming that the LF is described by the Schechter form (Schechter 1976): 

\begin{equation}
  \phi(M)\,dM=0.921\, \phi^{*}\, x^{\alpha+1}\, {\rm exp}^{-x}\, dM
  \label{eq:lf}
\end{equation}

\noindent
where $x=10^{0.4\,(M^{*}-M)}$.

  The Schechter function has three free parameters ($\phi^{*}$, $M^{*}$,
 and $\alpha$); $\phi^{*}$ is for the density normalization,
  $M^{*}$ indicates the characteristic luminosity of the distribution
 where the number density of bright galaxies starts
 to fall off, and $\alpha$ is the slope of the faint end of the 
 luminosity function.
 For the LF of all nearby galaxies, these 
 parameters are estimated to be $\phi^{*}=0.01$--$0.02~h^{3}$ Mpc$^{-3}$,
 $M^{*}=-19.0$ to $-19.7 + 5{\rm log}(h)$ B mag, and $\alpha=-0.8$ to $-1.4$
 (Marzke et al. 1998; Lin et al. 1996; Zucca et al. 1997) 

 In the STY method, the luminosity function parameters are found
 by maximizing the probability of the observed data, and hence
 by maximizing the following  likelihood function:

\begin{equation}
  {\rm ln}{\it L} = \Sigma {\rm ln}~p_{i},
\end{equation}

\noindent
 where $p_{i}$ is the normalized probability of finding galaxy $i$ with 
absolute magnitude $M_i$ at redshift $z$
in a magnitude-limited survey.  
The normalized probability is given by

\begin{equation}
 p_{i}=\phi(M_{i})/\int^{min(M_{max},M_{2})}_{max[M_{min},M_{1}]} \phi(M) dM.
 \label{eq:pi}
\end{equation}

\noindent
  Here, $M_{i}$ is the absolute magnitude of the object, 
 $M_{1}$ and $M_{2}$ are 
 the brightest and faintest absolute magnitude limits of the sample, 
 and $M_{max}$ and $M_{min}$ are
 the maximum and minimum absolute magnitudes observable at
 redshift $z$ given the apparent magnitude limits of the survey.  
  In our analysis, we do not restrict $M_{1}$ and $M_{2}$ so these 
 quantities are irrelevant here.
 Since the normalized likelihood function is independent of density,
 the STY method provides two of the three LF parameters 
 ($\alpha$ and $M^{*}$) and is furthermore free of the 
 problem of density inhomogeneities, provided 
 that there is no correlation between the LF and density.
 However, for the same reason the STY method does not provide 
 $\phi^{*}$, for which we need to resort to an independent method.

\subsubsection{Normalization for the STY-estimated LF}

   To estimate the number density parameter $\phi^{*}$, 
 we use the following unbiased estimator 
 for the mean number density $\bar{n}$  from
 Davis \& Huchra (1982)

\begin{equation}
  \bar{n} = \frac{\sum \, w_{i}(z)}
          {\int s(z)w(z)dV}.
  \label{eq:nbar}
\end{equation}

\noindent
  Here, $w_{i}(z)$ is a weighting function for galaxy $i$
at redshift $z$, and $s(z)$ is the selection 
 function, which we define in redshift space as

\begin{equation}
  s(z) =\frac{ \int_{max(L_{1},L_{min}(z))}^{min(L_{max}(z),L_{2})} \phi(L) dL}
           {\int_{L_{1}}^{L_{2}} \phi(L) dL}.
\end{equation}

\noindent
  In this equation, $L_{1}$ and $L_{2}$ are the minimum and maximum
 luminosities of the luminosity interval over which we would like to 
 determine $\bar{n}$, and  $L_{min}(z)$ and $L_{max}(z)$ are the minimum 
 and maximum luminosities observable at redshift $z$ for given 
 survey apparent-magnitude limits. 
 
  The variance of this estimator is 

\begin{equation}
  \delta \bar{n}^{2} =\frac{ \bar{n}^{2} \int dV_{1} dV_{2} s_{1} s_{2} w_1 w_2  \xi(r_{1},r_{2}) + \bar{n} \int dV s w }{ (\int dV s w)^{2} },
\label{eq:dn}
\end{equation}
   
\noindent
 where the integral is done over the survey volume, and $\xi(r_{1},r_{2},z)$
 is the two-point correlation function at redshift $z$. 

  The optimal weighting function that minimizes the variance is roughly

  \[ w=1/s(z) \],

\noindent
  and we use this weight and equation (\ref{eq:dn}) to estimate the number density and
 its error. 
  When the variance is minimized, the fractional error for the measurement
 of $\bar{n}$ is roughly (Davis \& Huchra 1982)

\begin{equation}
  \frac{\delta n}{\bar{n}} \simeq (J_{3}/V)^{0.5},
\label{eq:j3}
\end{equation}
    
\noindent
 where $J_{3}=4\pi \int r^{2} \xi(r) dr$. For nearby galaxies, 
 $J_{3} = 10^{4} h^{-3} {\rm Mpc}^{3}$ (Lin et al. 1998). 
 Note that this rough estimate based upon equation (\ref{eq:j3}) is accurate  
 only when the depth of the survey volume in each dimension
 is much greater than the correlation scale
 (i.e., $x >\!\!> r_{0}, y >\!\!> r_{0}$, and $z >\!\!>r_{0}$). 
 When the total volume is large
 but the depth of the volume in one or two dimensions 
 is comparable to or less than the correlation scale (e.g., $x \leq r_{0}$,
 and/or $y \leq r_{0}$, as is our case here),  
 equation (\ref{eq:j3}) overestimates the fractional error. 

  To obtain a more accurate error estimate,
  we integrate equation (\ref{eq:dn}) numerically.
  This requires knowledge of the clustering properties of E/S0s
 at the redshift of interest, which are not very well known. 
  Nevertheless, as shown in Appendix B, we find it plausible to use
 an E/S0 clustering evolution model
 with a spatial two-point correlation function that evolves as
 $\xi(r)=(r_{0,E/S0}/r)^{\gamma}/(1+z)^{3-\gamma+\epsilon}$, 
 with $r_{0,E/S0}=8\,h^{-1}$~Mpc (comoving coordinate),
 $\gamma=1.8$, and $\epsilon=0.8$ and that cuts off at a scale
 $r = 20\,h^{-1}~{\rm Mpc}$.

 This clustering model gives $J_{3} \simeq
 2 \times 10^{4} h^{-3}$ Mpc$^{3}$ from equation (\ref{eq:dn}) at $z=0$, while 
 for the galaxy population 
as a whole at $z=0$, the  recent observed value is 
 $J_{3}=10^{4} h^{-3}$ Mpc$^{3}$ (Lin et al. 1996), with some previous 
 estimates indicating a smaller $J_{3}$ (e.g., $J_{3}$=1700 from
 Davis \& Peebles 1983). 
  The fact that $J_{3} \simeq 10^{4} h^{-3}$ Mpc$^{3}$
 from the analysis of the power spectrum and two-point 
 correlation function of the Las Campanas redshift survey (LCRS) data
 indicates that the effect of the possible 130 $h^{-1}$ Mpc-scale structure
 (Landy et al. 1996) is negligible when estimating errors
 in the number density.
  Therefore the adopted cutoff at $r=20 h^{-1}$ Mpc is well justified
 (see also Peebles 1994).
 
  With the assumed clustering model, we find that the fractional error 
 of the number density of QS-E/S0s is about 20--25\% at $z\sim 0.8$,
 and about 35\% at $z \sim 0.4$, several times greater than  
 the error estimates based on the Poisson statistics alone.   
  These errors are comparable to the uncertainties and biases in the raw
 counts that were estimated in the previous section.
  This order of magnitude in the fractional error was also found
 by de Lapparent et al. (1989) for local galaxies.  
  We have also tried increasing the cutoff value 
 in the integral to $r=40~ h^{-1}$ Mpc
 but find that  the error estimate does not change significantly (at the
 10\% level).

  Finally, the number density parameter, $\phi^{*}$, is calculated as

\begin{equation}
 \phi^{*}=\frac{\bar{n}}{\int^{L_{2}}_{L_{1}} [\phi(M)/\phi^{*}]\,dM},
\end{equation}

where \[\phi(M)/\phi^{*}\] is the partial LF derived from the STY method.

  Note that, since the value of $\phi^{*}$ correlates with the estimate of 
 $M^{*}$, the uncertainty in $M^{*}$ introduces another source of
 error into $\phi^{*}$. We take this into account by adding 
 the error contributed from $M^{*}$ to the error estimated
 above in quadrature.

\subsection{Method 2}

  In this method, 
  we parametrize the luminosity evolution as 

\begin{equation}
 E(z)=Q\,z,  
 \label{eq:ez}
\end{equation}

\noindent
 or equivalently,

\begin{equation}
 M^{*}(z)=M^{*}(0) - E(z).
 \label{eq:m*}
\end{equation}
 
\noindent
  The function $E(z)$ is expressed in magnitude units, and 
  thus  galaxies get brighter as a function of redshift
 by $E(z)$ mag.  A similar formalism was used in
 the analysis of CNOC2 data by Lin et al. (1999). 
 The linear form of equation (\ref{eq:ez}) approximates
the passively evolving stellar populations of the BC96 models. 

  For the number density evolution, we use the following
 parameterization:

\begin{equation}
 \phi^{*}(z)=\phi^{*}(0)~(1+z)^{m}.
 \label{eq:phi*}
\end{equation}

\noindent
 The value of $m$ is claimed to lie between $-1.5$ and $-1$ 
 for a CDM-dominated Einstein-de Sitter universe with hierarchical clustering
 (Baugh et al. 1996; Kauffmann et al. 1996).

  The parameters $M^{*}$ and $\phi^{*}$ 
 in equation (\ref{eq:lf}) are now replaced by $M^{*}(z)$
 (equation  (\ref{eq:m*}))
 and $\phi^{*}(z)$ (equation  (\ref{eq:phi*})), respectively, to provide 
 the LF which incorporates the evolutionary change.

  In estimating the parameters of this LF 
 ($Q, m, M^{*}(0), \alpha$ and $\phi^{*}(0)$),
 we follow the procedure described in Lin et al. (1999).
 First, we use the STY method with $p_{i}$ defined in the same way
 as equation  (\ref{eq:pi}) to estimate $M^{*}(0)$, $Q$, and $\alpha$. 
  The density evolution parameter ($m$) is then estimated 
 by maximizing the likelihood that each galaxy will be at its 
 observed redshift, where the likelihood, $L^{'}$, is

\begin{equation}
 {\rm ln}~ {\it L^{'}} = \Sigma~{\rm ln}~(p^{'}_{i}) + constant,
\end{equation}

\noindent
and 

\begin{eqnarray}
 p^{'}_{i} = p(z_{i}|M_{i}(0),Q)
       = (1+z_{i})^{m}/\int^{min[z_{max}(M_{i}(0)),z_{2}]}_{max[z_{min}(M_{i}(0))],z1}
         (1+z)^{m}\frac{dV}{dz} dz.
\end{eqnarray}

  Finally, $\phi^{*}(0)$ is estimated analogously to the procedure in 
 Section 4.2.3 except that we now take the density and the luminosity evolution
 terms into consideration.  
 To do this, we modify  equation  (\ref{eq:nbar}) as follows:

\begin{equation}
  \bar{n} = \frac{\sum \, w_{i}/(1+z_{i})^{m}}
          {\int s(z)w(z)dV}.
\end{equation}

\noindent
  The absolute magnitudes in the selection function,
 $s(z)$, are now calculated taking the luminosity evolution, $E(z)$,
 into account.
  The error for $\phi^{*}(0)$ is estimated using the method described
 in Section 4.2.3.

\subsection{The luminosity function of E/S0s at $z < 1.2$}

\subsubsection{Results from Method 1}
 
   Figure  \ref{fig:lf_all.ps}
 shows the luminosity function of the Groth Strip QS-E/S0s in two 
 redshift intervals,  one at 
 $z=0.05$--$0.6$ ($z_{med}=0.45$; figures on left with blue points and 
 blue dashed lines) and the other at $z=0.6$--$1.2$ 
 ($z_{med}=0.83$; figures on right with red points and red solid lines).
 Three different cosmologies are shown: top, Einstein-de Sitter universe;
 middle, an open universe with $\Omega_{m}=0.2$;
 and bottom, a flat universe with $\Omega_{m}=0.3$ and $\Lambda=0.7$,
 which is currently favored  
 (Im, Griffiths, \& Ratnatunga 1997; Riess et al. 1998;
 Perlmutter et al. 1999; for a review, see Primack 2000).
   Also plotted is the local LF of E/S0s from Marzke et al. 
 (1998) as a dotted line, and the local LF of E/S0s from Marinoni et al. (1999)
 as a dot-dashed line.
  Table 3 summarizes the LF parameters for the local LFs.
  Note that $\alpha \simeq -1.0$ in both cases, which justifies the use of 
 a fixed $\alpha$ value for our fits.  However, the 
 $M^{*}$ values from the two works differ by about 0.5 magnitudes;
 the origin of this discrepancy may be due to differences
 in the magnitude systems,
 as discussed in Marinoni et al. (1999). 
  It is not clear which system is close to our magnitude system, thus
 we consider both values.   
  We also notice that the value of $\phi^{*}$ from Marinoni et al. (1999)
 is lower than the value from Markze et al. (1998) by nearly a factor of 2.
  This can be attributed to difference in classification schemes.
  Marinoni et al. (1999) select E/S0s as objects with $-5 \leq T < -1.5$,
 as opposed to Marzke et al. (1998), who take $T < 0$. 

  The data points for the Groth Strip QS-E/S0s come from the $1/V_{max}$ estimates, 
 while the colored curves are the LF functions fit to the same points using
 the STY method.
   The bin sizes for the QS-E/S0 LF points are determined so that 
 each bin contains at least 3 galaxies, with the minimum 
 bin size being
 0.2 magnitudes. Some LF points include as many
 as 15 E/S0s.

  To check for any strong biases in our sample (such as incompleteness),
 we calculate $\langle V/V_{max} \rangle$.
  The values of $\langle V/V_{max} \rangle$ are 
 found to be 0.56, 0.54, 0.53 ($\pm 0.03$) for the Einstein-de Sitter, open,
 and $\Lambda$ universes respectively. These values are close
 to those expected for passively evolving
 E/S0s ($\sim 0.55$), and thus imply both that 
 the number of galaxies at high redshift is fairly complete (as found
 previously) and that 
 the number density of E/S0s has not evolved 
 dramatically.  We return to the latter point after
 comparing $\phi^{*}$ at $z=0$ with $\phi^{*}$ at $z \sim 1$ quantitatively.

   Table 4 lists the best-fit LF parameters
 for the two distant redshift intervals using Method 1. 
   Since the faint end of the LF is not well determined
 and since the number of E/S0s in our sample is not large enough to provide 
 a meaningful constraint on all three LF parameters simultaneously, we 
 estimate $M^{*}_{B}$ by fixing $\alpha=-1.0$.
   This allows us to compare  $M^{*}_{B}$ and $\phi^{*}$
 of Groth Strip QS-E/S0s with the local values in Table 3.
   Results are summarized in Table 5,
 which lists the shift in $M^{*}_{B}$ with respect to the local values,
 $\Delta B \equiv M^{*}_{B}(z=0) - M^{*}_{B}(z)$.  Here, 
 the value of $\phi^{*}$ from Marzke et al.
 (1998) in Table 3
 has been multiplied by a rough correction factor of
0.7 in order to account for a possible difference in 
 E/S0 classification criteria (see below).

  A robust conclusion from Table 5 is that the bright end ($L > L^{*}$) of 
 the $z \sim 0.8$ LF is significantly {\it brighter}
 with respect to both  local LFs, by about 1.1--1.6$\pm 0.2$ mag
 relative to Marzke et al., and 0.7--1.2$\pm0.2$ mag, 
 relative to Marinoni et al.
  This is most naturally explained by  a
 luminosity brightening of all galaxies by
 these amounts from $z=0$ to $0.8$.
  The precise amount depends slightly on the local samples and the 
 assumed cosmology, but is largely independent of any biases or incompleteness
 in the actual counts. 
  Moreover, the derived luminosity evolution is internally
 consistent, with $\Delta M_{B}^{*}$ for the more distant redshift
 bin being roughly twice as bright as in the nearer bin.
   Finally, the observed degree of brightening is much larger than
 uncertainties in our own magnitudes or any differences between our own
 magnitude system and that of the local samples, which can amount to
 as much as 0.5 mag.

  We have calculated the expected magnitude brightening back to 
  $z \sim 0.8$ using  passive-evolution models and varying both
 $z_{for}$ and metallicity (CB96).  The predicted 
 amount of brightening at $z \sim 0.8$ from these models
 is comparable to the values listed in Table 5.
  In particular, for models involving
 short bursts of star formation followed by passive stellar aging, 
 we find that $z_{for} \sim 1.5$ for $\Delta B \simeq 1.5$ 
 while $z_{for} \gtrsim 3$ for $\Delta B \simeq 1.0$.
  In the $\Lambda$ universe, for which the brightening is largest, 
 the bulk of stars in field E/S0s may have formed at redshifts
 as low as $z_{for} \simeq 1.5$, especially if the local
 $M^*_B$ from Marzke et al. is used.
  For the open or  Einstein-de Sitter 
 universes, $z_{for} > 2$--$3$ is a better fit to the results.
  If alternative evolutionary models with extended 
 star-formation times are used,
  the initial redshift of star formation can be pushed either earlier
  or later depending on the precise history of star formation. 

   Related papers by our group discussing Groth Strip early-type galaxies
 estimate the amount of luminosity brightening 
 by studying the fundamental plane of field E/S0s 
 (Gebhardt et al. 2000) and the size-luminosity relation of luminous 
 bulges to $z \sim 0.8$ (Koo et al. 2000).   The fundamental plane shows
 a brightening of 1.4--1.8 mag in the $B$ band back to $z \sim 0.8$,
 while the size-luminosity relation 
 of luminous bulges shows a similar result; both are 
 consistent with the LF analysis here.
  Additional color information is used in these papers to further
 constrain spheroidal star-formation histories;
 the general conclusion is again that the bulk of star formation
 was quite early.

   The fits to $\phi^{*}$ in Table 5 can also be used to test for
  evolution in the number density, assuming that the luminosity 
  evolution is well represented by
  the above change in $M^{*}$ and that the Groth Strip QS-E/S0 sample counts
  identically the same kind of galaxies as the local samples.
   Our method tends to select E/S0s with $T \lesssim -2$ and may 
 miss $\sim 30$\% of E/S0s with $T < 0$ according to the tests
 performed in section 3. Visual inspection of
 red, non QS-E/S0s galaxies confirms that we may be missing roughly this many 
 E/S0s. On the other hand, the local E/S0 local sample of Marzke et al. 
 (1998) is chosen to have $T < 0$, while Marinoni et al. is based on
 E/S0s with $T < -1.5$. This suggests that the 
 Marzke et al. (1998) normalization
 may need to be multiplied by a factor of 0.7 to correct for
 classification mismatch.
  With this correction to Marzke et al., Table 5 shows that
 $\phi^{*}$ at $z \sim 0.8$ is 1.6--2.0 times greater than the value at $z=0$ 
 for the Einstein-de Sitter universe.  This result is implausible
 as it indicates {\it more} E/S0 galaxies at higher redshift than now,
 which is not predicted by any formation model.
 If it were well established, this trend  might even suffice to rule out the
 Einstein-de Sitter cosmology, which in any case is highly disfavored
 by numerous other evidence (Bahcall et al. 1999; Primack 2000).
  However, given the $\sim\pm$30\% uncertainty in both the nearby and
 distant counts, the discrepancy must be considered marginal.
  The above increase in E/S0s at high redshift disappears in the two 
 low-$\Omega_{m}$ 
 cosmologies, for which the number density remains flat 
 at all epochs within the errors.  Although uncertainties
 are large, the data seem most compatible with a picture in which the
 great majority of E/S0 galaxies were already in place by $z \sim 1$
 and their numbers have been roughly constant since then.

  The above analysis using Method 1 relies on
 comparison to the local E/S0 luminosity function, and thus admits numerous
 sources of systematic error, as we have noted.
  Another point not discussed so far
 is the highly correlated error between $M^{*}$ and $\phi^{*}$ in the fitting
 procedure, i.e.,
 when $M^{*}$ is estimated to be brighter, $\phi^{*}$ becomes smaller
 even for essentially similar data.
 Both of these problems can be minimized by using Method 2, 
  which estimates the luminosity and number density evolution 
 simultaneously from the Groth Strip QS-E/S0s alone.

\subsubsection{Results from Method 2}

  In deriving the evolution parameters $Q$ and $m$ using Method 2,
 we adopt two slightly different approaches.
  In the first approach, we place no constraints
 on any of the LF parameters except for
 fixing $\alpha = -1.0$. The derived $M^{*}(0)$ and $\phi^{*}(0)$ are then 
 compared to local values.
  In the second approach, we fix $M^{*}(0)$ to the value
 in Marzke et al. (1998)  or in Marinoni et al. (1999).  Fixing $M^{*}(0)$
 helps reduce the error in $Q$ and $m$, but the two local values of $M^{*}(0)$ 
 differ by almost 0.5 mag. We 
 consider both values and see what impact this has.
  Table 6 summarizes the resultant LF parameters of E/S0s, with 
  errors as indicated in parentheses.
  Figures \ref{fig:cont_om1}--\ref{fig:cont_lam07}
 show the significance of the measurements 
 in contour plots of $Q$ and $m$. 
  As expected, larger values of $Q$ are associated with
 more negative values of $m$.
   
  The first approach (solving for $M^{*}_{B}(0)$ simultaneously)
 gives parameters for the local function that
 are consistent with the local data.
  The best-fit value for $M^{*}_{B}(0)$ is $ \simeq -20.03 \pm 0.6$ mag
 for all three cosmologies, in almost perfect agreement
 with $-$20.14 mag as found by Marzke et al. (1998)
 but also consistent within the errors
 with the value $-$20.54 mag found by Marinoni et al. 
 (1999).  
 The estimates of $\phi^{*}(0)$ also match well with 
 the raw value of $\phi^{*}(0)$ in Marinoni et al. (1999), or with the 
 value from Marzke et al. (1998) corrected by the factor 0.70.

 The derived evolutionary parameters $Q$ and $m$ 
 are also plausible and consistent 
 with the results from Method 1.
 The fitted $Q$ values imply a luminosity brightening of 1.4--1.9 $\pm$0.7 mag 
 back to  $z=1$ in all models, showing definite evolution,
 as found also by Method 1.
 The fitted values of $m$ depend more strongly on cosmology, with
 $m$ being $\sim+0.5$ for the Einstein-de Sitter universe (more
 objects at high redshift), versus $-0.5$ and $-0.9$ for the open and $\Lambda$ universes (fewer objects at high redshift).  
 However, the  error bars on $m$ are $\pm$0.7, so the data
 are in fact statistically consistent with {\it no} number density evolution
 in all three cosmologies; again, this agrees with Method 1.

  Tighter constraints on all parameters can be obtained by
 fixing $M^{*}_{B}(0)$, as in the second approach.
  When we adopt $M^{*}_{B}(0) = -20.14$ from Marzke et al., 
 we obtain almost the same values for the other parameters
 because the Marzke et al. value of $M^{*}_{B}(0)$ is
 close to our own unconstrained value.  
  However, the smaller error on $Q$ 
 now makes the detection of luminosity evolution
 much more secure ($>5$-$\sigma$) under all three cosmologies.
  The error in the number density evolution also shrinks slightly, 
  but the Einstein-de Sitter and open universe are still consistent with no
  evolution in number density, while the $\Lambda$ universe favors
  a slight {\it decrease} in past number densities at about
  the 2-$\sigma$ level
  (however, this result would no longer be statistically significant if 
  our possible systematic undercounting of distant galaxies
  at the 30\% level were included).

  When $M^{*}_{B}(0) = -20.54$ is adopted from Marinoni et al. (1999), we find
 a smaller amount of luminosity brightening due to the fact that
 this value of $M^{*}_{B}(0)$ for local E/S0s is brighter by 0.4 mag than
 that of Marzke et al.
  Nevertheless, the significance of the luminosity brightening
 at high $z$ remains 
  high.  All values of $m$ are increased by about 0.6 from
  the previous fit using Marzke et al., indicating more galaxies
  at high $z$ than before.   
   Now, the Einstein-de Sitter universe shows an increase in number density
  at the 2.4-$\sigma$  level, while the other two universes are consistent
  with no number density evolution.

   To summarize, Methods 1 and 2 are basically consistent, indicating
  that the data are both internally consistent 
  with themselves (low vs. high $z$) and externally 
  consistent with the local samples.  If the Einstein-de Sitter
  universe is excluded as unlikely, a {\it brightening}
  in $M_{B}^{*}$ of 1.6--2.0 mag has been detected to $z \sim 1$.
   This result is robust and is largely independent of the count errors
  although it does depend on the relative definitions of the local
  and distant magnitude systems, which may differ by up to 0.5 mag.
  Evolution in the number density is much harder to determine given the errors;
  the data may favor a small decrease in number density
  at the level of a few tens of percent
  to $z \sim 1$, but they are also consistent with {\it no change}.
  Sources of error in the number densities include 
  raw and systematic errors in the distant
  counts, uncertainties in the local counts, and systematic classification
  differences between the local and distant samples.  Taken together,
  these inject errors of as much as a factor of two into the derived
  evolutionary trends in number density.

\subsection{Effect of changing redshift range}

 As pointed out before, our results are potentially sensitive to 
 a chosen redshift interval. We do not find bright E/S0s 
 at $z < 0.2$ due to
 our bright magnitude cut, and this could potentially lead to an 
 underestimate of the bright end of the nearby LF. 
  There may be a genuine lack of   
 bright E/S0s at $z \gtrsim 1$ as well, in which case 
 the bright end of LF ($z \sim 0.8$) may be also 
 be affected.
  For these reasons, we have tried using 
  different redshift ranges to see if there is any strong change
 in the estimated evolutionary parameters.
  Since this is merely a sensitivity check,
 the cosmology is restricted to 
 the open universe for simplicity.
  Table 7 lists the results of the tests, for four different 
 redshift limits: $0.2 < z < 1.2$, $0.05 < z < 0.8$, $0.05 < z < 1.0$,
 and $0.2 < z < 1.0$, all using Method 2.  
  Changing $z_{min}$ changes the results by only
 a small amount ($\Delta Q = 0.1$ and $\Delta m \simeq 0.2$), in a direction
 that makes $M^{*}$ at lower redshifts brighter. 
 However, the effect of changing the upper redshift limit is 
 more pronounced---reducing it makes
 the number density parameter more positive by $\Delta m \sim 1$ or more
 with respect to the values listed in Table 6, while not changing $Q$ 
 very much.
  The sense of this confirms the earlier suspicion
 from Figure \ref{fig:zi_zphot}
 and \ref{fig:zab_zphot} that  bright E/S0s may be missing
 beyond $z > 1.0$ in our data.    Possibilities for the cause of
 the apparent deficiency 
 include (i) the
 systematic underestimate of photometric redshifts at $z > 0.8$,
 and (ii) a genuine failure to detect bright E/S0s at $z > 1$. Future analysis 
 with near-IR data will be able to provide a firmer answer to
 hypothesis (i), as they enable 
 better estimates of photometric redshifts at $z > 0.8$ by sampling
 the light above the 4000 $\AA$ break.  Larger, deeper datasets
 would provide a check of hypothesis (ii). We also tried 
 increasing the upper redshift limit, but find only 
 negligible effects on the evolution parameter estimates. 
 This suggests 
 that observational selection effects act against detecting
 E/S0s at $z > 1.2$.

\subsection{Monte Carlo simulation and Malmquist bias}

  This section tests our results using a mock catalog 
 created by Monte Carlo simulation for other possible biases due 
 to random measurement errors.
  At $z > 0.8$, errors of E/S0 photometric redshifts become greater,
 and the number of E/S0s in our sample declines as  redshift increases. 
 Hence, we expect to lose more galaxies below $z_{max}$ than adding more 
 galaxies at above $z_{max}$. This biases the number of high-z E/S0s to 
 be underestimated.
    On the other hand, in a volume-limited 
 sample,  random, symmetric errors in redshift space make the measured 
 redshifts to be always slighlty underestimated due to the fact that 
 more galaxies scattered in from higher redshifts than from lower redshifts
 simply because of the volume effect. Again, this kind of Malmquist bias 
 leads to an underestimate of the absolute fluxes.
    The random error in apparent magnitude 
 creates a bias in another direction. At around $L^{*}$, as the number 
 of $L < L^{*}$ galaxies are greater than the number of 
 $L > L^{*}$ galaxies, the random scattering in the flux would put 
 more $L > L^{*}$ galaxies than there actually are, leading to   
 overestimating the bright end of the LF. 
   Our LF analysis takes into account some of these biases by incorporating
 measurement errors in the LF estimates. However, our procedure
 is not perfect. In order to test any biases left in the LF measurements 
 due to random measurement errors in apparent magnitudes and redshifts,
 we created mock catalogs of E/S0 galaxies using the Monte-Carlo simulation
 procedure similar to that described in Im et al. (1995a).  Into the 
 simulated catalogs, we introduced random measurement errors. Then, 
 we estimated the LF parameters from mock catalogs, and compared the outputs
 against input LF parameters. 
   By exploring a range of LF parameters and cosmologies, we find that
 $\Delta Q = Q_{output}-Q_{input} \lesssim 0.1$, and
 $0 < \Delta m = m_{output}-m_{input} \gtrsim -0.35$. This test shows that
 the random errors slightly overestimate the amount of luminosity and
 number density evolution.  To be precise, 
 the evolution parameters listed in Table 5--7 would 
 need to be corrected by this amount, but, since the amount of systematic 
 bias is smaller than 1-$\sigma$ errors, we neglect this effect. 

\section{Discussion}

\subsection{Implications for the merging rate since $z = 1$}

  If E/S0s are formed via mergers of disk-dominated galaxies  
 with similar masses (e.g., Kauffmann et al. 1993)
 and not destroyed or converted  to spirals afterwards, the number
 density of E/S0s can only decrease as we look back in time. 
  When the number density evolution is modeled as 
 $n(z)=n(0) \times (1+z)^{m}$ with $n(0)$ being the number density of 
 E/S0s at $z=0$, semi-analytic models based upon
 an $\Omega_{m}=1$ CDM-dominated universe
 predict $m = -1.5$ to $-1.0$, i.e,
 at least half of all massive E/S0s today were 
 formed via major mergers since $z \sim 1$
 (Baugh et al. 1996; Kauffmann et al. 1996). 
   Our results for the Einstein-de Sitter universe
 are clearly inconsistent with such model predictions,
 excluding the value $m=-1$ (doubling in numbers since $z=1$)
 at more than the 97--99.7\% confidence level. 

  A way to make the merging picture more consistent with an
  Einstein-de Sitter universe involves
 an alternative merging scenario 
 wherein early-type galaxies simply 
 increase their mass monotonically through {\it minor} 
 mergers with other early-type galaxies 
 or disk galaxies, leaving the overall number density 
 of large E/S0s unchanged.  This brightening
  would mimic $\Delta B$ evolution 
 in the opposite direction to passive aging.
  Van Dokkum et al. (1999) find that 
 close, bright pairs in one distant cluster mostly consist of red, early-type
 galaxies  rather than late-type galaxies. If such a trend can be found 
 to apply in low density environments,
 this would provide good observational support for 
 the alternative merging scenario.

  When an open or a flat non-zero $\Lambda$ universe   
 are assumed, 
 our count data admit the possibility of a moderate increase in number
 density from $z = 1$ to now.  
 Taken literally, the best-fit values of $m$ suggest an increase
 of a few tens of a percent in the numbers of E/S0s since $z=1$.
 Semi-analytical models tend to predict that number
 density evolution of early-type galaxies is weaker in 
 open or $\Lambda$ universes (e.g., Kauffmann \& Charlot
 1998);
 our results are quite compatible with such predictions.
 
  The little-or-no number density evolution seen for the mostly 
 ``red'' QS-E/S0s further implies that the expected number of
 blue luminous E/S0s  should be small.  
  If brief, episodic bursts of star formation
 make E/S0s bluer (e.g., Charlot \& Silk 1994;  
 Trager et al. 2000 for evidence from local ellipticals),
 the interval of time for finding blue colors combined with a smooth,
 undisturbed morphology is only of order $\lesssim 1$ Gyr.
  Assume that the duration of a burst is $\delta t$, that the fraction
 of present-day E/S0s that underwent a starburst since $z=1$
 is $f_{burst}$, that the average number of bursts per E/S0 since $z=1$ is 
 $n_{burst}$, and that the number of bursts is roughly constant in time
 since $z=1$ to now.
 The fraction of distant E/S0s that will appear blue ($f_{blue}$) is then 

\begin{equation}
  f_{blue}= 0.06~(\frac{n_{burst}}{1}) (\frac{f_{burst}}{0.5})
  \frac{(\delta t/1~{\rm Gyr})}{(t(z=1)/8~{\rm  Gyr})},
  \label{eq:fblue}
\end{equation}

\noindent
 where $t(z = 1)$ is the look-back time from $z=0$ to $z=1$.
 Our counts suggest that the fraction of blue E/S0s is not more 
 than 10\% of all E/S0s,
 which is consistent with the small percentage predicted by 
 equation  (\ref{eq:fblue}).  Note that this argument assumes 
 that episodic star formation occurs randomly from $z=1$ to $z=0$;
 if it occurs predominantly at a particular redshift,
 we should see an increased number of blue E/S0s at around that 
 redshift. Currently, our data are not sufficient to test this.

   An alternative interpretation of the count data is
 that distant E/S0s consist at least partially 
 of bulges of galaxies that will later accrete disks to
 become spirals.  
 Comparison of the number density of
 such objects at different redshifts would then be meaningless.
  However, this is quite unlikely, at least at the bright end
 of the LF, since the bulges of local spirals are not
 as luminous or massive as the brightest E/S0s.
  Recent works furthermore suggest that 
 large disk galaxies were already largely in place at $z \simeq 1$ 
 (Vogt et al. 1997; Lilly et al. 1998; Simard et al. 1999)
 and that normal spiral galaxies out to $z \simeq 1$ were as abundant 
 as those at $z=0$ (Im et al. 1999; Brinchmann et al. 1998;
 Driver et al. 1998).  Wholesale transformation of galaxy
 types does not look likely at the present time
 but is a possibility that clearly must be studied further.
 
  Overall, the LF of the Groth Strip QS-E/S0s is quite consistent with the view
 that the majority of luminous  E/S0s were already in place at $z \sim 1$
 and that their luminosities have evolved smoothly and quiescently 
 over time, with only a small number of 
 significant star-formation bursts per galaxy since that epoch.

\subsection{Comparison with other studies}

  The definition of our $Q$ parameter for the luminosity evolution is 
 identical to that in CNOC2 (Lin et al. 1999), which was defined
 for a sample at $z < 0.55$. Their sample 
 was selected based on color rather than morphology,
 but  our $Q$ values are in  good agreement with theirs 
 ($Q=1.58$ -- $2.00 \pm 0.49$), excepting the case where $M^{*}_{B}(0)$ in 
 Marinoni et al. (1999) is assumed.
  For a morphologically selected distant field elliptical sample,
 Schade et al. (1999) measured a luminosity brightening of 
 $\Delta M_{B} = 0.97 \pm 0.14$ mag from $z=0$ to $z\simeq 1$ 
 in an Einstein-de Sitter universe. This again is in reasonably 
 good agreement with our values. Using another morphologically selected 
 field early-type galaxy sample, Im et al. (1996) reported 
 a luminosity brightening of
 $\Delta M_{B} \simeq (0.6$ - $1.5) \pm 0.5$ mag
 back to $z \simeq 1$ for a flat universe with 
 or without $\Lambda$, again consistent with the present results.
 As noted, the fundamental plane (FP) of high redshift early-type
 galaxies provides yet another 
 independent estimate of luminosity evolution. For cluster early-types,
 the luminosity brightening is only about $0.75$--$1.0$ magnitudes
 in rest-frame $B$ back to $z \simeq 0.83$ (van Dokkum et al. 1998). 
 The brightening we find here for field early-types is larger than this
 but is consistent with the stronger luminosity evolution 
 found for {\it field} E/S0s using the fundamental plane
 (Treu et al. 1999; Gebhardt et al. 2000).
 Thus all evidence seems to agree in implying significant
 luminosity brightening of field E/S0s back to $z \sim 1$.

 Results on number density evolution do not agree nearly as well.
 The CNOC2 group found
 rapid number density evolution proportional to 
 $10^{0.4\, P\, z}$ with $P=-1.07 \pm 0.49$ and $P=-1.79 \pm 0.49$ 
 for the Einstein-de Sitter and open universes, respectively.
 If extrapolated, 
these values correspond to only 0.4 and 0.2 times the present number density
 of galaxies at $z \sim 1$. 
 Kauffmann, Charlot, \& White (KCW; 1996) likewise found that
 number density decreased strongly back in time,
 as $(1+z)^{-1.5}$ in an Einstein-de Sitter universe.
 These studies both disagree with our estimtes of
 low or non-existent evolution in Table 6.
 We have noted that both the CNOC2 and KCW samples were selected
 based purely on colors; we suspect this might be the cause of 
 at least a part of the discrepancy, for the following reason.
 At low redshift ($z \lesssim 0.4$),
 the spread in colors is small 
 for the different galaxy types, and it is easy
 for photometric errors to bump blue galaxies into a red-galaxy sample.
 The precise choice of color boundary also matters greatly.
 Both effects could lead to an increase in 
 in the apparent number density of red galaxies at  low redshift.
 The CFRS sample used by KCW is furthermore known to be deficient
 in red objects beyond $z > 0.8$ (Totani \& Yoshii 1998; Im \& Casertano 2000).

  Results from the morphologically selected samples of
 Schade et al. (1999) and Menanteau et al. (1999) 
 appear superficially to agree better with the present work,
 but there is an important
 difference. 
  Both works showed little evolution in claimed E/S0 number
 density, but their samples include a substantial number of 
 {\it blue} E/S0s.
  Such blue objects are not present in our more tightly
 selected morphological sample, and our red objects by
 themselves are steady.

   The only previous data in genuine agreement with ours is the
  morphologically selected sample of Im et al. (1996).  
  The work by Im et al. (1996) shows a luminosity brightening
 of order $\Delta B =1.73$ mag out to $z \simeq 1$
 for a flat non-zero $\Lambda$ 
 universe. They also found little decline in number density
 to $z=1$, and concluded that more than $70\%$ of $z=0$
 E/S0 galaxies seemed to be formed before $z = 1$. These numbers match well
 with our results, although their sample may have
 been slightly more loosely selected
 than ours (they found 6.7 E/S0s per WFPC2 field vs.~ 
 5.1 E/S0s per WFPC2 field 
 in the present study).

\subsection{Is the density of E/S0 galaxies in the Groth Strip typical?}

  Since E/S0 galaxies preferentially live in high-density environments,
 our counts could be biased too high if the Groth Strip contains superclusters.
  We review here briefly some tests of this hypothesis in 
 fields that are not adjacent to the Groth Strip.
  One such test looked at E/S0s in the Hubble Deep Field flanking fields
 (hereafter HDFF; Williams et al. 1996);
 no significant difference was found 
 with regard to number density or other properties of E/S0s 
 below $z = 1$.
  Another test checked 
 the number of E/S0s in the GSS versus the number in the HST WFPC2 
 fields that were used for the studies of Im et al. (1996; 1999).
  The mean surface density of E/S0s in the GSS is actually
 slightly smaller (a factor of 0.8) than
 the mean surface  density of E/S0s in these  30 other HST WFPC2 fields,
 showing that the Groth Strip QS-E/S0s are not dominated by 
 populations in high-density regions. 
  These two tests suggest that E/S0s in the Groth Strip can be considered 
 as representative of field E/S0s.

  Also note that Fig. 8a shows several prominent peaks in the redshift 
 distribution of GSS galaxies. A large fraction of E/S0s at $z > 0.6$
 are associated  with these peaks at $z \simeq 0.8$ and $z\simeq 1.0$. 
  Thus, the derived amount of the evolution could be potentially biased 
 due to the existence of these peaks.
  One test for this is to derive the evolution parameters
 using different redshift cuts.
  We estimate the evolution parameters 
 adopting higher redshift cuts which  excludes 
 these peaks (e.g., $0.05 < z < 0.78$ and $0.05 < 0.95$). 
  We find that output $m$ and $Q$ values derived with 
 these redshifts cuts are similar to the values listed in Table 7,  
 suggesting that the peaky distribution itself does not affect the results
 significantly.   We also point out that 
 the rather wide adopted redshift interval 
 $0.6 < z < 1.2$ for the LF (derived with method 1)
 averages out the peaky distribution.
  The amount of the evolution estimated with the ``averaged-out'' LF  
 matches very well that derived with an alternative method (method 2),
 yet another indication that the peaky distribution is not 
 affecting significantly our results. 
  Certainly, a more decisive statement can be made for this issue 
 by analyzing a larger set of data.

\subsection{Uncertainty in the local number density of E/S0s}

  We have already mentioned various uncertainties regarding 
 the normalization of the local LF,  $\phi^{*}$.  One issue
 is the precise range of Hubble types in the various samples, 
 which may account for the
 much of the difference of a factor of two between Marinoni et al. (1999)
 and Marzke et al. (1998). 
   An additional question is 
  the normalization of the local LF for {\it all} types of galaxies together
 (see Marzke et al. 1998 and references therein),
 with the local universe appearing to be underdense in some studies
 by as much as a factor of 1.8 (e.g., Ellis et al. 1996).
  Driver et al. (1996, 1998) fit the observed number counts of E/S0s
  using the LF of Marzke et al. (1998),
  but with the local normalization boosted up by this factor. 
  If such a renormalization is necessary, 
 our conclusions regarding modest number density evolution would change
 drastically---the number density of QS-E/S0s at $0.05 < z < 0.6$ 
 would be about two times smaller than the local value, meaning that 
 about 50\% of present-day E/S0s would had to have formed 
 {\it very} recently!

   There is independent evidence that 
 the high normalization adopted by Driver et al. (1996)
 is not necessary;
 they invoked it to fit the 
 bright end of their E/S0 number counts in Driver et al. (1995). However,
  Im et al. (2000) studied E/S0 number counts using an expanded 
 sample of 56 HST WFPC2 fields,
 which contained the 6 fields used by Driver et al.
 (1996, 1998) as a subset.  
  Im et al. find that E/S0s in the 6 fields of Driver et al. are 
 1.5 times more frequent than average, eliminating most of the discrepancy
 with the local normalization.
  We conclude that the normalization of the overall local LF by Marzke et al. 
 is consistent and that the numbers of E/S0s in this function are reasonable
 after correction by the previously justified factor of 0.7 to match our Hubble types.

\subsection{Colors}

 Figure  8 indicates that the color-redshift relation of Groth Strip 
QS-E/S0s is in 
 a reasonably good agreement with predictions of passive luminosity 
 evolution models. However, we find that the
 $V-I$ colors of Groth Strip QS-E/S0s at $z > 0.8$ are somewhat redder than 
 the model predictions. Koo et al. (2000) and Gebhardt et al. (2000)
 analyze colors of early-type galaxies and luminous bulges 
 and find that the rest-frame $U-B$ colors (roughly, observed 
 $V-I$) of these objects at $z \sim 0.8$ are 0.2--0.3 mag redder
 than what passively evolving models predict for
 a luminosity evolution $\Delta B$ of
 $\sim$ --1.8 mag or more.
  The problem is lessened by using a high-metallicity model, but
 it does not go away entirely.
  Other possible solutions include dust-extinction,
 more complicated star formation histories,  and 
 uncertainties in the stellar evolutionary models. 
  For more discussion,
 see Gebhardt et al. (2000)
 and Koo et al. (2000).

\section{Conclusions}

   Using the residual parameter ($R$) and bulge-to-total light
 ratio ($B/T$), we successfully separate distant E/S0s from 
 other types of galaxies. With this  quantitative classification scheme,
 we identify 145 E/S0s at $16.5 < I < 22$ in the Groth Strip that lie 
 at $z \lesssim 1.2$, termed ``quantitatively selected'' QS-ES0s.
  Spectroscopic redshifts are available for 44 of these QS-E/S0s, and 
  we find a very tight correlation between $z_{spec}$ and $V-I$ in
  the sense that their colors are the reddest among 
 field galaxies at each redshift over the redshift range 
 $0 < z < 1.2$.
 We use this tight correlation of color and $z_{spec}$ 
 to estimate redshifts for the remaining Groth Strip QS-E/S0s
 without spectroscopy and find that
 these photometric redshifts, $z_{phot}$, 
 are accurate to $\sim 10$\% for $z_{phot} \lesssim 1$.

  Using the full sample of 145 Groth Strip QS-E/S0s, 
 we construct their luminosity function  at $z \simeq 0.8$ ($0.6 < z < 1.2$)
 and at $z \simeq 0.4$ ($0.05 < z < 0.6$) for three cosmological
 models: an Einstein-de Sitter universe with $\Omega_{m} = 1$,
 an open universe with $\Omega_{m} = 0.2$, and a flat universe with
 $\Omega_{m} = 0.3$ and $\Lambda = 0.7$.
 A robust result is that rest-frame B magnitudes have brightened by 
 $\Delta B \simeq 1.1 - 1.9$ mag 
 since $z = 1$, with larger evolution taking place in open and $\Lambda$ universes.
 This result is consistent with previous studies of the distant E/S0 luminosity
 function and with our own studies of distant bulges and the field E/S0 fundamental plane (in preparation).
  Evolution in the number density of E/S0s is less well constrained, and
 pushing all errors to their maximum values yields uncertainties of a factor of two in number density to $z \sim 1$.
 For the open and $\Lambda$ universes, 
 the data are most 
 consistent with roughly constant numbers of E/S0s back in time,
 perhaps favoring a moderate decline in numbers by a few tens of percent
 at $z \sim 1$.
 For the Einstein-de Sitter universe, the data favor 
 an {\it increase}, and any drop in galaxies
 by as much as a factor of two at $z \sim 1$
 is strongly ruled out.  

 The amount of luminosity evolution estimated from $M^{*}_{B}$ 
 at $z \sim 0.8$ implies that the major formation epoch  
 of stars in E/S0s occurred at  $z_{for} \gtrsim$ 2--3 
 in the Einstein-de Sitter and open universe models, shifted down to
 as recently as 
 $z_{for} \sim 1.5$ in the $\Lambda$ model.  The large amount of
 evolution coupled with the modest change in number density 
 is consistent with a picture in which the majority of luminous 
 E/S0s galaxies in the field today ($ \gtrsim 70$\%) already existed at
 $z \sim 1$
 and that they have not undergone dramatic evolution other than
 steady dimming of their stellar populations since then.  If the major
 merging of {\it disk} 
 galaxies is responsible for the formation of luminous  
 field E/S0s, such a process must have happened predominantly
 before $z \sim 1$.
 
\acknowledgements

 This paper is based on observations with the NASA/ESA Hubble Space Telescope,
 obtained at the Space Telescope Science Institute, which is operated by
 the Association of Universities for Research in Astronomy, Inc., under
 NASA contract NAS5-26555. Funding for DEEP was provided by NSF grant
 AST-9529098.  
 This work was also supported by the STScI grants GO-07895.02-96A,
 AR-6402.01-95A, AR-8767, and AR-07532.01-96. 
 We would like to thank Katherine Wu for providing background
 subtracted images of local galaxies whose raw images were originally 
 made available by Zsolt Frei. We also thank Ron Marzke, Jon Loveday,
 and Christien Marinoni for discussions on the local luminosity function, 
 and Phil Choi, Raja Guhathakurta, Ricardo Schiavon, and Nicolas Cardiel  
 for their careful reading of the manuscript. 
  We are also grateful to the referee, Simon Driver, for his useful comments.

\input{ref}
\input{figures}

\newpage

\input{tab1}

\input{table}

\newpage

\input{tab3-7}

\clearpage

\appendix

\section{Error in $R$ from Background Noise}

  Several effects contribute to the error estimates of the residual 
 and asymmetry parameters (see Section 3,
 also Conselice et al. 1999 and Wu 1999). 
  Here, we quantify the error due to the background noise with the 
 caveat that it is only a single component of the total error and therefore
 only a lower limit.

  We assume that the noise in all pixels is Gaussian, with true rms
 value $\sigma_{b}$. This includes the object pixels, as background
 noise (from, for example, readout noise and sky photon statistics)
 is assumed to dominate everywhere. 
  $R$ is defined as 

\begin{equation}
 R=R_{T}+R_{A},
\end{equation}

\noindent
 where $R_{T}$ and $R_{A}$ are given in 
 Eqs. (\ref{eq:rt}) and (\ref{eq:ra}).
 We further assume that both $R_A$ and $R_T$ are intrinsically
zero, i.e., that the object is inherently symmetric.  It may
be shown that only under this condition is the assumed background
correction strictly valid.  Furthermore, we are most interested
in estimating errors for the case $R$ is small, i.e., near zero.
 
  The $\sigma_{b}$ of the background is estimated locally because it may vary 
 over the image. An error is introduced into $R_{T}$ and $R_{A}$ 
 if $\sigma_{b}$ is mis-estimated, as then the background correction,

\begin{equation}
 \Sigma \frac{1}{2} |B_{ij} \pm B_{ij}^{180}|,
\end{equation}

\noindent
  will be slightly wrong.  However, with the noise being purely
  Gaussian locally,
  the only source of error in our estimate of
 $\sigma_{b}$ comes from the fact 
 that only a finite number of background pixels is available over which
 to estimate it. For this background estimate, we use  the same 
 number of pixels  
 as the number of pixels under the object, namely,
 $N_{pix}$.  We calculate below the error in Eq. A2 due to the
 finite number of pixels in the background estimate.

 We assume that a mean background level has been subtracted, so
that the distribution function of $(B_{ij} \pm B_{ij}^{180})$ is
Gaussian with zero mean and $\sigma = \sqrt{2} \sigma_b$.
 Given that $R_A$ and $R_T$ are both zero, it follows from the
above assumptions that the distribution function of 
$(R_{ij}\pm R_{ij}^{180})$ is the same as that of  $(B_{ij} \pm B_{ij}^{180})$.
Hence, there will be an identical error in the estimate of the quantity

\begin{equation}
 \Sigma \frac{1}{2} |R_{ij}\pm R_{ij}^{180}|.
\end{equation}

\noindent
  We can therefore add the two errors in quadrature to find the final total 
 error in $R_A$ and $R_T$ due to Gaussian background noise.

  The error in the background correction, 
$\delta(\Sigma\frac{1}{2}|B_{ij} \pm B_{ij}^{180}|)$,
is simply $N_{pix} \times \delta b$, where $b$ is the estimated 
value of the background, 
$\langle \frac{1}{2}|B_{ij} \pm B_{ij}^{180}| \rangle$,
and $\delta b$ is its error.  The distribution
function of $\frac{1}{2}|B_{ij} \pm B_{ij}^{180}| $ is
not Gaussian, but the error of its estimated mean can nevertheless
be calculated
using the central limit theorem, provided that the number of
samples (pixels) is large enough.
The central limit theorem
says that the computed mean using $N$ samples from a distribution $F(x)$
 approaches $<x>$ with an error $\sigma_{x}/\sqrt{N}$, where
 $\sigma_{x}$ is the variance of $x$.
   The minimum $N$ for this to hold with some accuracy is typically
 $N=100$,  which is generally true in our case since $N_{pix} > 100$.
The error in the background correction can then be shown to be

\begin{equation}
\delta(\Sigma\frac{1}{2}|B_{ij} \pm B_{ij}^{180}|)
 = \sqrt{\Sigma (\delta |B_{ij}|)^2},
\end{equation}

\noindent 
  where $(\delta |B_{ij}|)^2$ is the variance
of $|B_{ij}|$.  
In deriving Eq. (A4) we have taken into account that the left side sums 
 over all pairs of pixels twice.

 The formula for the variance is  

    \[ \sigma_{x}^{2} = <x^{2}> - <x>^2. \]

\noindent
Applying this to the variance $(\delta |B_{ij}|)^2$ we have

\begin{equation}
 (\delta|B_{ij}|)^2 = \langle |B_{ij}|^{2} \rangle  
  - \langle |B_{ij}| \rangle^{2}.   
\end{equation}

\noindent
 The first term in Eq. (A5) is

\begin{equation}
 \langle |B_{ij}|^{2} \rangle = 
 \int_{0}^{\infty} x^{2} {\rm exp}^{-\frac{x^{2}}{2 \sigma_{b}^{2}}} dx / 
 \sqrt{\pi/2} ~\sigma_{b}  
 = \sigma_{b}^{2},
\end{equation}

\noindent
 and the second term in Eq. (A6) is

\begin{equation}
 \langle |B_{ij}| \rangle = 
 \int_{0}^{\infty} x {\rm exp}^{-\frac{x^{2}}{2 \sigma_{b}^{2}}} dx / 
 \sqrt{\pi/2}~ \sigma_{b}  
 = \sqrt{\frac{2}{\pi}}~\sigma_{b}.
\end{equation}

Thus, the rms value of  $|B_{ij}|$ is

\begin{equation}
 \delta|B_{ij}| = \sqrt{\frac{\pi-2}{\pi}} \sigma_{b}.
\end{equation}

\noindent
 Using Eq. (A4) and Eq. (A7),
 the uncertainty in the background correction is then,

\begin{equation}
 \delta(\frac{1}{2}\Sigma|B_{ij}\pm B_{ij}^{180}|)
 = \sqrt{N_{pix}} \delta|B_{ij}| 
 = \sqrt{\frac{\pi-2}{\pi} N_{pix}} \times \sigma_{b}.
\end{equation}

\noindent
 Adding the identical term in quadrature for the residual sum,
 we obtain the total error in $R_{T}$ and $R_{A}$, 

\begin{equation}
 \delta R_{T~or~A}
 =\frac{\sqrt{\frac{2(\pi-2)}{\pi}N_{pix}} \times \sigma_{b}}{\Sigma I_{ij}}.
\end{equation}

\noindent
We ignore the error in the denominator ---
 the sum in the denominator is a sum of total intensities
 whereas both sums in the numerator are sums of residuals, and they are 
 furthermore subtracted, so it is fair to assume that the fractional
 error in the denominator is small and can be ignored.

  Since the signal-to-noise of an object image is defined as 
 \[S/N = {\Sigma \, I_{ij}} / \sqrt{N_{pix}} \, \sigma_{b} \]

\noindent
  we have

\begin{equation}
 \delta R_{T~ or~ A} \simeq 0.85 (S/N)^{-1}.
\end{equation}
\noindent
 However, $R$ is the sum of $R_A+R_T$, and the pixels used are common to both 
$R_{T}$ and $R_{A}$. Hence it is 
 reasonable to assume that the errors $\delta R_A$ and
$\delta R_T$ add arithmetically rather than 
 in quadrature, and we finally get

\begin{equation}
 \delta R \simeq 1.7 (S/N)^{-1}.
\end{equation}

  One can also derive a similar relation for the asymmetry parameter, $A$,
defined as 
\[ A =\frac{\Sigma |X_{ij} - X_{ij}^{180}| - \Sigma |B_{ij}-B_{ij}^{180}|}
{\Sigma I_{ij}}, \]

\noindent
 where $X_{ij}$ is the flux or a flux related quantity at $(i,j)$. 
  In that case,  we get an almost identical formula to Eq. A12, namely,

\begin{equation}
 \delta A = 1.7 (S/N)^{-1}.
\end{equation}

\section{Simulated images of Frei et al. galaxies}

\begin{figure*}[hb]
\vskip 0.0cm
\psfig{figure=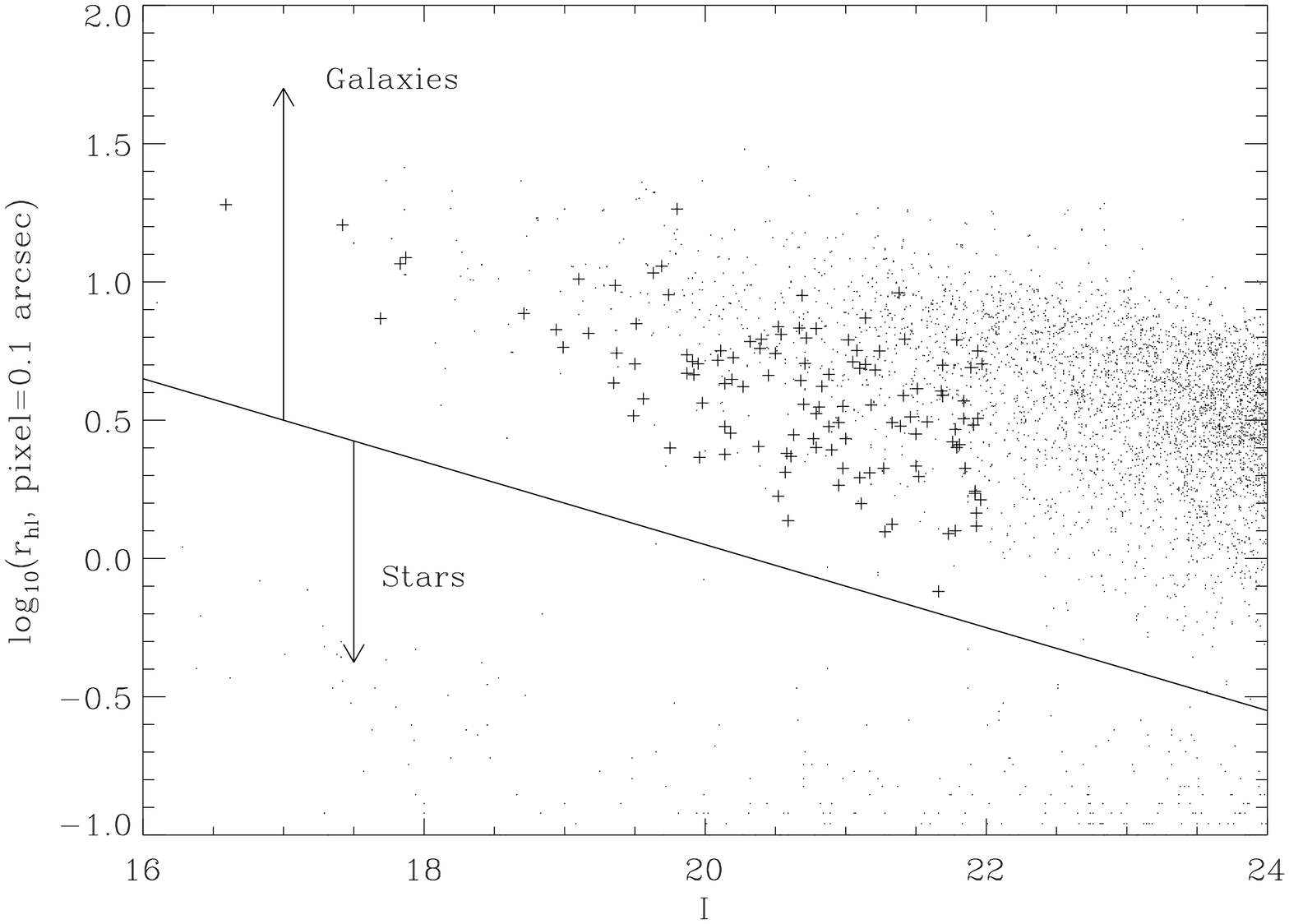,width=12.0cm,angle=90}
\figcaption[f25.ps]{
 Size-magnitude relation of Groth Strip QS-E/S0s (plusses) superposed on 
 other galaxies in GSS. The solid line divides 
 stellar objects and galaxies.
 \label{fig:smag_eso}
}
\end{figure*}

  As we explained in Section 3, we rescaled and added noise to
 the images of local galaxies from Frei et al.(1996),
 in order to test our quantitative scheme of 
 morphological classification when the image quality is close 
 to the GSS data. 
  The size-magnitude relation of Groth Strip QS-E/S0s is plotted
 in Figure \ref{fig:smag_eso} 
 together with the size-magnitude relation of other types of galaxies.
  Like nearby galaxies,
  E/S0s tend to be more compact than other types of galaxies at a given
 magnitude since E/S0s have centrally concentrated light profiles.
  The $r_{hl}\simeq 5$ pixels corresponds to $r_{hl,med}$ of $I < 21$ 
 E/S0s, and $r_{hl}\simeq 3$ pixels is roughly equal to 
 $r_{hl,med}$ of $21 < I < 22$ E/S0s. 
  Hence, images are rebinned and box-averaged so that the simulated galaxies 
 have $r_{hl}\simeq~5,3,2,1.5$,and $1$ pixels.   
  Noise is added so that the simulated galaxy images
 have $S/N \sim 30-70$, comparable to GSS galaxies at $20 < I < 22$
 (see Figure \ref{fig:sn}).
  We also add an additional background area to the simulated image to help  
 determining the background correction for $R$ more accurately, since
 some of the Frei et al. galaxies almost fill up the original image,  
 making it difficult to estimate the background noise.
  We have not convolved the images with the WFPC2 PSF;  
 however, this does not affect significantly our results since the 
 Wide Field Camera portion of the WFPC2 undersamples the PSF. 
  Simulated images of representative galaxies 
 are shown in Figure  \ref{fig:frei_monex}.
  When $r_{hl} \sim 3$ pixels, the morphological details of 
 late-type spiral galaxies are still somewhat visible, and $R$ gives us
 a good measure of the complexity in their morphology. 
  When $r_{hl} \simeq 2$ -- $3$ pixels, it becomes challenging to
 classify galaxies by eye: Non-smooth, and asymmetric features of 
 galaxy morphology are visible for late-type galaxies, but more detailed
 morphological features such as spiral arms are almost completely wiped out.
  For those objects, $R$ still provides a reasonable measure of morphological
 detail from the non-smoothness and asymmetry in galaxy SB profile. 
  When $r_{hl} \simeq 1$ pixels, it is almost impossible to
 classify galaxies by eye, but $R$ can still be used 
 to identify the late-type galaxies with large $R$.

\vskip 0.2cm

\begin{figure*}[ht]
\vskip 0.2cm
\figcaption{
  Images of 16 representative Frei et al. galaxies,
 simulated to have $r_{hl}\simeq~5, 3,
 2$, and 1 pixels and $S/N \sim $30--70.
  Images are ordered by morphological type (T-type).
  Numbers indicated in each panel are (from upper left corner in clockwise
 direction): (1) galaxy name (NGC number), (2) T-type, (3) output
 $r_{hl}$ (pixels), (4) $R$, and (5) $B/T$.
\label{fig:frei_monex}
}
\end{figure*}

\section{Clustering properties of E/S0s out to $z \simeq 1$}

  Studies of local galaxies show that bright E/S0s are more strongly
 clustered than other types of galaxies, and that 
  E/S0s live preferentially in high density environments 
 (Willmer, da Costa, \& Pellegrini  1998; Loveday et al. 1995). 
  However, there is no study which estimates the clustering 
 of morphologically selected E/S0s at high redshift. 
 In order to estimate the clustering of E/S0s, we 
 can take three approaches: 
 (1) use the clustering properties of bulge-dominated galaxies 
 (e.g., Neuschaefer et al. 1997),  
 (2) use the clustering properties of red galaxies  as estimated from
 other redshift surveys (e.g., LeFevre et al. 1996), 
 or (3) estimate the clustering properties of the  
 Groth Strip QS-E/S0s directly from our GSS data.
  The first two approaches may suffer from contamination: 
 in the first from late-type galaxies with high $B/T$ ($>0.3$),
 and in the second from early-type spirals as well as dusty late-type 
 galaxies.

  According to Neuschaefer et al. (1997),
 the amplitude of the two-point angular correlation 
 function  of bulge dominated galaxies
 ($A_{w}$ in $\omega \simeq A_{w} \, \theta^{-0.8}$) normalized
 at 1$\arcmin$ is about 0.1 at $18 < I < 22$.
  Note that this value is several times greater than the value of $A_{w}$
 for the rest of the population in Neuschaefer et al. (1997),
 consistent with the clustering of E/S0s at $z=0$ relative to
 the other types of galaxies. 
   Neuschaefer et al. (1997) also find that the $A_{w}$ of MDS bulge
 dominated galaxies is well fit by a clustering model with
 a clustering scale of $r_{0} = 5.5~ h^{-1}$ Mpc fixed
 in  physical (or proper) coordinates.
  If we model the evolution of the spatial correlation function as
  $\xi(r)=(r_{0}/r)^{\gamma}/(1+z)^{3-\gamma+\epsilon}$ with respect to
 the comoving coordinate system, the above case corresponds to $\epsilon = 0$.
  With this model, $r_{0}(z=0.75)=4~ h^{-1}$ Mpc in comoving coordinates;  
  however,  since studies of local E/S0s suggest 
 that $r_{0,E/S0} \sim 6$ -- $8~ h^{-1}$ Mpc (Loveday et al. 1995; 
 Guzzo et al. 1997; Willmer et al. 1998), it seems more reasonable 
 to model the clustering evolution as $r_{0}(z)=r_{0,E/S0}/(1+z)^{(3-\gamma
 +\epsilon)/\gamma}$ with $\epsilon$ = 0.8. With this model, we also get
 the observed amplitude of the two-point angular correlation function,
 $A_{w} \simeq 0.1$. Furthermore, this model is consistent with the expectation
 of the linear growth of the density perturbation.

  As for the second approach, LeFevre et al. (1996) estimate that 
 $r_{0} = 2.5~h^{-1}$ Mpc for red galaxies in the CFRS.
  This value appears to be much lower than the value quoted for
  bulge-dominated galaxies.
  In LeFevre et al. (1996), red galaxies are selected as 
 galaxies with colors redder than those of Sbc type SED; 
 it might be that more weakly clustered galaxies 
 are introduced into the red galaxy sample by this somewhat less conservative
 color selection. 
 
  The third approach seems quite attractive, since the 
 clustering property is internally determined from the data we are studying.
  The variance of the number density of galaxies for a given survey
 geometry for  which $V^{1/3} < r_{0}$ is

\begin{equation}
 \delta n^{2} = n^{2} (\langle \xi \rangle  + 1/n)
\end{equation}

\noindent
  Here $<\xi>$ is the average of the two-point correlation function
 over the volume, and can be approximated for a slab geometry as

\begin{equation}
 \langle \xi \rangle \simeq  3.9 \frac{r_{0}^{\gamma}}{V^{\gamma/3}}.
\end{equation}

\noindent
 To obtain  the mean and the variance of the number density of GSS spheroids
 for a small volume element V,
  we divide our GSS volume into 10 equal volumes from $z=0.5$ to $z=1$ 
 in the radial direction. From these volume elements,
 we find the $\delta n^2 \simeq 340$,
 and $\bar{n} = 8.2$. The clustering evolution as a function of redshift
 is ignored here for this order of magnitude calculation.   With these 
 values, we get $r_{0} \simeq 4\, h^{-1}$ Mpc (in comoving scale)
 at $z \sim 0.75$, 
 consistent with the value preferred by the study of the angular correlation 
 function of bulges (Neuschaefer et al. 1997). 
  Note that these values are 
 estimated from the Groth Strip QS-E/S0s with spectroscopic redshifts by  
 applying a selection function similar to that in Koo et al.(2000).
  The photometric redshift
 sample can not be used for this particular estimate of 
 $\langle \xi \rangle$ since the uncertainty in the
 photometric redshift ($\delta z \sim 0.1$) is big enough to smooth out 
 the true clustering property along the radial direction. 

  The excellent agreement between the first and the third methods demonstrates 
 that it is reasonable to assume the clustering of E/S0s as
 $\xi(r,z)=(r_{0}/r)^{\gamma}/(1+z)^{3-\gamma+\epsilon}$ where we adopt 
 $r_{0} = 8 \,h^{-1}$ Mpc, $\gamma=1.8$ and $\epsilon=0.8$. 
 Therefore, we adopt this clustering model for the 
 Groth Strip QS-E/S0s to estimate
 the error in their number density. 

\end{document}

%% file: intro.tex
  In the local universe,
 E and S0 galaxies are overwhelmingly found to be 
 red galaxies that have smooth,
 centrally concentrated surface brightness profiles and
 a high degree of elliptical symmetry.  
  Giant 
 E/S0s have surface brightness profiles dominated by the $r^{1/4}$
 law, which also characterizes luminous bulges of spiral galaxies 
(see Mihalas \& Binney 1981 and references therein). 

 The formation of E/S0 galaxies is of interest because it is a sensitive
 barometer for the timescale of the formation of galaxies
 and structure in the universe.
  There are two extreme theories 
 for the formation of E/S0s. One is that E/S0s formed very early
 via monolithic collapse of the protogalactic gas at high redshift
 (the so-called monolithic collapse model; Eggen,
 Lynden-Bell, \& Sandage 1962; Larson 1975).
   The other is that E/S0s formed via a continuous process of merging,
 as occurs
 in hierarchical models of structure formation in which
 the matter density is dominated by cold dark matter (CDM; Blumenthal et al.
 1984; Baron \& White 1987).
 For example, in a flat universe with $\Omega_{m} = 1$, 
 some hierarchical merger models predict that 
 as many as 50--70\% of present-day
 ellipticals assembled later than  $z = 1$
 (Kauffmann, White, \& Guiderdoni 1993; Baugh, Cole, \& Frenk 1996a).
   However, this formation 
epoch is sensitive to the assumed values of 
 cosmological parameters, with the major merger epoch 
 occurring at higher redshift in universes with lower matter density
 (Kauffmann \& Charlot 1998).
  Other physical processes, such as more frequent merging at higher 
 redshifts, could also influence the prediction (e.g., Somerville, Primack, \&
 Faber 2000).  

   A key observational test 
 of whether E/S0 galaxies have evolved strongly at recent epochs
 is to measure the number density and colors of 
 carefully selected E/S0 galaxies  
 out to $z \sim 1$. An $L^{*}$ early-type galaxy ($M_{B} \simeq -19.4\,
 +5\,{\rm log}(h)$;   Marzke et al. 1998) would have an apparent magnitude of
 $I \simeq 20$-$21$ at $z\simeq 1$ depending on the details of its  
 luminosity evolution. Currently, the determination of redshifts of 
 $I \sim 22$ early-type galaxies is feasible
 with ground-based telescopes of aperture $\gtrsim 4$ m
 (Koo et al. 1996; Lilly et al. 1995; Cowie et al. 1996;
 Cohen et al. 1996, 2000). 
   Moreover, previous studies show that  
  morphological classification of faint galaxies is possible 
  down to $I \simeq 22.5$ using 
  HST images with exposure times $\gtrsim$1 hour
 (Griffiths et al. 1994a, 1995b; Driver, Windhorst, \& Griffiths 1995; 
  Glazebrook et al. 1995; Im et al. 1996, 1995a;
  Abraham et al. 1996; Brinchmann et al. 1998; Schade et al. 1999; 
  Im, Griffiths, \& Ratnatunga 2000; Roche et al. 1996, 1998).
  Thus, in principle, a combination of HST imaging and 
 spectroscopy using large ground-based telescopes can provide the data 
 to constrain the number density evolution of E/S0s at $z < 1$.

   Testing for evolution in the comoving number density
 of any galaxy population requires selecting exactly the
 same kinds of objects at higher redshifts that comprise
 the target sample at the current epoch.
  In this paper, we aim to select 
 galaxies that
 meet {\it quantitative} morphological criteria for
 being E/S0s.  Galaxy colors redden rapidly
 after a starburst, approaching within 0.2 mag of their
 asymptotic $B-V$ values in less than 2 Gyr (e.g., Worthey 1994).
 This is also roughly the time scale for morphological peculiarities
 induced by starbursts to smooth out and disappear (Mihos 1995).
   Essentially,  our selection criteria 
 choose just those galaxies at each redshift that are morphologically
 the most symmetric and smooth. We  expect that
 galaxies selected this way  have had the least star formation
 over the last few Gyr well past their last merger, and, hence,
 should occupy the \emph{red envelope} of galaxies at each redshift.
  If galaxies reach the red envelope
 and stay there, our method measures how their numbers accumulate
 over time.

  Previous studies of the evolution of field E/S0 number densities
 utilize various data sets and reach contradictory conclusions.
  To study distant early-type galaxies that did not have HST images,
 Kauffmann, Charlot, \& White (KCW; 1996) use the data from the
 Canada France Redshift Survey (CFRS; Lilly et al. 1995)
 and select early-type galaxies 
 from the red envelope based on evolving model color curves
 in $z$ vs. $V-I$.  Based on $\langle V/V_{max} \rangle$ statistics
 (Schmidt 1968),  they claim a significant deficit
 of red early-type galaxies at $z \sim 1$, with the number density
 there being roughly one-third of the value at $z = 0$.  
 They interpret this strong evolution as evidence
 for recent assembly and star formation in E/S0 galaxies.
  However, subsequent works have shown that the CFRS sample is 
 deficient in red galaxies beyond $z = 0.8$, probably 
 due to incompleteness in the redshift measurements, and that cutting
 the sample lower than that redshift removes all evidence 
 for evolution (Totani \& Yoshii 1998; Im and Casertano 2000).
 These works also note that using model color curves to select
 galaxies from the red envelope alone is risky; 
 models by different authors vary, and
 the region identified as the red envelope  depends heavily on the history of
 star formation assumed in the models. Possible systematic errors in $V-I$ can 
 also affect sample selection significantly.

  Both qualitative and quantitative classification 
 methods have been applied to deep HST images to isolate samples of
 morphologically selected early-type galaxies. 
   Schade et al. (1999) study CFRS galaxies that also have HST images.
   They choose ellipticals based on a good fit to an
   $R^{1/4}$ law together with an asymmetry parameter;
   no color cut is used.  They see no drop-off
  in numbers out to $z = 1$, in apparent contradiction to 
  KCW, but many of their most distant E/S0s are blue and
  lie below the color cut adopted by KCW---it is not clear
  that these objects would be classified as E/S0s if seen locally.
    A similar conclusion regarding the lack of large number-density evolution
 of E/S0s is also reached by Im et al. (1999), who combine spectroscopic 
 redshifts available in the literature with existing HST images.
 E/S0s in Im et al. (1999) are selected as galaxies that have  
 a significant bulge ($B/T \gtrsim 0.3$) {\it and} appear morphologically
 featureless when visually classified. They present redshift distributions
 of various galaxy types and find that those of
 E/S0s is consistent with no number density evolution out to 
 $z \sim 1$ (for similar works, see Roche et al. 1998; Driver et al. 1998;
 Brinchmann et al. 1998). 
  However, the number of early-type galaxies in both of these studies
 is small (about 40), and thus it is difficult to draw 
 firm conclusions.

   Results from the number counts of larger sets of 
 morphologically selected early-type galaxies are also consistent
 with little number-density evolution out to $z=1$
 (Im, Griffiths, \& Ratnatunga 2000; Driver et al. 1996, 1998;
 Menanteau et al. 1999).
  For example, Menanteau et al. (1999) study galaxies in
 deep HST archive exposures, supplemented by ground-based
 infrared H+K' near-infrared imaging.  They classify
 galaxies morphologically using both quantitative
 (A/C method; Abraham et al. 1996) and qualitative
 (visual) methods.  Lacking redshifts, they model galaxy
 counts versus color and conclude that the number density
 of distant spheroids has not changed substantially since $z \sim 1$.
 However, this conclusion applies only if blue spheroids
 are included; a major deficit of red spheroids with $V-(H+K') > 2.0$ 
 is seen.  Since models predict that these are precisely the 
 colors of red spheroids at $z \gtrsim 1$, this could imply a strong 
 decline in the number of galaxies in the red envelope at $z \gtrsim 1$,
 although this does not put strong constraints on evolution at $z < 1$
 (for similar works but different views,
 see Barger et al. 1999; Bershady, Lowenthal \& Koo 1998; Broadhurst \&
 Bouwens 2000).
  However, modeling counts is notoriously
  sensitive to the assumed luminosity function
  and star formation history,  and
  essentially similar count data
  can also be fitted to models 
  with substantial number density evolution (Baugh et al. 1996b; 
  Im et al. 2000). Redshift information remains essential to break 
  the degeneracy in the model predictions.

 Our method in this paper uses the comoving luminosity function
to characterize the number density of distant E/S0 galaxies.
 This method yields both a characteristic magnitude $M^*$ 
as well as the local number density normalization $\phi^*$.
Variations in $M^*$ versus redshift are an additional measure of galaxy 
evolution; if ellipticals are just forming at $z \sim 1$,
we would expect their luminosities to be very bright there
(and their colors to be very blue).  Luminosity evolution can
also be tracked in other ways, for example, 
by studying zeropoint offsets in the size-luminosity
relation or the fundamental plane, and color offsets
can be measured using residuals from the color-magnitude
relation.  The absolute stellar ages of E/S0 stellar populations
can also be measured by fitting to broadband colors.  
Unlike the luminosity function method, these techniques
all yield useful information with just a few objects,
but their conclusions may in consequence be less general.

 As a group, such studies are converging on the consensus
that distant E/S0 galaxies show both luminosity and color
evolution, and that these effects are 
are consistent with models in which
the bulk of stars in E/S0 galaxies formed before,
in some cases well before, $z \sim 1$.
 Many studies 
of both field and cluster E/S0s indicate that the rest-frame B-band 
 surface brightness of these galaxies is brighter 
 at higher redshifts, consistent with the expected brightening
 of passively evolving  stellar populations formed well previously
(Schade et al. 1997, 1999; van Dokkum et al. 1998; 
 Bender et al. 1998; Kelson et al. 1997; J{\o}rgensen et al. 1999; 
 Treu et al. 1999; Gebhardt et al. 2000).
  Cluster colors are basically consistent with this picture (e.g.,
Stanford, Dickinson, \& Eisenhardt 1998), 
 but some surveys 
 find that roughly 30--50\% of morphologically normal field E/S0 galaxies 
 are quite blue due to recent star formation
 (Schade et al. 1999; Menanteau et al. 1999; also see Abraham et al. 1999).
  Other field studies
 do not find  many blue, bright early-type galaxies 
 out to $z\sim 1$ (Im et al. 1996; Im, Griffiths, \& Ratnatunga 2000;
 Kodama et al. 1999; Franceschini 1998). The discrepancies
 between these studies 
 appear to be caused by differences in morphological classification
 and sample-selection criteria.  Quantitative classification of E/S0s,
 such as we apply here,
 can help resolve these by offering an objective way of 
 selecting E/S0s.
       
 The previous study of the E/S0 luminosity function
that most closely resembles ours is by
Im et al. (1996).  These authors use 376 visually classified field
E/S0s at $I<22$ from the HST Medium Deep Survey and
other HST fields to construct a luminosity function (LF)
to $z \sim 1.2$.  They find that 
$L^*$ field E/S0 galaxies brighten by 1--1.5 magnitudes back to $z \sim 1$,
consistent with passive evolution models.  They also find no
significant number density evolution from $z \sim 0.2$ to $z \sim 1$.
 The sample used by Im et al. (1996) is the largest to date with
complete redshifts and Hubble-type classifications from HST images.
However, the redshifts are based on $V-I$ colors with
only a limited number (23) of spectroscopic calibrators.  The morphological
classifications are also mostly based on qualitative visual methods.

 The present paper follows the basic plan of Im et al. (1996) but
with several important additions.  Like that paper, we start with
a sample of galaxies with HST images (in the Groth
strip) and with spectroscopic redshifts ($z_{spec}$, from Keck) that are
used to calibrate our photometric redshifts based on $V-I$.
 However, our sample of spectroscopic $z$'s is much 
larger (44) and extends more uniformly to $z\sim 1$.
 Second, we invest considerable effort in developing morphological
classification criteria that isolate red-envelope galaxies
to high accuracy.  We show that the selection procedure works using
the Keck redshift sample, which
allows us to plot whether candidate galaxies actually lie on the
red envelope; we tune the procedure so that they do. 
  A final check by color cut in 
 $V-I$ vs. $I$ succeeds in identifying 
 and correcting for a small fraction of blue
 interlopers.
  For the resultant red-envelope sample,
 $V-I$ is an excellent photometric redshift indicator,
 again demonstrated by the test sample with Keck redshifts.
  For the blue interlopers, the photometric redshifts will be underestimated,
 but we find that they do not bias our results ($< 10$\%)
 since their number is small, and they affect
 mostly the very faint end of the LF. 
 For that reason, we do not try to eliminate the blue interlopers 
 from our sample by the color criteria. This is useful for keeping
 our selection criteria simple. 
  The net result is a technique that can identify E/S0 candidates that
 predominantly occupy the red envelope
 to a given magnitude limit \emph{using only $V$ and $I$ HST images}.
 This is used to expand the final sample, which numbers 145 GSS galaxies,
 by including galaxies that lack spectroscopic redshifts.
  With this significant sample size, our study places a
 significant constraint on the number density and
 luminosity evolution of luminous field E/S0 galaxies
 to $z \sim 1$.

  This paper is one of three 
 papers on the evolution of early-type galaxies
 based on DEEP data.
  Results on the study of luminous bulges will be presented in
 Koo et al. (2000), and the fundamental plane of field E/S0s out to
 $z \sim 1$ will be explored by Gebhardt et al. (2000). 
   Section 2 presents the basic observational data for the present study.
  Section 3 develops the method for selecting
 E/S0s using HST images and explains how it is applied.
  Section 4 shows our fits for 
 the luminosity function, and
 Section 5  provides a short discussion of the results.
 Conclusions are given in Section 6.
 
   The Hubble constant, $H_{0}$, is quoted as
 $h=H_{0}/$(100 km sec$^{-1}$ Mpc$^{-1}$), and we adopt the 
 value $h=0.7$ throughout this paper.

%% file: figures.tex
\begin{figure*}[ht]
\figcaption{
 Images of nearby E (top), Sbc (middle), and Im (bottom) galaxies
 and their residual images after subtracting the best-fit model image.
 The azimuthally averaged one-dimensional surface brightness profile
 along the major axis (points with errors) is also plotted
 and compared to the model-fit 1-D profile
 for the bulge (dashed line),
 the disk (dotted line), and the sum of bulge plus disk (solid line).
 Numbers below the residual images indicate the bulge-to-total light 
 ratio ($B/T$) and the residual parameter ($R$).
 \label{fig:resimage}
}
\end{figure*}

\begin{figure*}[hbt]
\psfig{figure=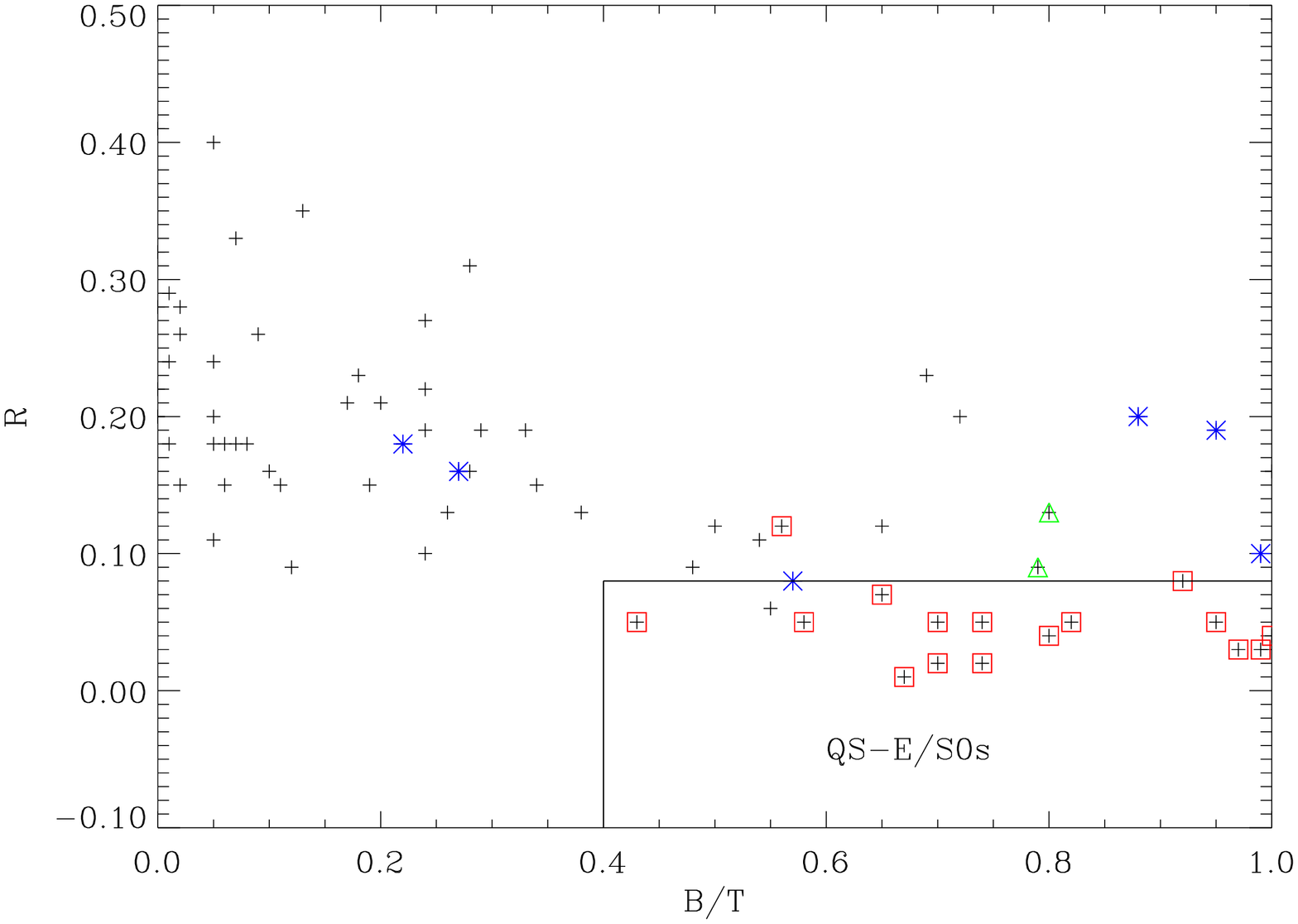,width=16.0cm,angle=90}
\figcaption[f2.ps]{ 
 The quantity $R$ vs. $B/T$ for 80 nearby galaxies from Frei et al. (1999).
 Red squares are for T-type, $T \leq -3$, green triangles
 are for $T = -2$, and blue stars are for $-2 < T \leq 0$.
 Other types of galaxies are marked with crosses. The lower-right
 corner of the figure surrounded by a box represents the region
 where Es and S0s are found without much contamination 
 from other types of galaxies. \label{fig:a-bt}
}
\end{figure*}

\begin{figure*}[th]
\psfig{figure=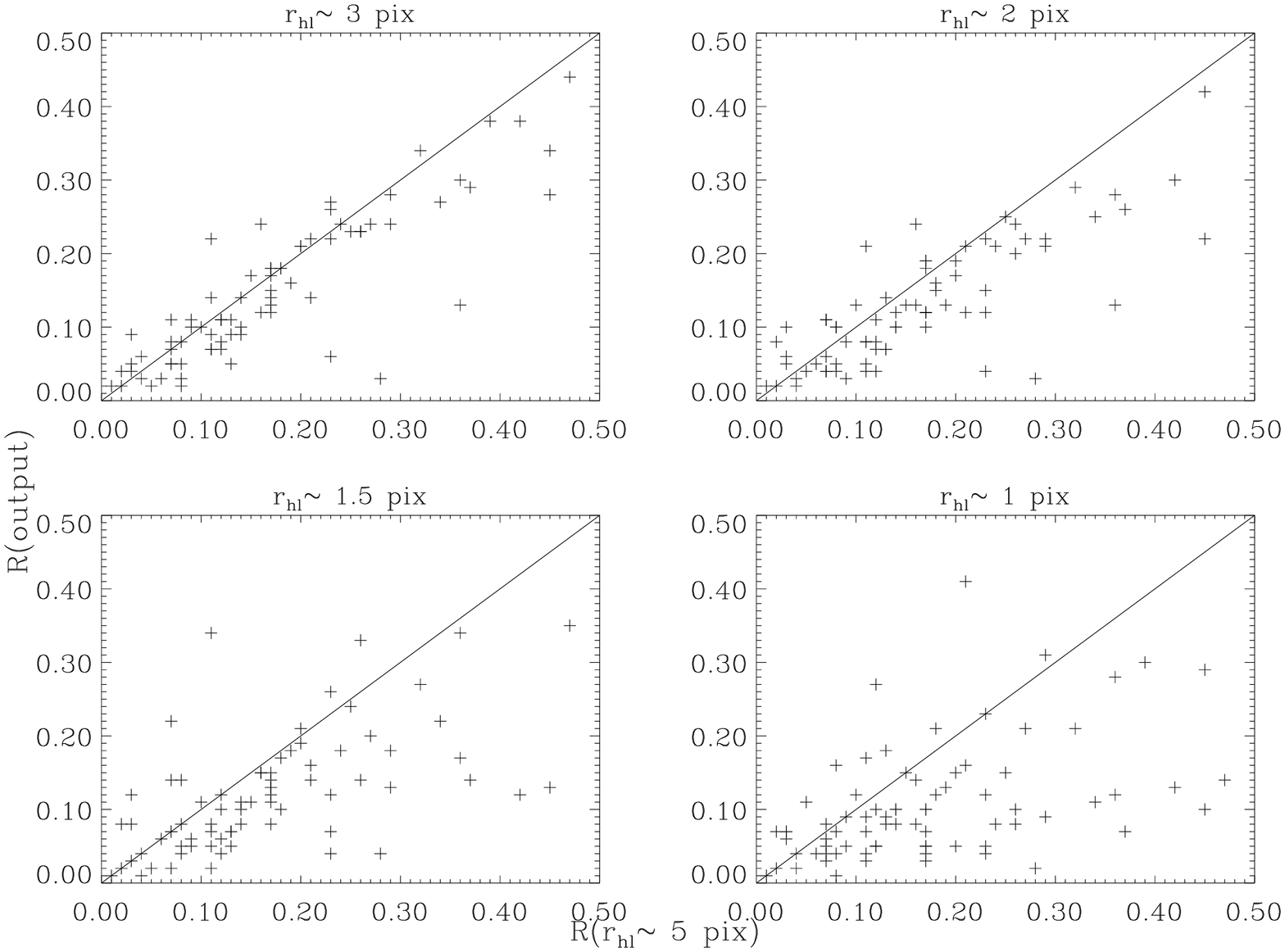,width=16.0cm,angle=90}
\figcaption[f3.ps]{
 The asymmetry parameter $R$ of Frei et al. galaxies 
 shrunk by different amounts,
 vs. input  $R$ values ($r_{hl}=5$ pix).
 Galaxies are binned by output $r_{hl}$.
Output values of $R$ are reasonably well correlated with
input values down to $r_{hl} = 2$--3 px.
 \label{fig:rrcom}
}
\end{figure*}

\begin{figure*}[bh]
\psfig{figure=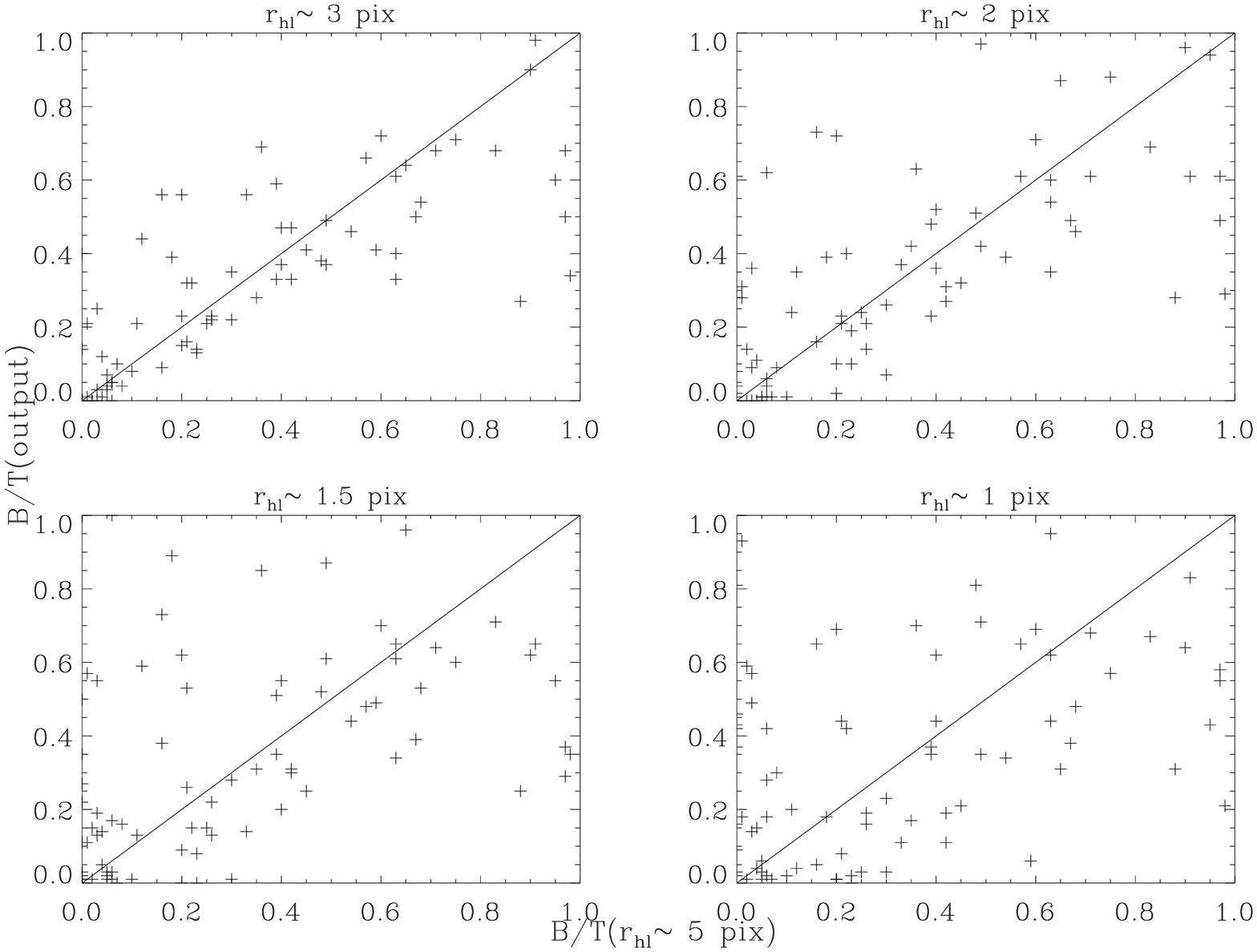,width=16.0cm,angle=90}
\figcaption[f4.ps]{
  $B/T$ of Frei et al. galaxies 
 shrunk to  different sizes, vs. input $B/T$ values ($r_{hl}=5$ pix).
 Galaxies are binned by output $r_{hl}$.
Output values of $B/T$ are reasonably well correlated with
input values down to $r_{hl} = 2$--3 px.
 \label{fig:btcom}
}
\end{figure*}

\begin{figure*}[htb]
\centerline{\psfig{figure=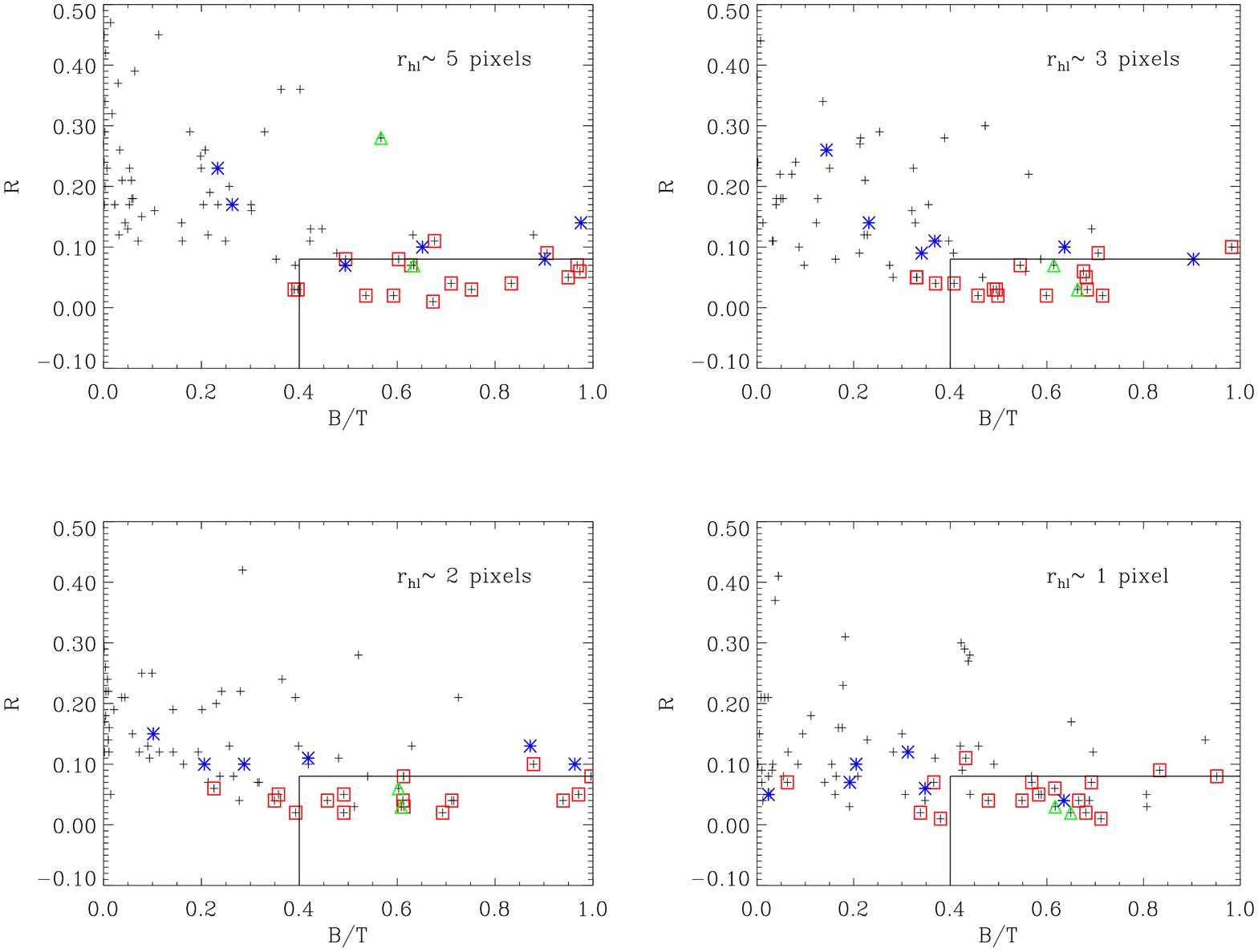,width=18.0cm,angle=-90}}
\figcaption[f5.ps]{
  $R$ vs. $B/T$ for simulated Frei et al. galaxies shrunk to 
 different sizes. Red squares are for  galaxy types $T \leq -3$,
 green triangles are for $-3 < T \leq -2$, and
 blue stars are for  $-2 < T \leq 0$. Other types of
 galaxies are marked with black crosses. 
 Galaxies are binned by output $r_{hl}$.
 The lines at lower right 
 corner of each plot represent a morphological cut at 
 $R = 0.08$ and $B/T=0.4$. 
 Within this boundary, the sample of QS-E/S0s is somewhat contaminated
 by spiral interlopers for $r_{hl} \sim 1$--2 pix.  However, this contamination
 is reduced in practice by reducing the $R$ boundary for  
 galaxies with very small apparent sizes,
as described in the text.
\label{fig:rall}
}
\end{figure*}

\begin{figure*}[bh]
\psfig{figure=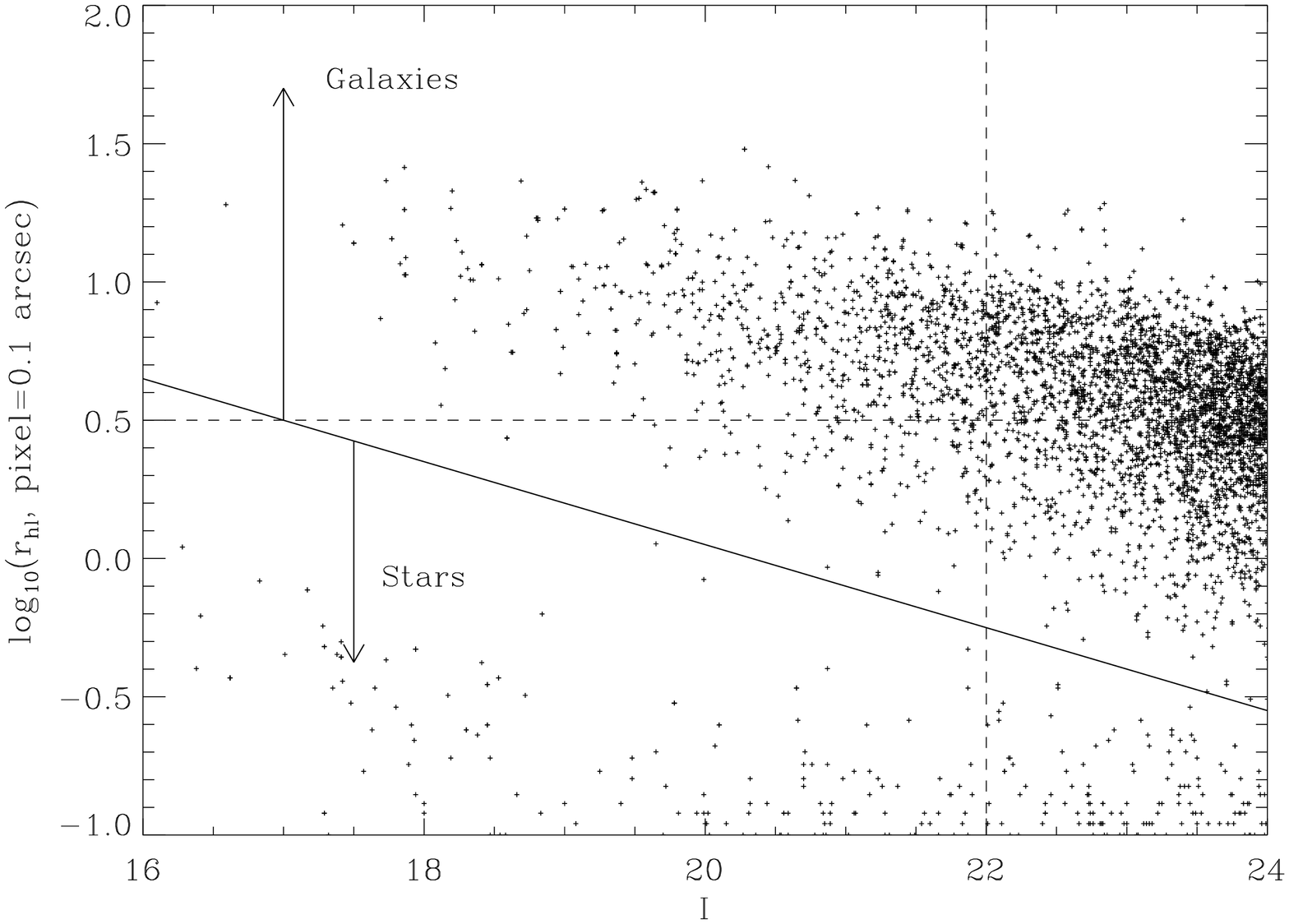,width=15.0cm,angle=90}
\figcaption[f6.ps]{
 Size-magnitude relation of objects in the GSS. The solid line divides 
 stellar objects and galaxies.
 The vertical dashed line marks the magnitude limit of the GSS
 sample.  The horizontal dashed line marks the limit above which both
 $R$ and $B/T$ are well measured without strong bias (3 pix).  Nearly all of the sample is in this
 regime.
 \label{fig:smag}
}
\end{figure*}

\begin{figure*}[bh]
\psfig{figure=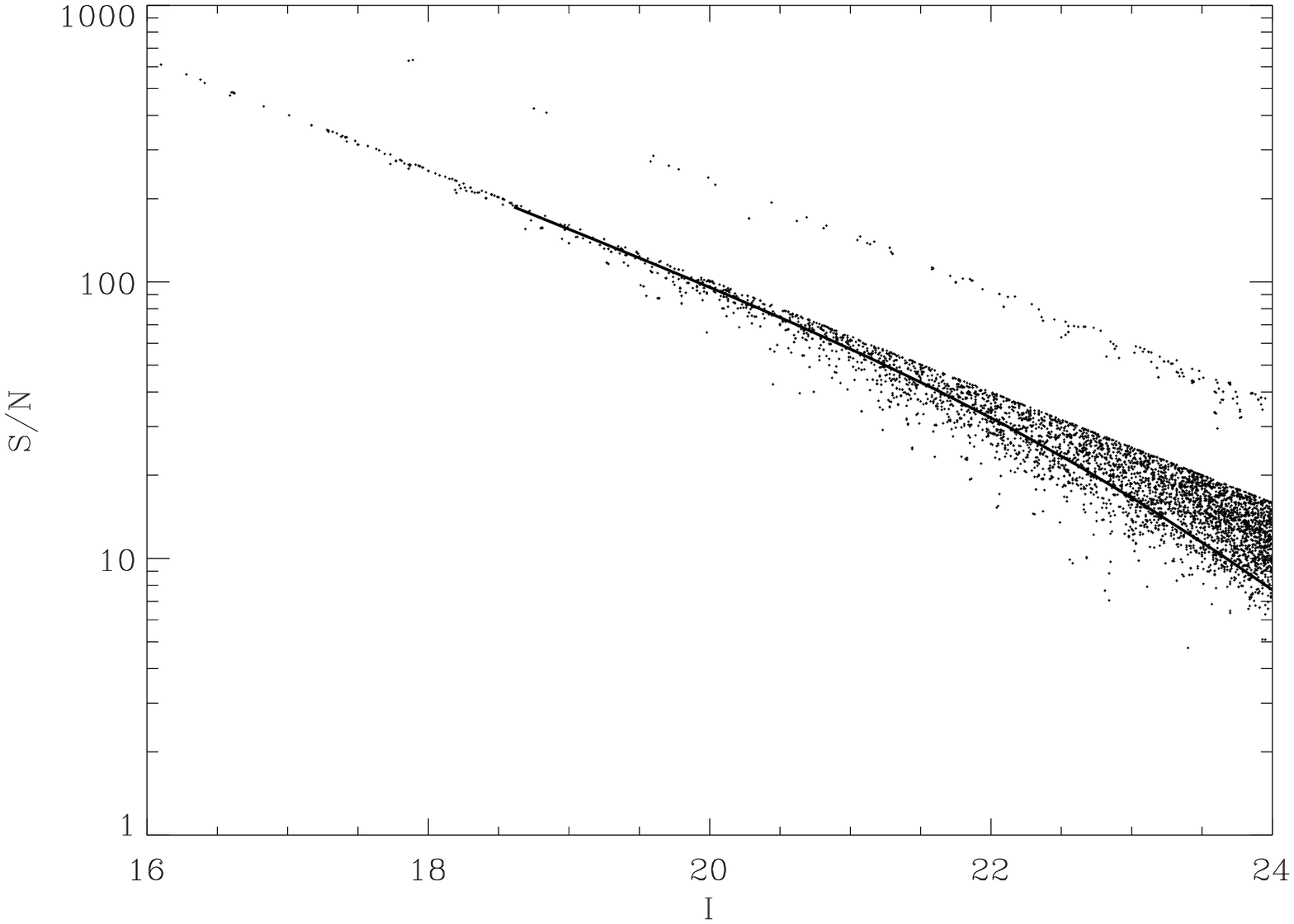,width=15.0cm,angle=90}
\figcaption[f7.ps]{
  Signal-to-noise ($S/N$) of objects in the Groth Strip. The $S/N$ is calculated within 
 a circle of radius $= r_{hl}$. 
  The solid line is a theoretical
 line for a galaxy with $r_{hl}=0.3$ arcsec (3 pix).
  Objects that form the higher $S/N$ sequence are from the deep GSS pointing 
 with long exposure time.
 \label{fig:sn}
}
\end{figure*}

\begin{figure*}[ht]
\centerline{\psfig{figure=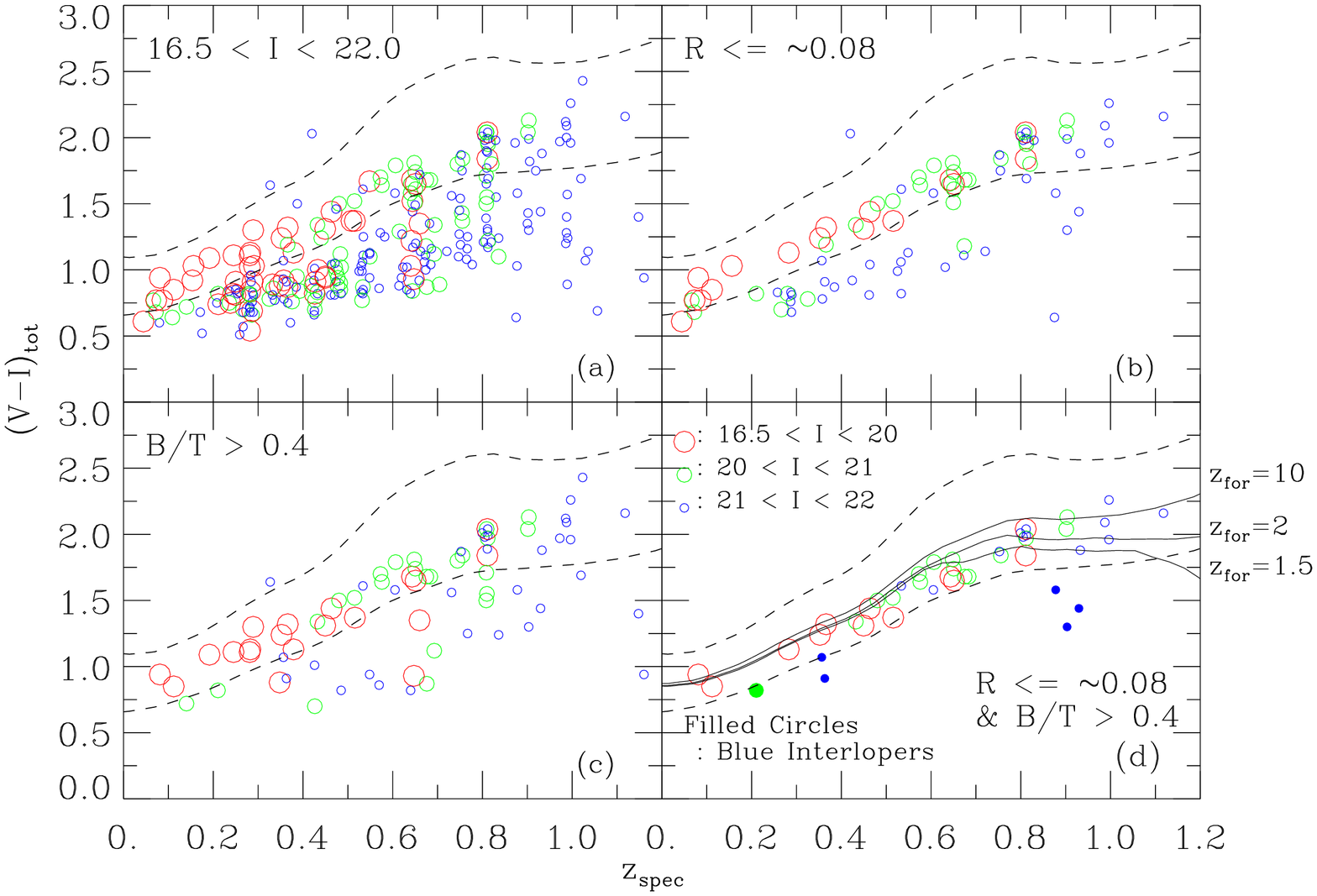,width=18.0cm,angle=90}}
\figcaption[f8.ps]{
  This figure shows the effect of each selection cut applied separately
 to isolate red-envelope E/S0s.
 $V-I$ colors are plotted vs. spectroscopic redshifts 
 for GSS galaxies with spectroscopic redshifts (the $z_{spec}$ sample) and
 $16.5 < I < 22$. 
  Figure (a): All 262 objects with $z_{spec}$ are plotted. 
           The large, red circles represent objects with $I \leq 20$,
       the mid-sized, green circles show objects with $21 < I \leq 21.0$,
       and the small, blue circles show objects with $21.0 < I \leq 22.0$.
        Also plotted are the theoretical upper and lower color boundaries for 
        passively evolving E-galaxy models (see text). 
  Figure (b): Objects that survive the residual parameter ($R$) cut 
       are  plotted. Note that the  $R$ cut used is a function 
       of the apparent half-light radius, as explained in the text.
  Figure (c): Objects that survive the bulge-to-total light ratio ($B/T$) cut  
       are plotted. 
  Figure (d): Final, ``quantitatively selected (QS) E/S0s'' are plotted,
       which consist of 44 objects that survive both the $R$ 
       and  $B/T$ cuts. The blue interlopers
       (filled circles)
       represent a failure of the selection technique.  
       However, their number is small (15\%), and their photometric
       redshifts based on $V-I$ place them at low redshift, so they
       do not contaminate distant bins.
       Solid lines represent passively evolving solar-metallicity
       models with the formation times indicated.
  \label{fig:i22sel}}
\end{figure*}
\vskip 0.1cm

\begin{figure*}[ht]
\centerline{\psfig{figure=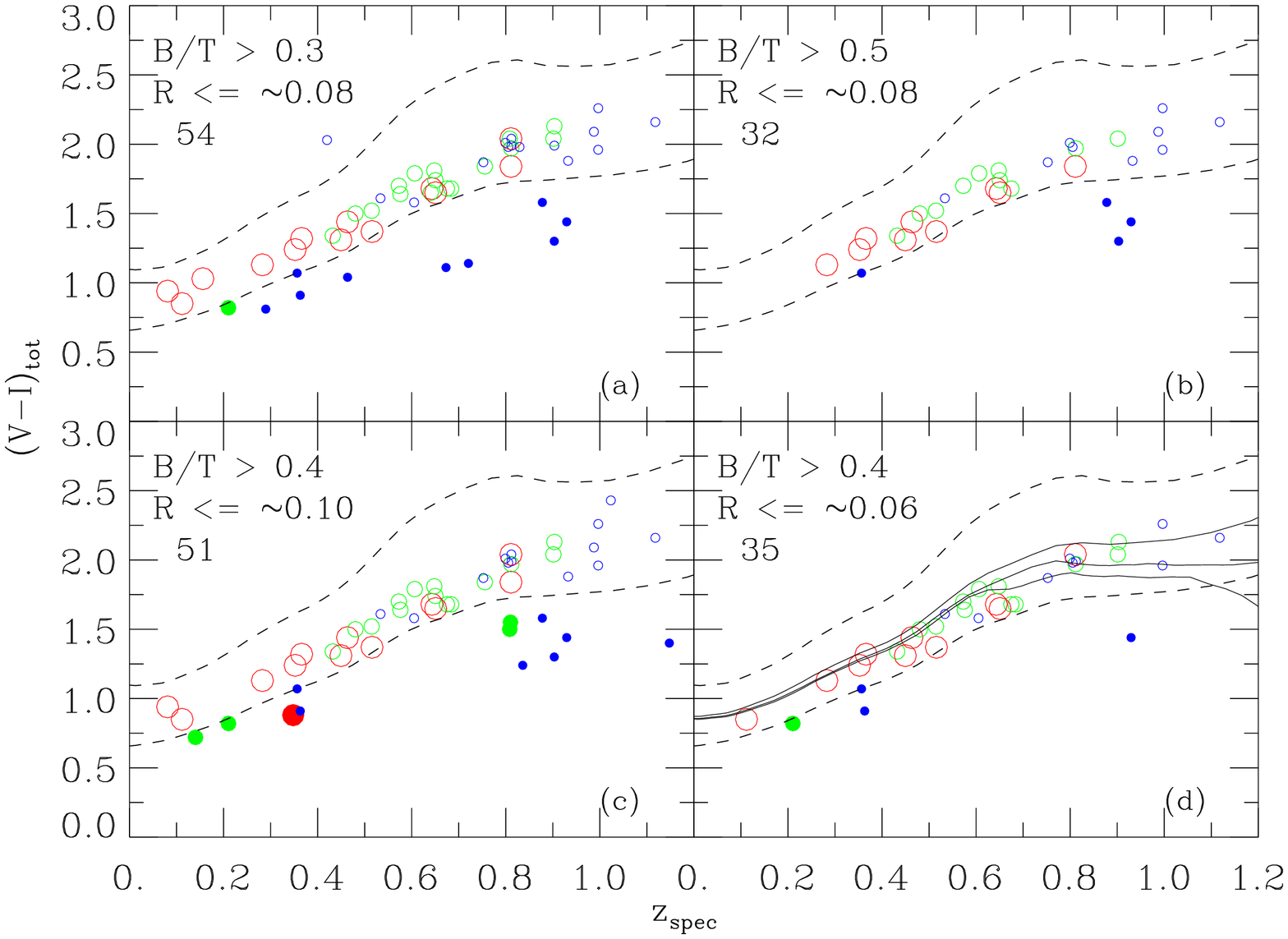,width=18.0cm,angle=90}}
\figcaption[f9.ps]{
 This figure shows how the QS-E/S0 sample changes when the
 selection criteria are varied by small amounts. The total number
 of selected galaxies is shown in each panel; these are to be compared
 to the actual number of 44 galaxies selected in Figure 8.  Figures (a) and (b) show how
 the sample changes when the $B/T$ cut is changed, and
 Figures (c) and (d) show the effect of changing the $R$ cut.
 Loosening the cuts mainly increases the number of interlopers, while tightening
 them mainly decreases the red target sample.  Thus, the adopted cuts
 in Figure 8 are close to optimum.
  For details, see text. 
\label{fig:i22check}
}
\end{figure*}

\clearpage

\begin{figure*}[htb]
\figcaption{ 
   Images of GSS QS-E/S0s in the $z_{spec}$ sample,
 ordered by redshift.
   Numbers indicated in each panel are (from upper-left corner in clockwise
 direction): (1) GSS ID, (2) $I$, (3) $V-I$, (4) $B/T$,
 (5) residual parameter, $R$,
 (6) rest-frame B-band absolute magnitude ($M_{B}$), and (7) $z_{spec}$.
   The 6 blue interlopers are plotted separately  at the bottom 
 of the panel.
  Displayed intensity values are roughly square-rooted 
 (($I$)$^{1/2.2}$ to be precise)
 to show morphological features as clearly as possible (see text).
  The box size corresponds to 30 kpc in physical coordinates,
 assuming an open universe with $\Omega=0.3$ and $h=0.7$.
  To compare with local samples,  
 images of local E/S0s taken from Frei et al. (1996) 
 are shown directly with the same pixel scaling. 
 \label{fig:eso}
}
\end{figure*}

\clearpage

\begin{figure*}[htb]
\vskip 0.2cm
\figcaption{
  Images of galaxies in the $z_{spec}$ sample 
 that are {\it not} classified as E/S0s (high $R$ or 
 low $B/T$), although their colors are as red as QS-E/S0s.
  Galaxy images are ordered by redshift.
  Numbers indicated in each panel are, from upper-left corner in clockwise
 direction: (1) GSS ID, (2) $I$, (3) $V-I$, (4) $B/T$, (5) $R$, 
 (6) rest-frame B-band absolute magnitude ($M_B$), and 
 (7) $z_{spec}$.
\label{fig:redim}
}
\end{figure*}

\clearpage

\vskip 0.0cm
\begin{figure}
\psfig{figure=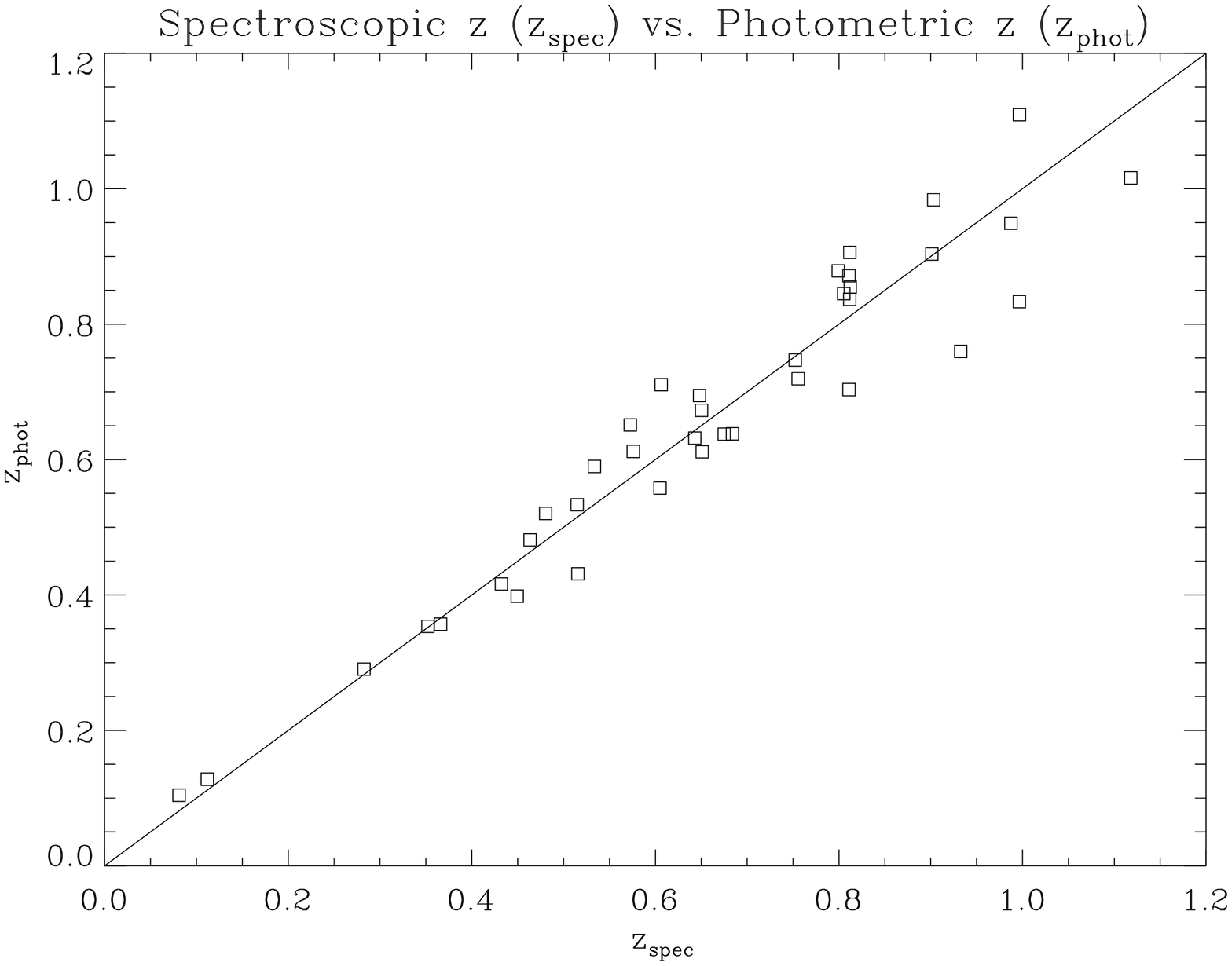,width=15.1cm,angle=90}
\figcaption[f12.ps]{
 Spectroscopic redshifts ($z_{spec}$) vs.
 photometric redshifts ($z_{phot}$) for ``red'' QS-E/S0s.
\label{fig:zspeczphot}
}
\end{figure}

\vskip 0.0cm
\begin{figure}
\psfig{figure=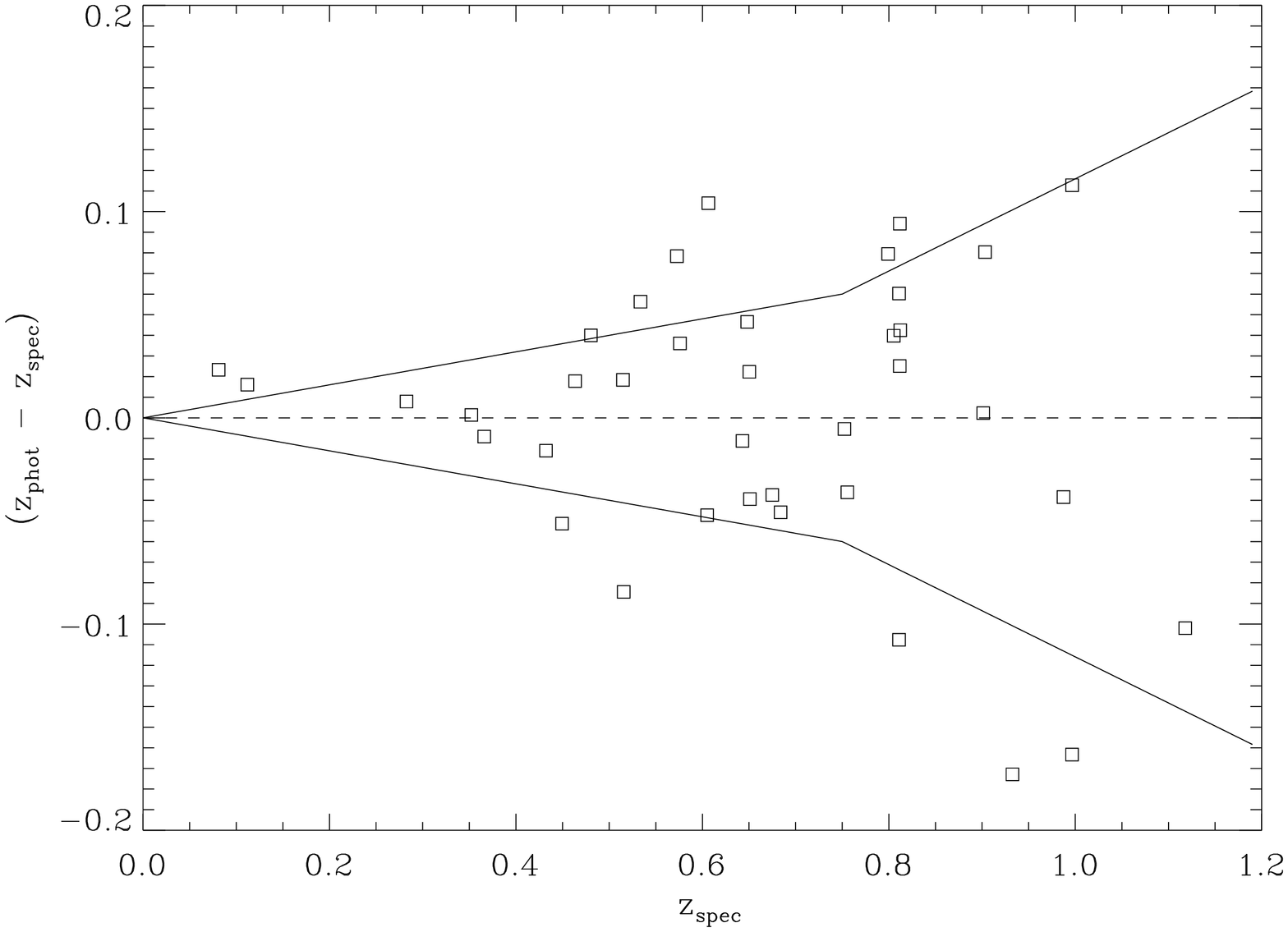,width=15.1cm,angle=90}
\figcaption[f13.ps]{
 The quantity $z_{spec}$ vs. ($z_{phot}$-$z_{spec}$) for the ``red'' 
 QS-E/S0s of Figure 12. The solid lines indicate the estimated rms error,  
 $\delta z_{phot}$, as a function of $z_{spec}$. Note that
 $\delta z_{phot}$ increases when $z_{spec} > 0.8$ but stays within
 $\sim 15$\% of $z_{spec}$ out to $z_{spec}=1.2$.
 \label{fig:dz}
}
\end{figure}

\begin{figure}
\psfig{figure=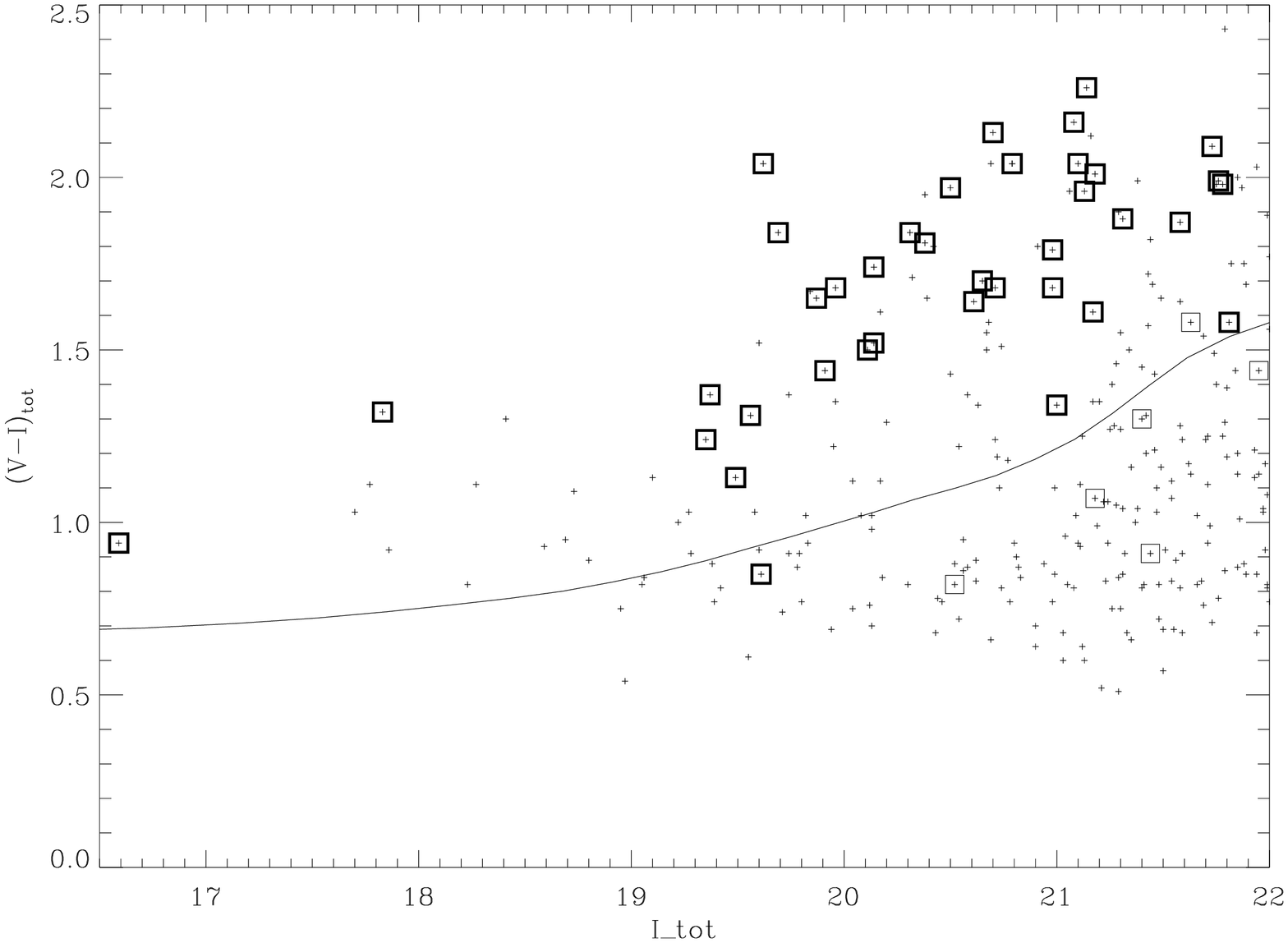,width=15.1cm,angle=90}
\figcaption[f14.ps]{
 $I_{814}$ vs. $(V-I)$ of 
 Groth Strip QS-E/S0s (squares) compared to all
 galaxies in GSS with $z_{spec}$.
 The thick squares are the ``red'' QS-E/S0s, thin
 squares are ``blue interlopers,''
 and crosses indicate remaining galaxies. 
 Also plotted 
 is a fiducial color cut 
 that might be used for 
 eliminating blue interlopers even without spectroscopic redshifts. 
 We have not actually pruned the sample using this cut but have kept track of 
 candidate blue
 interlopers using it.  \label{fig:ccut}}
\end{figure}

\clearpage

\begin{figure*}[ht]
\vskip 0.2cm
\figcaption{
  Images of the 98 GSS E/S0s selected in the $z_{phot}$ sample.
  Images are ordered by $z_{phot}$, excepting  
  the last 11 objects, which are candidate blue interlopers.
  Numbers indicated in each panel are (from upper-left corner in clockwise
 direction): (1) GSS ID, (2) $I$, (3) $(V-I)$, (4) $B/T$, 
 (5) rest-frame B-band absolute magnitude ($M_{B}$), and (6) $z_{phot}$.
\label{fig:esop}
}
\end{figure*}

\clearpage

\begin{figure*}[ht]
\vskip 0.2cm
\figurenum{\ref{fig:esop}}
\figcaption{--- Continued.}
\end{figure*}

\clearpage

\vskip 0.0cm
\begin{figure}
\psfig{figure=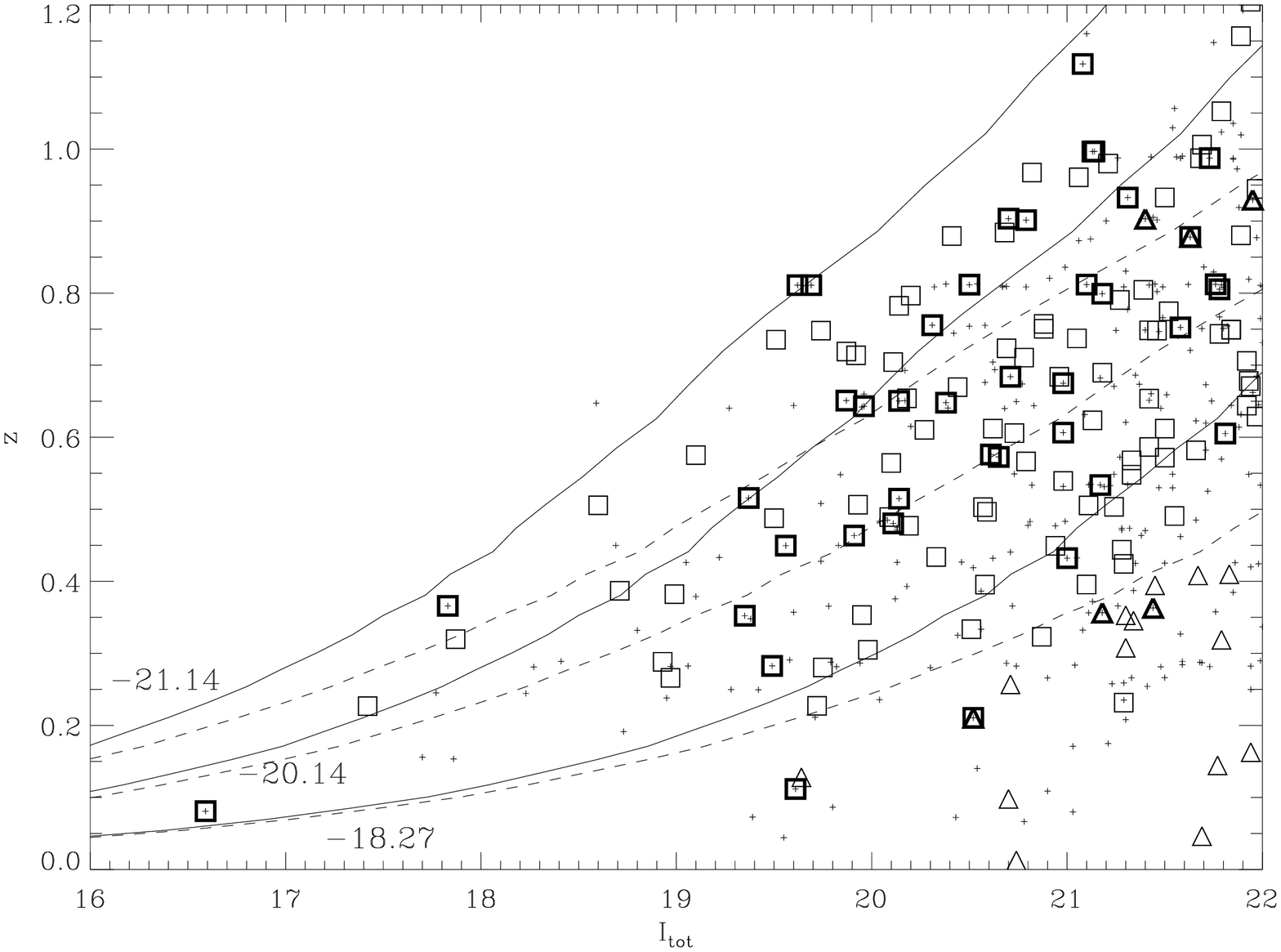,width=15.1cm,angle=90}
\figcaption[f16.ps]{
 $I_{tot}$ vs. $z$ of GSS galaxies.
 QS-E/S0s are marked with squares, and blue interlopers (including candidates)
 are shown with triangles. Thick symbols represent QS-E/S0s in 
 the spectroscopic sample, and thin symbols are QS-E/S0s in the  
 photometric redshift sample. Small crosses are non QS-E/S0 galaxies 
 with spectroscopic redshifts. Also plotted are lines of constant 
 absolute magnitude ($M_{B}=$--21.14, --20.14, and --18.27). Solid lines
 include luminosity evolution of $1.7 \times z$, while dashed lines 
 assume no luminosity evolution. Note that $M_{B}=-20.14$ corresponds 
 to $M^{*}$ for the Marzke et al. local E/S0 LF.
\label{fig:zi_zphot}
}
\end{figure}

\vskip 0.2cm
\begin{figure}
\psfig{figure=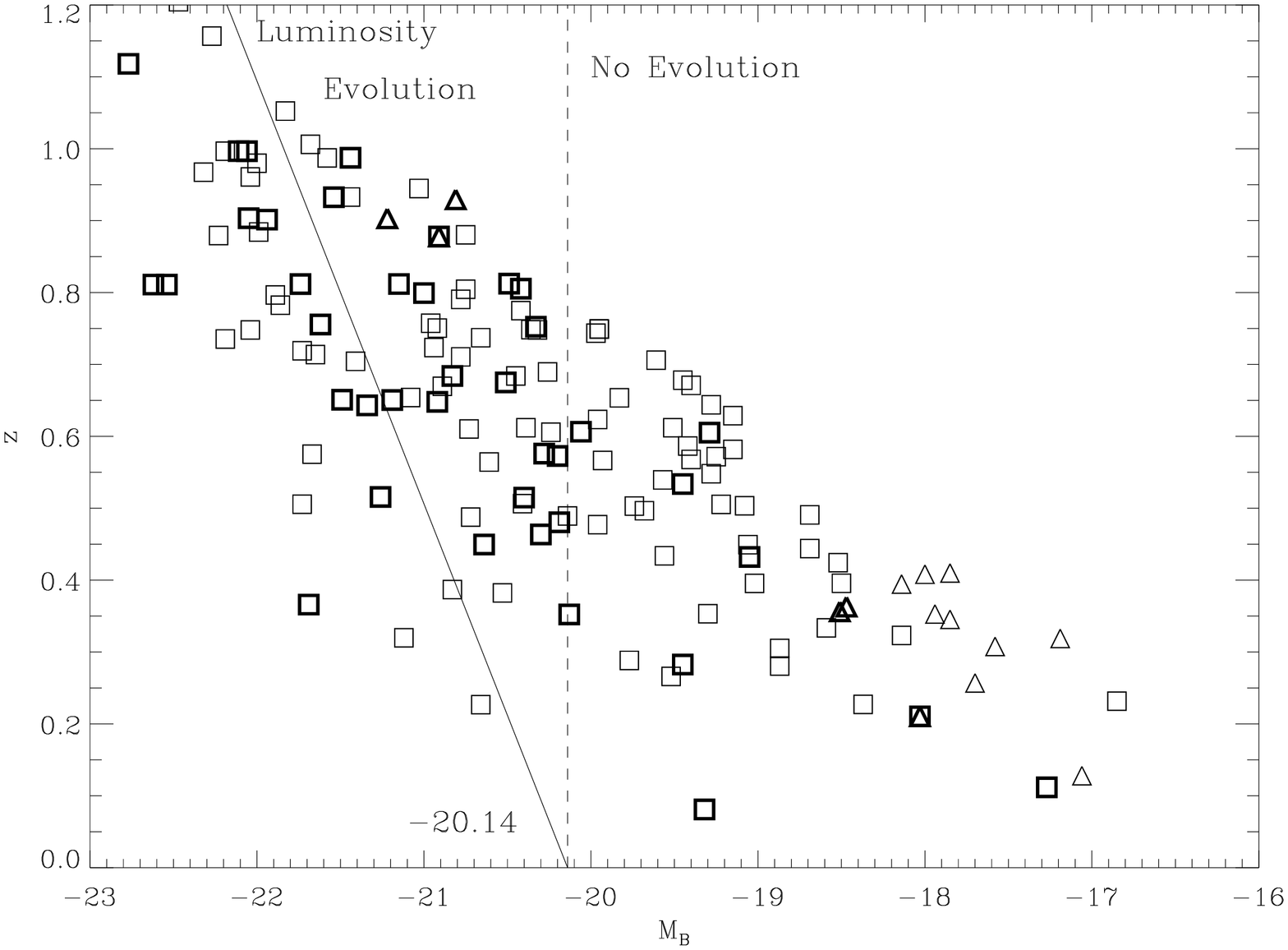,width=15.1cm,angle=90}
\figcaption[f17.ps]{
 $M_{B}$ vs. $z$ of QS-E/S0s. 
 Meanings of symbols and lines are the same as in Fig.\ref{fig:zi_zphot}.
 Values of $M_{B}$s here are derived assuming no luminosity evolution.
\label{fig:zab_zphot}
}
\end{figure}
\vskip 0.2cm

\vskip 0.2cm
\begin{figure}
\psfig{figure=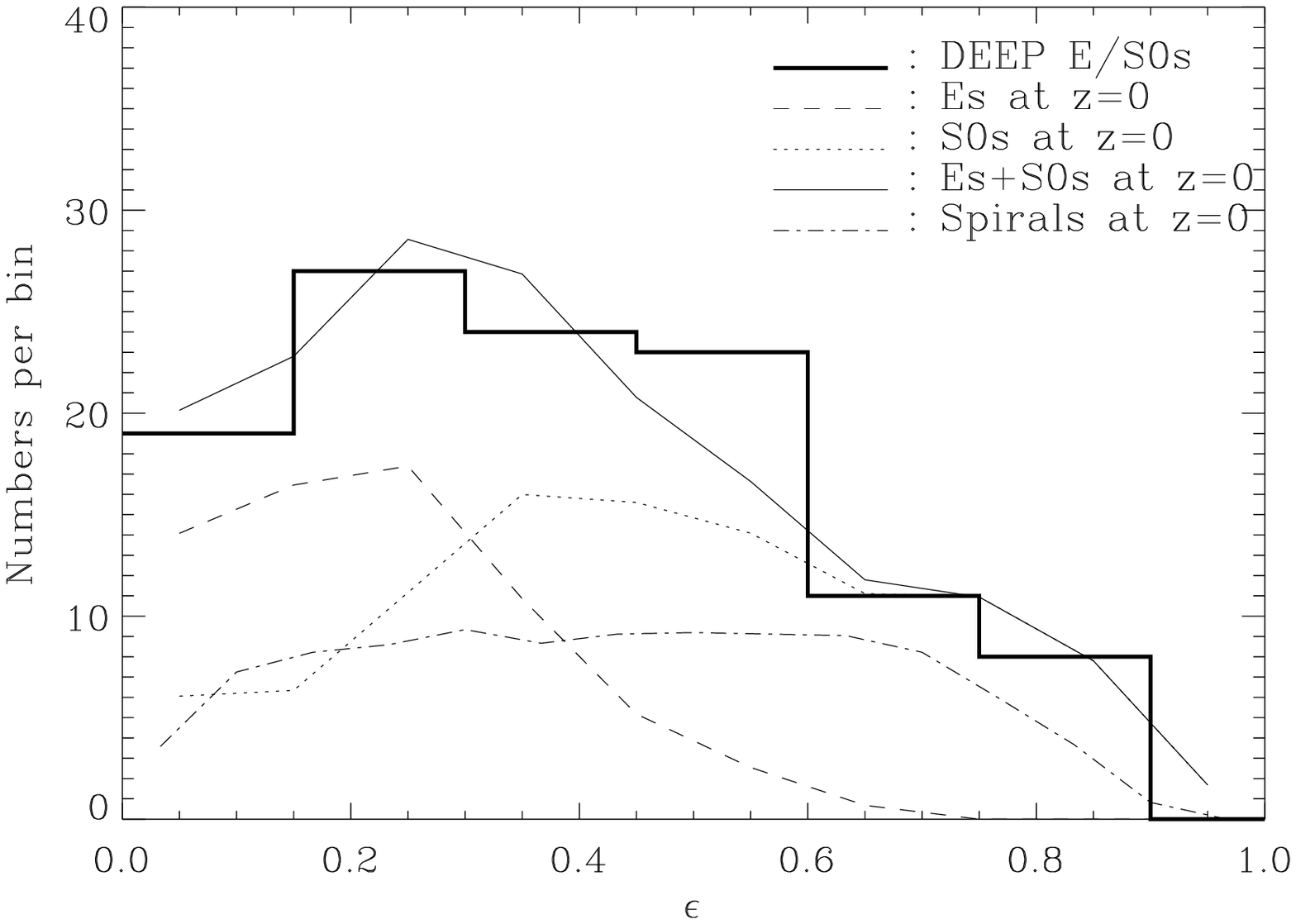,width=15.1cm,angle=0}
\figcaption[f18.ps]{
 The ellipticity distribution of Groth Strip QS-E/S0s (heavy line).
 Also plotted (thin line) is a model distribution combining local 
 cluster Es and S0s in which the fraction of Es is assumed to be
 40\%.  The dashed line represents the contribution from Es,
 and the dotted line represents the contribution from S0s.
  The local sample of Es and S0s is taken from Dressler et al. (1980).
  The ellipticity distribution of spiral galaxies from Lambas et al.
 (1992) is also plotted as the dot-dashed line, for reference.
\label{fig:abtotal}
}
\end{figure}

\begin{figure}
\psfig{figure=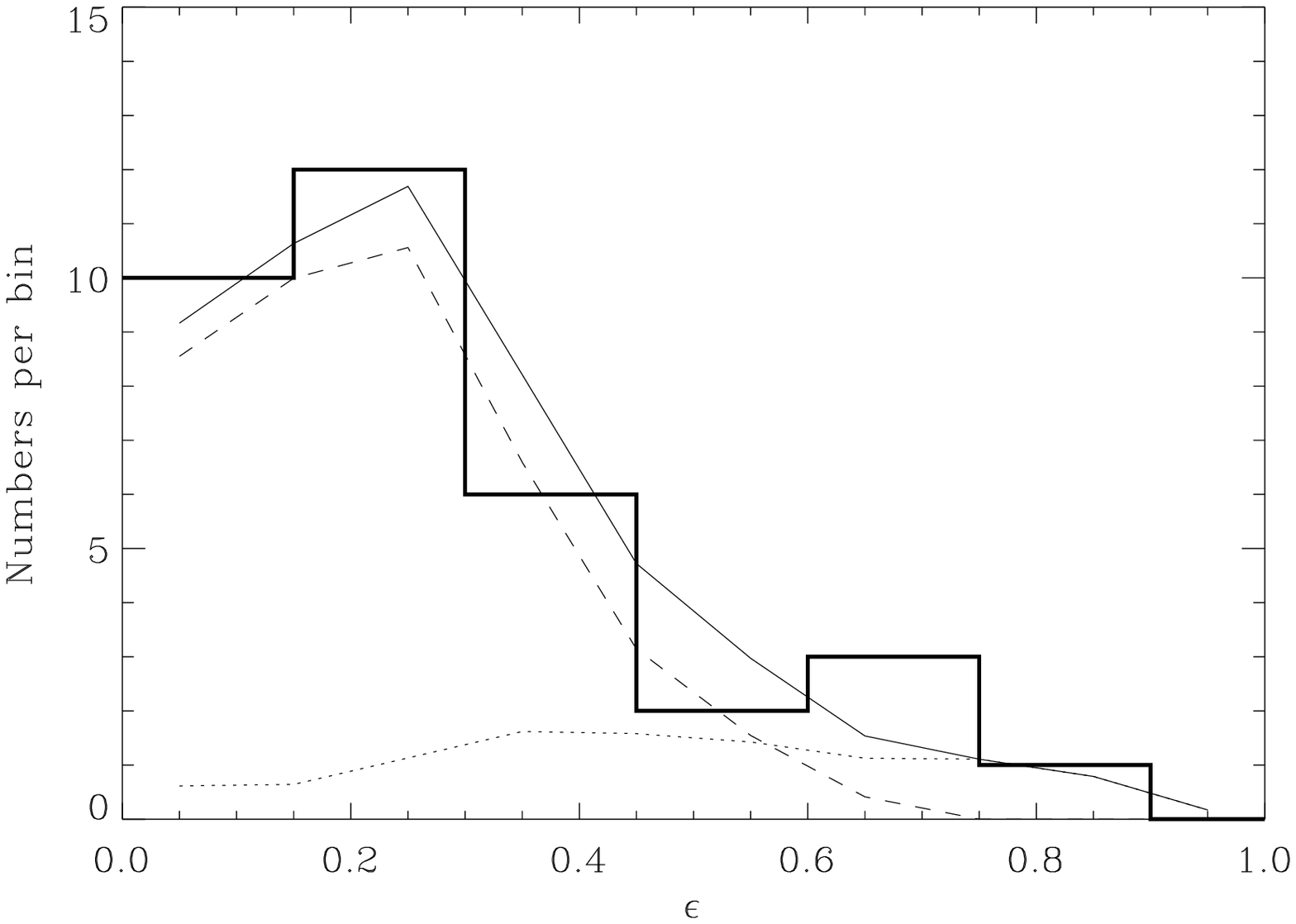,width=15.1cm,angle=0}
\figcaption[f19.ps]{
 The ellipticity distribution of bright Groth Strip 
 QS-E/S0s with $M_{B} < -20$ (heavy line).
 Also plotted is the ellipticity distribution combining local
 cluster Es and S0s in which the fraction of Es is assumed to be
 80\% (thin line).  The dashed line represents the contribution from Es,
 and the dotted line represents the contribution from S0s.
 \label{fig:abbright}}
\end{figure}

\begin{figure}
\psfig{figure=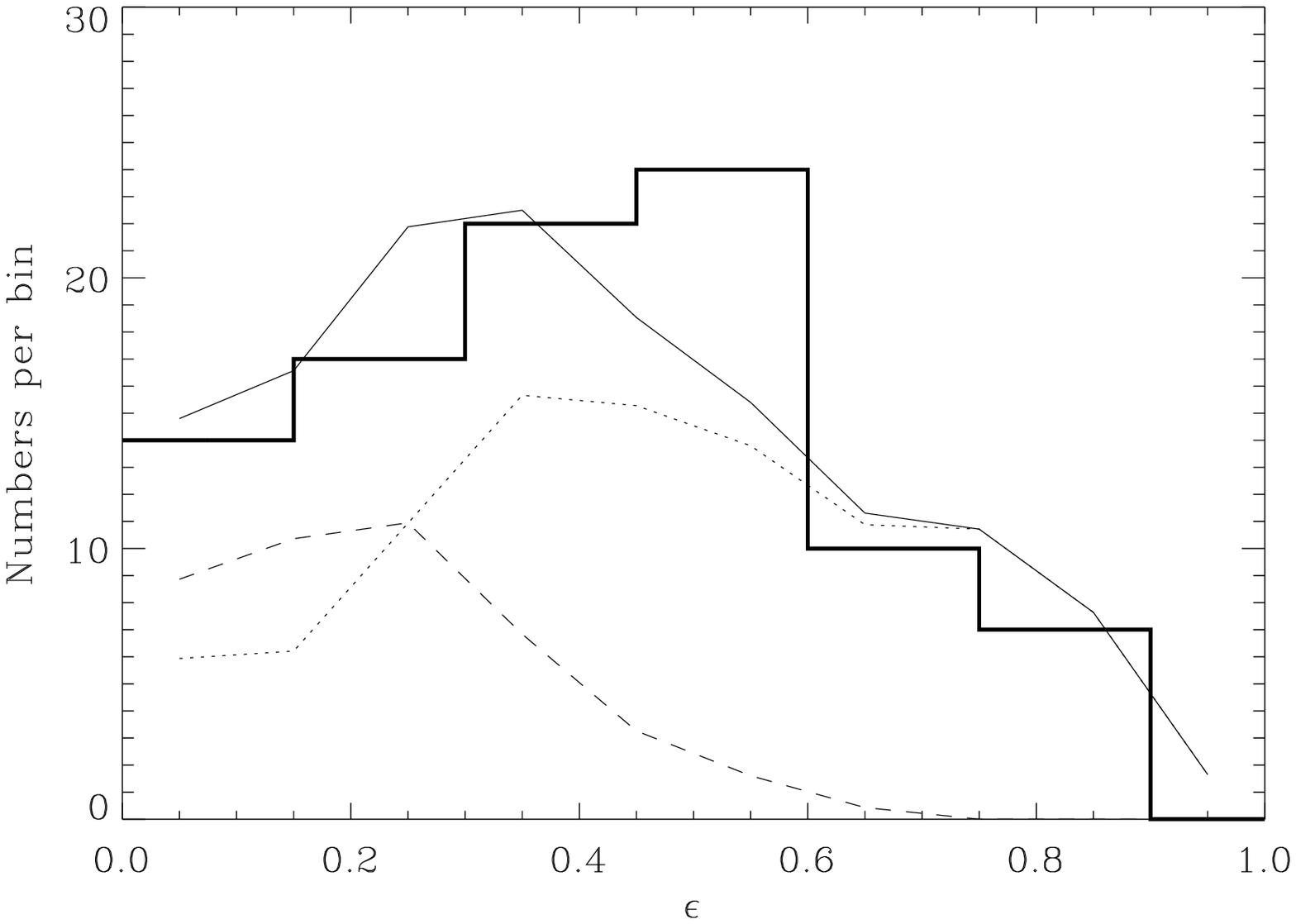,width=15.1cm,angle=0}
\figcaption[f20.ps]{
 The ellipticity distribution of faint 
 Groth Strip QS-E/S0s with $M_{B} > -20$
 (heavy line).
 Also plotted is the ellipticity distribution combining local
 cluster Es and S0s in which the fraction of Es is assumed to be
 30\% (thin line).  The dashed line represents the contribution from Es,
 and the dotted line represents the contribution from S0s.
 \label{fig:abfaint}}
\end{figure}

\begin{figure}[ht]
\psfig{figure=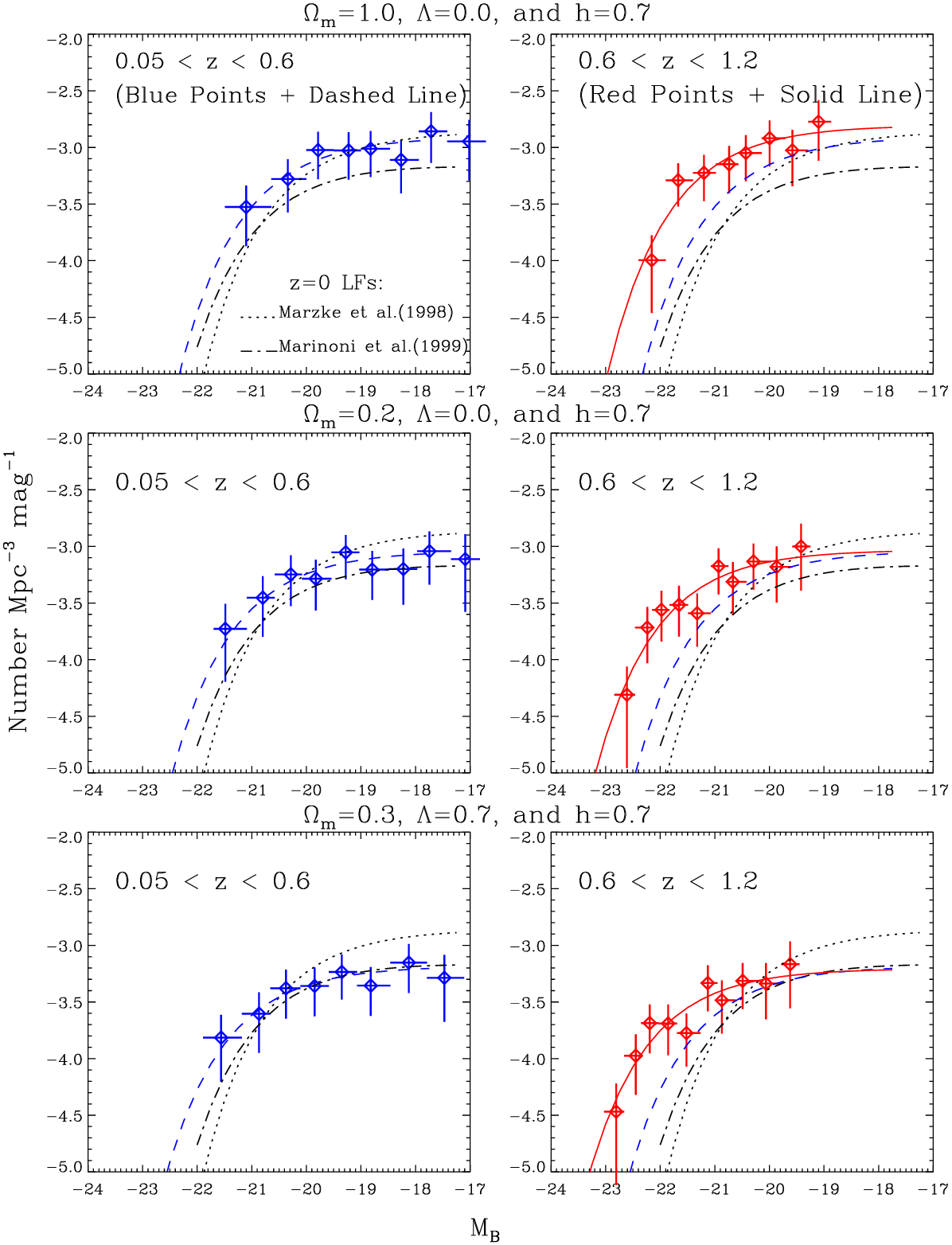,width=16.cm,angle=0}
\figcaption[f21.ps]{
   The luminosity function of GSS E/S0s
 at two different redshift intervals for three different cosmologies.
  The LFs at $0.05 <z < 0.6$ ($z_{med} \simeq 0.4$) are indicated
 by blue points, dashed lines, and errors in the figures on the left.
  The LFs at $0.6 < z < 1.2$ ($z_{med} \simeq 0.8$) are shown 
 by red points, dashed lines, and errors in the figures on the right.
   The red solid lines and the blue dashed lines are drawn using
 the LF parameters in Table 4, but we get virtually 
 identical lines using the parameters in
 Table 5. 
   To compare to $z=0$, two local E/S0 LFs are plotted---that of 
 Marzke et al. (1998)  with the dotted line,
 and that of Marinoni et al. (1999)  with the dot-dashed line.
\label{fig:lf_all.ps}}
\end{figure}

\begin{figure}[ht]
\psfig{figure=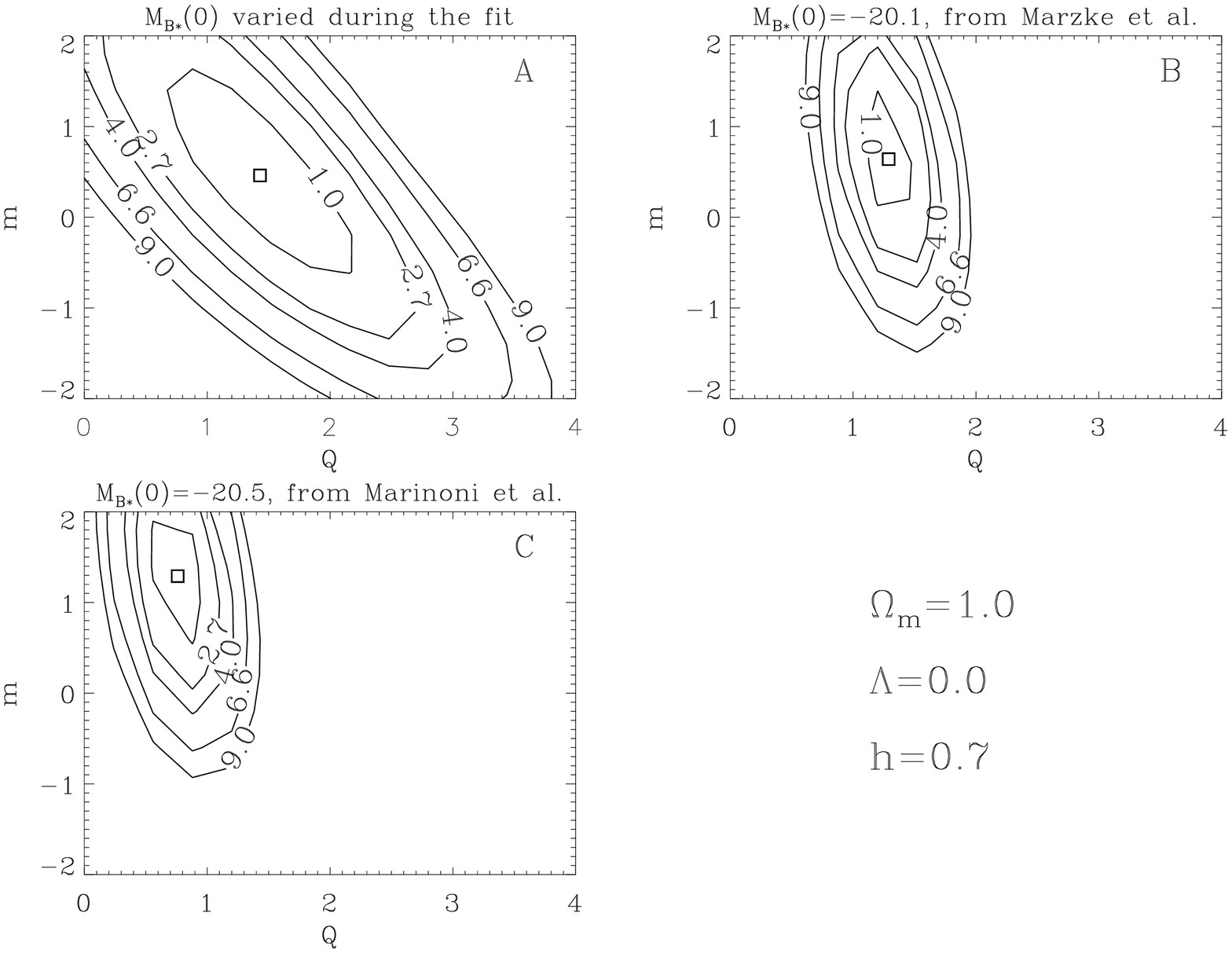,width=16.cm,angle=90}
\figcaption[f22.ps]
{ Error contours in $m$ (number density evolution parameter) vs.
 $Q$ (luminosity evolution parameter)
 for three different approaches for the Einstein-de Sitter universe. 
 In panel A, $Q$ and $m$  are derived along with $M_{B*}(0)$.
 In B and C, they  are derived assuming fixed 
 values of $M_{B*}(0)$ from the local samples of Marzke et al. (1998) and
 Marinoni et al. (1999), respectively.
  The numbers indicated on each contour are $\Delta \chi^{2}$ values.
  When projected onto the one-dimensional intervals either on 
 $m$ or $Q$, 
 $\Delta \chi^{2}=1.00, 2.71, 4.00, 6.63$, and 9.00 correspond to
 confidence levels of 68.3\%, 90\%, 95.4\%, 99\%, and 
 99.73\%.   
\label{fig:cont_om1}}
\end{figure}

\begin{figure}[ht]
\psfig{figure=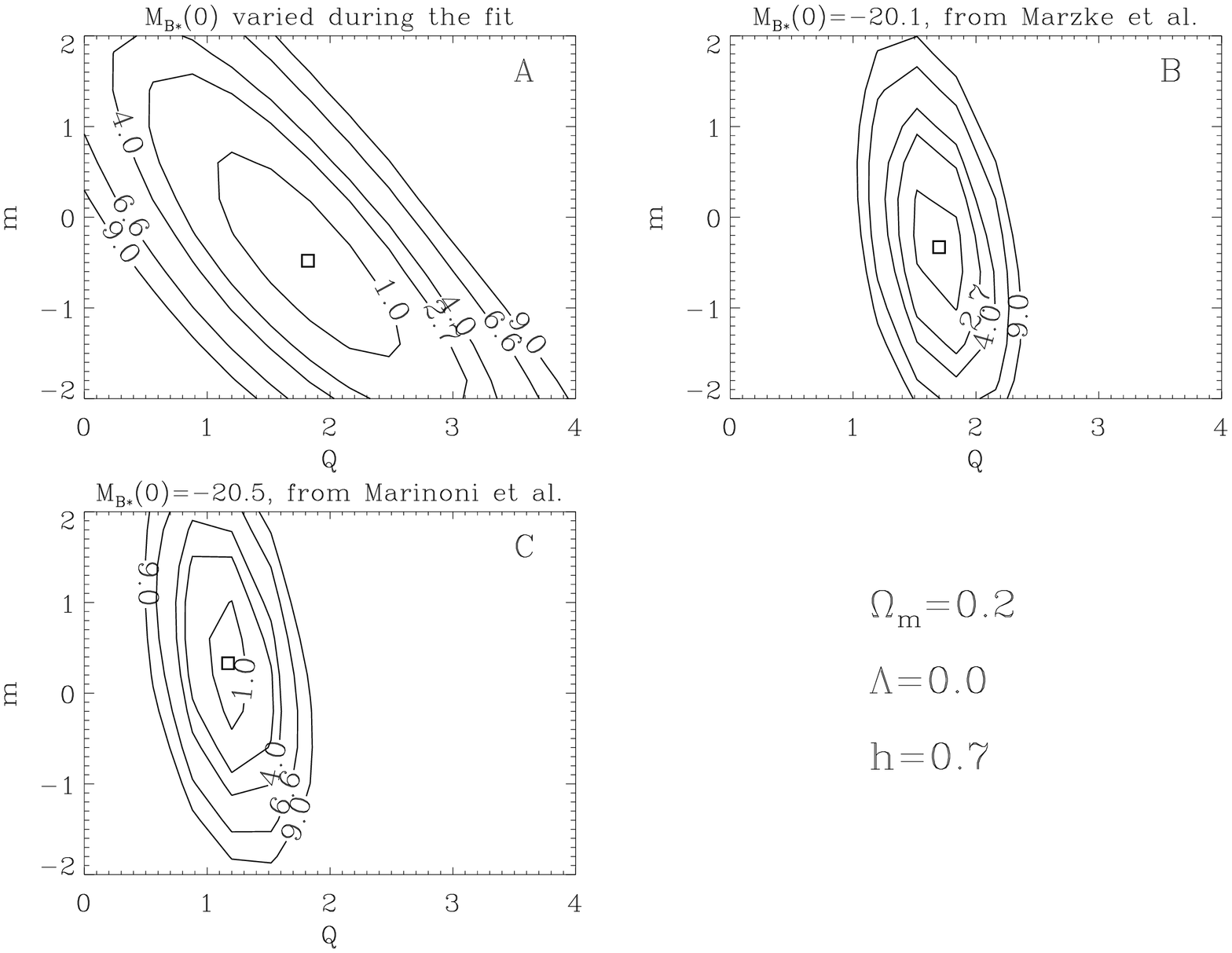,width=16.cm,angle=90}
\figcaption[f23.ps]
{
Same as Figure 22, for an open universe. 
}
\end{figure}

\begin{figure}[ht]
\psfig{figure=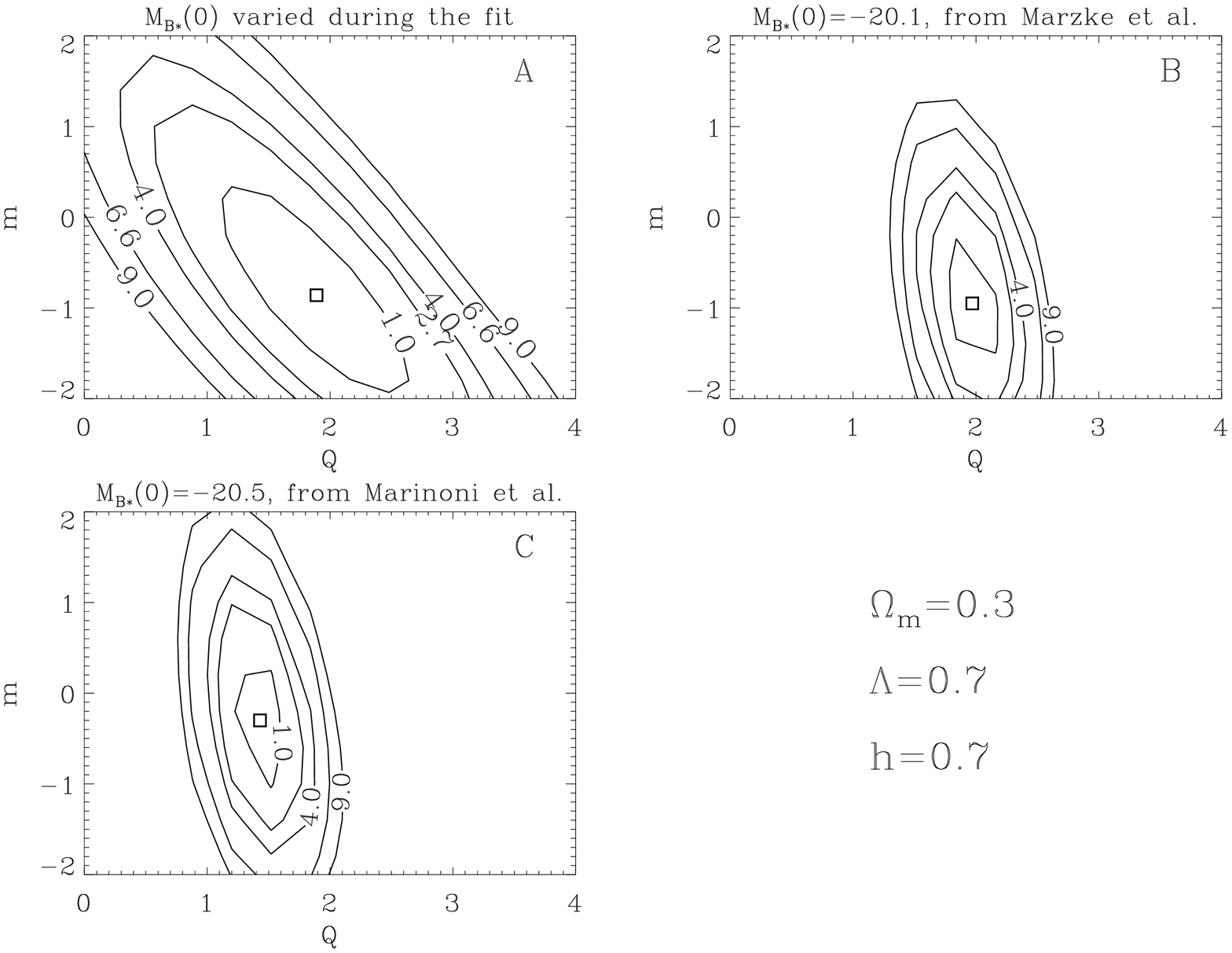,width=16.cm,angle=90}
\figcaption[f24.ps]
{ 
Same as Figure 22, for a non-zero $\Lambda$ universe.
\label{fig:cont_lam07}
}
\end{figure}

%% file: tab1.tex
\begin{deluxetable}{c c c c c}
\tablecolumns{5}
\tablecaption{Adopted $R$ cuts for selecting E/S0s from HST images}
\tablewidth{0pt}
\tablehead{  & $r_{hl} < 1$ & $1 < r_{hl} < 2$ & $2 < r_{hl} < 3$ &
 $r_{hl} > 3$ 
}
\startdata
 $I < 21$        & $\leq 0.05$ & $\leq 0.06$ & $\leq 0.07$ & $\leq 0.08$ \nl
 $21 < I < 21.5$ & $\leq 0.05$ & $\leq 0.06$ & $\leq 0.06$ & $\leq 0.06$ \nl  
 $21.5 < I < 22$ & $\leq 0.05$ & $\leq 0.05$ & $\leq 0.05$ & $\leq 0.05$ \nl
\enddata 
\tablecomments{$r_{hl}$s are in units of pixels.}
\end{deluxetable}

%% file: table.tex
\begin{deluxetable}{c c c   c c c  c c c  c c c}
\tablecolumns{11}
\tablecaption{Quantitatively Selected E/S0s in the Groth Strip}
\tablewidth{0pt}
\tablehead{\colhead{ID} & \colhead{$z$} & \colhead{$z$ code} & \colhead{$I$} & 
\colhead{$V$} & \colhead{$(V-I)$} & \colhead{$M_{B}$} & \colhead{$B/T$} & 
\colhead{$R$} & \colhead{$r_{hl} (0\farcs1)$} & \colhead{$\epsilon$}}
\startdata
052\_1555 &   0.9607 &  0 &    21.06 &    23.16 &     2.22 & $  -22.04$ &     0.75 &     0.06 &     6.42 &     0.29 \\
052\_6543 &   0.7555 &  1 &    20.31 &    22.15 &     1.89 & $  -21.52$ &     0.44 &     0.08 &     6.34 &     0.31 \\
053\_4446 &   0.5480 &  0 &    21.33 &    22.88 &     1.44 & $  -19.28$ &     0.82 &     0.04 &     3.04 &     0.39 \\
053\_6418 &   0.6750 &  1 &    20.98 &    22.66 &     1.81 & $  -20.38$ &     0.72 &     0.04 &     3.52 &     0.60 \\
053\_7537 &   0.6430 &  1 &    19.96 &    21.64 &     1.70 & $  -21.24$ &     0.92 &     0.03 &     2.31 &     0.66 \\
054\_2510 &   0.0810 &  1 &    16.59 &    17.53 &     0.93 & $  -19.05$ &     0.41 &     0.07 &    19.04 &     0.26 \\
062\_6859 &   0.9874 &  1 &    21.90 &    24.29 &     2.17 & $  -21.36$ &     0.56 &     0.02 &     3.01 &     0.35 \\
064\_3021 &   0.9965 &  0 &    21.13 &    23.09 &     2.06 & $  -22.19$ &     0.64 &     0.02 &     5.00 &     0.41 \\
074\_6044 &   0.9966 &  1 &    21.14 &    23.40 &     2.29 & $  -22.18$ &     0.70 &     0.05 &     7.42 &     0.40 \\
082\_6533 &   0.5029 &  0 &    20.57 &    22.04 &     1.46 & $  -19.74$ &     0.41 &     0.06 &     2.07 &     0.37 \\
083\_3771 &   1.2044 &  0 &    21.94 &    24.30 &     2.11 & $  -22.47$ &     0.86 &     0.01 &     3.28 &     0.34 \\
083\_6766 &   0.4322 &  1 &    21.00 &    22.34 &     1.32 & $  -18.88$ &     0.53 &     0.03 &     2.63 &     0.64 \\
084\_2525 &   0.8123 &  1 &    21.76 &    23.75 &     2.08 & $  -20.43$ &     0.42 &     0.03 &     3.79 &     0.68 \\
092\_1339 &   0.9031 &  1 &    21.40 &    22.70 &     1.23 & $  -21.38$ &     0.87 &     0.05 &     2.67 &     0.18 \\
092\_4957 &   0.2105 &  1 &    20.52 &    21.34 &     0.86 & $  -17.38$ &     0.41 &     0.03 &     6.88 &     0.40 \\
093\_2470 &   0.8110 &  1 &    19.69 &    21.53 &     2.09 & $  -22.49$ &     0.52 &     0.08 &    11.30 &     0.30 \\
094\_2660 &   0.9033 &  1 &    20.70 &    22.83 &     2.22 & $  -22.08$ &     0.42 &     0.04 &     8.85 &     0.25 \\
094\_2762 &   0.9326 &  1 &    21.31 &    23.19 &     2.00 & $  -21.63$ &     0.55 &     0.05 &     4.70 &     0.34 \\
094\_6234 &   0.8053 &  1 &    21.78 &    23.76 &     2.04 & $  -20.37$ &     0.97 &     0.03 &     2.91 &     0.58 \\
102\_2157 &   0.1445 &  0 &    21.77 &    22.72 &     0.99 & $  -15.22$ &     0.46 &     0.02 &     7.64 &     0.47 \\
102\_3649 &   0.5336 &  1 &    21.17 &    22.78 &     1.62 & $  -19.35$ &     0.76 &     0.04 &     2.03 &     0.52 \\
102\_5148 &   0.5665 &  0 &    20.79 &    22.36 &     1.47 & $  -19.93$ &     0.91 &     0.05 &     3.25 &     0.35 \\
102\_5177 &   0.6839 &  1 &    20.71 &    22.39 &     1.91 & $  -20.70$ &     0.49 &     0.06 &     5.00 &     0.53 \\
102\_5358 &   0.7190 &  0 &    19.87 &    21.69 &     1.92 & $  -21.73$ &     0.74 &     0.05 &     4.69 &     0.05 \\
102\_5558 &   0.7371 &  0 &    21.05 &    22.88 &     2.02 & $  -20.66$ &     0.65 &     0.05 &     5.21 &     0.57 \\
102\_6549 &   0.5752 &  0 &    19.10 &    20.72 &     1.69 & $  -21.67$ &     1.00 &     0.03 &    10.26 &     0.22 \\
103\_1115 &   0.4635 &  1 &    19.91 &    21.35 &     1.46 & $  -20.17$ &     0.76 &     0.04 &     5.13 &     0.40 \\
103\_4766 &   0.8118 &  1 &    21.10 &    23.14 &     2.09 & $  -21.09$ &     0.45 &     0.05 &     1.96 &     0.64 \\
103\_6061 &   0.3631 &  1 &    21.44 &    22.35 &     0.80 & $  -17.90$ &     0.49 &     0.04 &     1.84 &     0.20 \\
103\_7221 &   0.9013 &  1 &    20.79 &    22.83 &     2.31 & $  -21.98$ &     0.54 &     0.04 &     6.75 &     0.11 \\
112\_7644 &   0.5818 &  0 &    21.66 &    23.27 &     1.60 & $  -19.15$ &     0.79 &     0.04 &     0.74 &     0.54 \\
113\_3136 &   0.9447 &  0 &    21.97 &    24.06 &     1.91 & $  -21.03$ &     0.57 &     0.01 &     4.99 &     0.48 \\
113\_3311 &   0.8117 &  1 &    20.50 &    22.47 &     1.97 & $  -21.69$ &     0.77 &     0.05 &     5.51 &     0.54 \\
114\_1866 &   0.9328 &  0 &    21.50 &    23.57 &     2.05 & $  -21.44$ &     0.54 &     0.01 &     2.78 &     0.19 \\
114\_3760 &   0.7234 &  0 &    20.69 &    22.50 &     1.89 & $  -20.94$ &     0.68 &     0.02 &     3.52 &     0.43 \\
122\_7569 &   0.6896 &  0 &    21.18 &    22.94 &     1.76 & $  -20.26$ &     0.47 &     0.06 &     2.56 &     0.63 \\
132\_2723 &   0.6102 &  0 &    20.27 &    21.91 &     1.62 & $  -20.73$ &     0.77 &     0.02 &     4.15 &     0.23 \\
133\_1016 &   0.7572 &  0 &    20.88 &    22.74 &     2.00 & $  -20.96$ &     0.76 &     0.06 &     4.62 &     0.48 \\
133\_4463 &   0.6120 &  0 &    20.62 &    22.26 &     1.62 & $  -20.39$ &     0.72 &     0.01 &     2.79 &     0.62 \\
133\_6041 &   0.9674 &  0 &    20.82 &    22.93 &     2.13 & $  -22.32$ &     0.64 &     0.07 &     4.18 &     0.00 \\
134\_1633 &   0.3230 &  0 &    20.87 &    22.06 &     1.21 & $  -18.14$ &     0.69 &     0.04 &     0.93 &     0.24 \\
134\_4363 &   0.6287 &  0 &    21.97 &    23.66 &     1.60 & $  -19.15$ &     0.92 &     0.01 &     2.68 &     0.25 \\
142\_2752 &   0.3454 &  0 &    21.34 &    22.58 &     1.08 & $  -17.85$ &     0.84 &     0.06 &     1.76 &     0.13 \\
142\_3329 &   0.4891 &  0 &    20.09 &    21.54 &     1.55 & $  -20.14$ &     0.51 &     0.07 &     5.24 &     0.25 \\
142\_7077 &   0.0127 &  0 &    20.74 &    21.42 &     0.69 & $  -10.78$ &     0.52 &     0.05 &     2.37 &     0.30 \\
142\_7764 &   0.3188 &  0 &    21.79 &    23.01 &     1.04 & $  -17.19$ &     0.67 &     0.01 &     2.56 &     0.29 \\
142\_7871 &   0.2267 &  0 &    17.42 &    18.53 &     1.22 & $  -20.66$ &     1.00 &     0.04 &    16.05 &     0.20 \\
143\_2770 &   0.8045 &  0 &    21.39 &    23.32 &     2.06 & $  -20.75$ &     0.60 &     0.03 &     2.99 &     0.25 \\
143\_7957 &   0.4772 &  0 &    20.19 &    21.62 &     1.60 & $  -19.96$ &     0.91 &     0.04 &     4.44 &     0.23 \\
144\_3353 &   0.4492 &  0 &    20.94 &    22.33 &     1.36 & $  -19.06$ &     0.50 &     0.03 &     1.82 &     0.25 \\
153\_0711 &   0.3078 &  0 &    21.30 &    22.48 &     1.19 & $  -17.58$ &     0.45 &     0.03 &     2.72 &     0.31 \\
153\_2471 &   0.9873 &  0 &    21.68 &    23.81 &     2.05 & $  -21.58$ &     0.67 &     0.04 &     4.01 &     0.67 \\
154\_6532 &   0.7434 &  0 &    21.78 &    23.63 &     1.88 & $  -19.97$ &     0.64 &     0.04 &     1.23 &     0.45 \\
\tablebreak
162\_5349 &   0.4804 &  1 &    20.11 &    21.61 &     1.48 & $  -20.06$ &     0.73 &     0.03 &     3.65 &     0.59 \\
163\_5528 &   1.0523 &  0 &    21.79 &    23.99 &     1.97 & $  -21.83$ &     0.65 &     0.01 &     2.52 &     0.13 \\
164\_1176 &   0.0980 &  0 &    20.70 &    21.52 &     0.76 & $  -15.40$ &     0.46 &     0.05 &     4.10 &     0.69 \\
164\_4063 &   1.0060 &  0 &    21.69 &    23.84 &     2.12 & $  -21.68$ &     0.82 &     0.03 &     3.88 &     0.49 \\
164\_4638 &   0.3198 &  0 &    17.87 &    19.12 &     1.32 & $  -21.12$ &     0.59 &     0.05 &    12.23 &     0.38 \\
164\_7956 &   0.3868 &  0 &    18.71 &    20.03 &     1.39 & $  -20.83$ &     0.63 &     0.04 &     7.74 &     0.21 \\
172\_5049 &   0.3564 &  1 &    21.18 &    22.25 &     1.01 & $  -18.10$ &     0.73 &     0.03 &     1.87 &     0.21 \\
172\_7753 &   0.6480 &  1 &    20.38 &    22.19 &     1.75 & $  -20.84$ &     0.86 &     0.04 &     2.55 &     0.54 \\
173\_3911 &   0.9798 &  0 &    21.21 &    23.33 &     2.06 & $  -22.00$ &     0.59 &     0.04 &     4.78 &     0.25 \\
173\_4039 &   0.5055 &  0 &    18.60 &    20.13 &     1.63 & $  -21.73$ &     0.58 &     0.08 &     7.01 &     0.05 \\
173\_4131 &   0.2316 &  0 &    21.29 &    22.35 &     1.06 & $  -16.85$ &     0.69 &     0.06 &     7.88 &     0.74 \\
173\_6766 &   0.3954 &  0 &    20.58 &    21.88 &     1.30 & $  -19.02$ &     0.46 &     0.04 &     2.38 &     0.53 \\
174\_2027 &   0.1279 &  0 &    19.64 &    20.49 &     0.78 & $  -17.06$ &     0.55 &     0.08 &     3.02 &     0.20 \\
174\_7829 &   0.8803 &  0 &    21.89 &    23.91 &     2.04 & $  -20.75$ &     0.71 &     0.04 &     2.88 &     0.78 \\
182\_4830 &   0.6438 &  0 &    21.92 &    23.63 &     1.76 & $  -19.28$ &     0.84 &     0.02 &     1.77 &     0.65 \\
183\_1868 &   0.8793 &  0 &    20.41 &    22.43 &     1.96 & $  -22.23$ &     0.54 &     0.08 &     6.17 &     0.15 \\
183\_2653 &   0.7479 &  0 &    19.74 &    21.61 &     1.98 & $  -22.04$ &     0.53 &     0.03 &     8.98 &     0.27 \\
183\_3056 &   0.7748 &  0 &    21.52 &    23.41 &     1.92 & $  -20.42$ &     0.65 &     0.04 &     1.96 &     0.27 \\
183\_4478 &   0.5865 &  0 &    21.42 &    23.03 &     1.84 & $  -19.42$ &     0.68 &     0.03 &     3.87 &     0.41 \\
192\_2330 &   0.5727 &  1 &    20.65 &    22.35 &     1.73 & $  -20.11$ &     0.63 &     0.06 &     6.82 &     0.17 \\
192\_3330 &   0.5759 &  1 &    20.61 &    22.25 &     1.66 & $  -20.17$ &     0.41 &     0.04 &     2.29 &     0.68 \\
192\_5343 &   0.3532 &  0 &    19.95 &    21.18 &     1.21 & $  -19.30$ &     0.90 &     0.05 &     5.10 &     0.58 \\
193\_1227 &   0.7992 &  1 &    21.18 &    23.19 &     2.10 & $  -20.93$ &     0.73 &     0.01 &     3.64 &     0.34 \\
193\_1616 &   0.3821 &  0 &    18.99 &    20.29 &     1.28 & $  -20.53$ &     0.82 &     0.04 &     5.77 &     0.00 \\
194\_6444 &   0.2271 &  0 &    19.72 &    20.74 &     1.01 & $  -18.37$ &     0.88 &     0.07 &     2.15 &     0.33 \\
203\_0833 &   1.1566 &  0 &    21.89 &    24.20 &     2.25 & $  -22.27$ &     0.70 &     0.02 &     4.95 &     0.77 \\
203\_1921 &   0.5717 &  0 &    21.50 &    23.09 &     1.65 & $  -19.25$ &     0.86 &     0.00 &     2.23 &     0.46 \\
203\_2022 &   0.6708 &  0 &    21.94 &    23.69 &     1.80 & $  -19.40$ &     0.62 &     0.00 &     1.29 &     0.41 \\
203\_2622 &   0.7505 &  0 &    20.88 &    22.73 &     1.88 & $  -20.92$ &     0.93 &     0.05 &     3.02 &     0.07 \\
203\_2634 &   0.3660 &  1 &    17.83 &    19.15 &     1.37 & $  -21.54$ &     0.83 &     0.05 &    11.64 &     0.41 \\
203\_3311 &   0.7492 &  0 &    21.84 &    23.70 &     1.71 & $  -19.95$ &     0.51 &     0.01 &     3.14 &     0.39 \\
203\_3418 &   0.6840 &  0 &    20.96 &    22.71 &     1.85 & $  -20.45$ &     0.87 &     0.04 &     3.10 &     0.49 \\
203\_5166 &   0.3336 &  0 &    20.51 &    21.71 &     1.15 & $  -18.59$ &     0.94 &     0.04 &     1.70 &     0.47 \\
203\_5720 &   0.6056 &  0 &    20.73 &    22.36 &     1.51 & $  -20.24$ &     0.70 &     0.08 &     4.73 &     0.33 \\
203\_7714 &   0.6535 &  0 &    21.42 &    23.13 &     1.62 & $  -19.83$ &     0.50 &     0.04 &     6.20 &     0.29 \\
212\_4836 &   0.3522 &  1 &    19.35 &    20.59 &     1.25 & $  -19.89$ &     0.51 &     0.03 &     4.31 &     0.70 \\
213\_1764 &   0.7904 &  0 &    21.27 &    23.18 &     1.89 & $  -20.78$ &     0.44 &     0.02 &     2.09 &     0.26 \\
213\_2362 &   0.4439 &  0 &    21.28 &    22.67 &     1.39 & $  -18.69$ &     0.76 &     0.04 &     1.26 &     0.16 \\
213\_6222 &   0.1118 &  1 &    19.61 &    20.46 &     0.82 & $  -16.79$ &     0.41 &     0.04 &     6.38 &     0.14 \\
213\_7564 &   0.7105 &  0 &    20.78 &    22.57 &     1.79 & $  -20.78$ &     0.83 &     0.03 &     2.71 &     0.52 \\
214\_2761 &   0.7354 &  0 &    19.51 &    21.37 &     1.90 & $  -22.19$ &     0.56 &     0.04 &     7.06 &     0.33 \\
223\_7276 &   0.4338 &  0 &    20.33 &    21.69 &     1.51 & $  -19.56$ &     0.73 &     0.04 &     5.98 &     0.16 \\
224\_3977 &   0.7824 &  0 &    20.14 &    22.05 &     1.88 & $  -21.86$ &     0.71 &     0.06 &     2.36 &     0.43 \\
224\_4363 &   0.4877 &  0 &    19.50 &    20.96 &     1.50 & $  -20.72$ &     0.56 &     0.05 &     5.04 &     0.60 \\
224\_4413 &   0.6695 &  0 &    20.44 &    22.17 &     1.81 & $  -20.89$ &     0.64 &     0.07 &     4.58 &     0.56 \\
232\_4429 &   0.0463 &  0 &    21.69 &    22.48 &     0.77 & $  -12.69$ &     0.71 &     0.04 &     6.54 &     0.70 \\
233\_5568 &   0.4242 &  0 &    21.29 &    22.65 &     1.30 & $  -18.52$ &     0.44 &     0.05 &     2.52 &     0.40 \\
233\_6431 &   1.1180 &  1 &    21.08 &    23.24 &     2.07 & $  -22.87$ &     0.79 &     0.03 &     5.60 &     0.36 \\
234\_4218 &   0.5393 &  0 &    20.98 &    22.51 &     1.68 & $  -19.57$ &     0.48 &     0.05 &     4.42 &     0.20 \\
242\_5538 &   0.4907 &  0 &    21.55 &    23.02 &     1.46 & $  -18.69$ &     0.61 &     0.04 &     1.91 &     0.66 \\
244\_4850 &   0.3049 &  0 &    19.98 &    21.13 &     1.31 & $  -18.87$ &     0.53 &     0.05 &     3.65 &     0.75 \\
253\_1150 &   0.5642 &  0 &    20.10 &    21.67 &     1.51 & $  -20.61$ &     0.60 &     0.07 &     4.25 &     0.23 \\
253\_2334 &   0.5679 &  0 &    21.33 &    22.91 &     1.66 & $  -19.40$ &     0.77 &     0.02 &     1.39 &     0.65 \\
253\_4345 &   0.7043 &  0 &    20.11 &    21.90 &     1.83 & $  -21.41$ &     0.53 &     0.02 &     5.64 &     0.13 \\
253\_4452 &   0.6119 &  0 &    21.50 &    23.15 &     1.66 & $  -19.51$ &     0.96 &     0.04 &     4.00 &     0.03 \\
254\_1671 &   0.7492 &  0 &    21.84 &    23.70 &     1.99 & $  -19.95$ &     0.61 &     0.02 &     3.67 &     0.06 \\
262\_2560 &   0.6234 &  0 &    21.13 &    22.79 &     1.52 & $  -19.96$ &     0.98 &     0.06 &     3.22 &     0.27 \\
262\_7430 &   0.6064 &  1 &    20.98 &    22.77 &     1.76 & $  -20.00$ &     0.83 &     0.04 &     2.12 &     0.60 \\
263\_6262 &   0.4098 &  0 &    21.83 &    23.19 &     1.33 & $  -17.85$ &     0.82 &     0.04 &     2.03 &     0.55 \\
263\_6340 &   0.3528 &  0 &    21.30 &    22.55 &     1.20 & $  -17.94$ &     0.46 &     0.06 &     3.77 &     0.39 \\
263\_6417 &   0.4082 &  0 &    21.67 &    23.02 &     1.46 & $  -18.00$ &     0.57 &     0.03 &     1.27 &     0.27 \\
264\_0931 &   0.6780 &  0 &    21.93 &    23.69 &     1.90 & $  -19.45$ &     0.57 &     0.04 &     1.41 &     0.34 \\
264\_6053 &   0.3955 &  0 &    21.10 &    22.41 &     1.36 & $  -18.50$ &     0.66 &     0.04 &     4.82 &     0.27 \\
\tablebreak
272\_2255 &   0.8842 &  0 &    20.68 &    22.70 &     2.11 & $  -21.99$ &     0.87 &     0.04 &     4.37 &     0.60 \\
272\_2871 &   0.5063 &  0 &    19.93 &    21.41 &     1.60 & $  -20.41$ &     0.51 &     0.08 &     7.44 &     0.54 \\
272\_5241 &   0.5053 &  0 &    21.11 &    22.59 &     1.41 & $  -19.22$ &     0.52 &     0.03 &     1.55 &     0.63 \\
273\_2671 &   0.3945 &  0 &    21.45 &    22.77 &     1.26 & $  -18.14$ &     0.59 &     0.04 &     2.77 &     0.65 \\
273\_5617 &   0.4965 &  0 &    20.59 &    22.05 &     1.48 & $  -19.68$ &     0.90 &     0.03 &     1.36 &     0.70 \\
274\_0837 &   0.7059 &  0 &    21.92 &    23.72 &     1.90 & $  -19.61$ &     0.87 &     0.04 &     1.85 &     0.13 \\
274\_3875 &   0.2826 &  1 &    19.49 &    20.62 &     1.12 & $  -19.15$ &     0.84 &     0.04 &     3.29 &     0.67 \\
274\_4341 &   0.7137 &  0 &    19.92 &    21.73 &     1.82 & $  -21.65$ &     0.50 &     0.03 &     4.56 &     0.39 \\
274\_5142 &   0.6051 &  1 &    21.81 &    23.39 &     1.62 & $  -19.16$ &     0.49 &     0.01 &     2.62 &     0.33 \\
274\_5920 &   0.8110 &  1 &    19.62 &    21.66 &     2.17 & $  -22.56$ &     0.48 &     0.04 &    10.74 &     0.14 \\
282\_5737 &   0.7524 &  1 &    21.58 &    23.45 &     1.92 & $  -20.23$ &     0.75 &     0.03 &     3.17 &     0.56 \\
283\_2254 &   0.6504 &  1 &    20.14 &    21.88 &     1.77 & $  -21.10$ &     0.82 &     0.06 &     2.99 &     0.72 \\
283\_3250 &   0.6509 &  1 &    19.87 &    21.52 &     1.87 & $  -21.37$ &     0.61 &     0.05 &     5.49 &     0.11 \\
284\_3854 &   0.4495 &  1 &    19.56 &    20.87 &     1.38 & $  -20.44$ &     0.79 &     0.04 &     3.76 &     0.53 \\
284\_5154 &   0.2883 &  0 &    18.93 &    20.07 &     1.18 & $  -19.77$ &     0.41 &     0.06 &     6.74 &     0.32 \\
284\_6253 &   0.2804 &  0 &    19.75 &    20.86 &     1.14 & $  -18.87$ &     0.74 &     0.05 &     2.51 &     0.10 \\
284\_7275 &   0.2661 &  0 &    18.97 &    20.07 &     1.12 & $  -19.52$ &     0.53 &     0.08 &     4.68 &     0.39 \\
292\_3076 &   0.6540 &  0 &    20.18 &    21.89 &     1.72 & $  -21.08$ &     0.88 &     0.04 &     2.86 &     0.23 \\
292\_7235 &   0.5034 &  0 &    21.24 &    22.72 &     1.64 & $  -19.08$ &     0.80 &     0.00 &     5.67 &     0.26 \\
294\_0718 &   0.5156 &  1 &    19.37 &    20.74 &     1.52 & $  -21.03$ &     0.73 &     0.03 &     5.48 &     0.56 \\
294\_2078 &   0.9295 &  1 &    21.95 &    23.39 &     1.40 & $  -20.97$ &     0.59 &     0.00 &     5.38 &     0.47 \\
294\_4544 &   0.7484 &  0 &    21.42 &    23.27 &     1.71 & $  -20.36$ &     0.46 &     0.06 &     4.26 &     0.16 \\
302\_3631 &   0.2569 &  0 &    20.71 &    21.79 &     1.04 & $  -17.70$ &     0.92 &     0.03 &     2.21 &     0.26 \\
312\_6405 &   0.7965 &  0 &    20.20 &    22.13 &     2.06 & $  -21.89$ &     0.66 &     0.04 &     5.31 &     0.04 \\
313\_1250 &   0.5148 &  1 &    20.14 &    21.66 &     1.58 & $  -20.25$ &     0.87 &     0.05 &     4.31 &     0.37 \\
313\_3515 &   0.7480 &  0 &    21.46 &    23.31 &     1.78 & $  -20.32$ &     0.53 &     0.03 &     3.26 &     0.43 \\
314\_2845 &   0.1628 &  0 &    21.94 &    22.93 &     1.05 & $  -15.35$ &     0.41 &     0.04 &     2.28 &     0.25
\enddata
\tablecomments{(1) ID: ID\# of the GSS E/S0s; (2) $z$ code:0
 if $z_{phot}$, 1 if $z_{spec}$; (3) $I$: $I$-band total magnitude;
 (4) $V$: $V$-band total magnitude;
 (5) $(V-I)$: $V-I$ color; (6) $M_{B}$: rest frame absolute magnitude in B-band
 (see section 4.2.1); (7) $B/T$: bulge-to-total light ratio
 measured in the $I$-band from our fits; (8) $R$:  
 residual parameter measured in the $I$-band; (9) $r_{hl}$: half light
 radius measured in the $I$-band in units of pixels (one pixel=$0\farcs1$);
 (10) $\epsilon$: ellipticity (see section 3.8)}
\end{deluxetable}

%% file: tab3-7.tex
\begin{deluxetable}{c c c c}
\tablecolumns{4}
\tablecaption{Parameters of Local E/S0 LFs}
\tablewidth{0pt}
\tablehead{ Sample & $\alpha$ & $M^{*}_{B}$ & $\phi^{*}~(10^{-3}$ Mpc$^{-3})$ 
}
\startdata
 Marzke et al. (1998)   & $-1.00 \pm 0.09 $ & $-20.14 \pm 0.10$ &
                          $1.51 \pm 0.31$ \nl
 Marinoni et al. (1999) & $-0.97 \pm 0.14$ & $-20.54 \pm 0.18$ &
                          $0.84 \pm 0.20$ \nl
\enddata 
\tablecomments{Parameters are adjusted assuming $H_{0}=70$ km sec$^{-1}$
 Mpc$^{-1}$.}
\end{deluxetable}

\begin{deluxetable}{c  c c   c  c c  c c c}
\tablefontsize{\tiny}
\tablecolumns{9}
\tablecaption{Parameters of LFs from Method 1}
\tablewidth{0pt}
\tablehead{ 
 \colhead{}   &  \multicolumn{2}{c}{$\Omega_{m}=1, \Lambda=0$}  & \colhead{} & 
  \multicolumn{2}{c}{$\Omega_{m}=0.2, \Lambda=0.0$} & \colhead{} &
  \multicolumn{2}{c}{$\Omega_{m}=0.3, \Lambda=0.7$}  \\ 
\cline{2-3} \cline{5-6} \cline{8-9} \\
 \colhead{}   & \colhead{$0.05<z<0.6$} & \colhead{$0.6<z<1.2$} & 
 \colhead{} & \colhead{$0.05<z<0.6$} & 
\colhead{$0.6<z<1.2$}  & 
 \colhead{} & \colhead{$0.05<z<0.6$} & 
\colhead{$0.6<z<1.2$} 
  } 
\startdata
 $\alpha$ &  -1.0 & -1.0  &    &  -1.0 & -1.0 &  & -1.0 & -1.0 \nl
 $ M_{I*}$ & $-22.79 \pm 0.19$ & $-23.38 \pm 0.15$ 
 &  & $-22.99 \pm 0.19$  & $-23.72 \pm 0.15$ 
 &  & $-23.17 \pm 0.19$  & $-23.92 \pm 0.15$ \nl 
 $ \phi^{*} $ & $(1.35 \pm 0.42) \times 10^{-3}$ & $(1.70 \pm 0.45) \times 10^{-3}$ 
 &  & $(0.99 \pm 0.31) \times 10^{-3}$ &  $(1.01 \pm 0.29) \times 10^{-3}$ 
 &  & $(0.71 \pm 0.22) \times 10^{-3}$ &  $(0.68 \pm 0.19) \times 10^{-3}$   \nl
 $ M^{*}_{B}$ & $-20.62 \pm 0.19$ & $-21.21 \pm 0.15$ 
 &  & $-20.82 \pm 0.19$  & $-21.55 \pm 0.15$ 
 &  & $-21.00 \pm 0.19$  & $-21.75 \pm 0.15$  
\enddata
\tablecomments{Parameters are calculated assuming $H_{0}=70$ km sec$^{-1}$
 Mpc$^{-1}$.  Errors include those caused by Poisson counting statistics 
 and clustering statistics
 but not those caused by incompleteness or galaxy misclassifications.}
\end{deluxetable}

\begin{deluxetable}{c  c c   c  c c  c c c}
\tiny
\tablecolumns{9}
\tablecaption{Luminosity and density evolution from Method 1}
\tablewidth{0pt}
\tablehead{ 
 \colhead{}   &  \multicolumn{2}{c}{$\Omega_{m}=1, \Lambda=0$}  & \colhead{} & 
  \multicolumn{2}{c}{$\Omega_{m}=0.2, \Lambda=0$} & \colhead{} &
  \multicolumn{2}{c}{$\Omega_{m}=0.3, \Lambda=0.7$}  \\ 
\cline{2-3} \cline{5-6} \cline{8-9} \\
 \colhead{}   & \colhead{$0.05<z<0.6$} & \colhead{$0.6<z<1.2$} & 
 \colhead{} & \colhead{$0.05<z<0.6$} & 
\colhead{$0.6<z<1.2$}  & 
 \colhead{} & \colhead{$0.05<z<0.6$} & 
\colhead{$0.6<z<1.2$} 
  } 
\startdata
 $\Delta B({\rm vs.~ Marzke})$
 & $0.48 \pm 0.21$  & $1.07 \pm 0.18$ & 
 & $0.68 \pm 0.21$  & $1.41 \pm 0.18$ & 
 & $0.86 \pm 0.21$  & $1.61 \pm 0.18$ \nl 
 $ \phi^{*}(z)/(70\%~\phi^{*}({\rm Marzke}))$
 & $1.28 \pm 0.48$    & $1.61 \pm 0.54$  & 
 & $0.94 \pm 0.37$ &  $0.96 \pm 0.34$   & 
 & $0.71 \pm 0.26$ &  $0.64 \pm 0.22$   \nl
 $\Delta B({\rm vs.~ Marinoni})$
 & $0.08 \pm 0.26$  & $0.67 \pm 0.23$ & 
 & $0.28 \pm 0.26$  & $1.01 \pm 0.23$ & 
 & $0.46 \pm 0.26$  & $1.21 \pm 0.23$ \nl 
 $ \phi^{*}(z)/\phi^{*}({\rm Marinoni})$
 & $1.61 \pm 0.63$ &  $2.03 \pm 0.72$  & 
 & $1.18 \pm 0.46$ &  $1.21 \pm 0.45$   & 
 & $0.85 \pm 0.33$ &  $0.81 \pm 0.30$   \nl
\enddata
\tablecomments{The quantity  $ \phi^{*}(0)$ from Marzke et al. (1998)
has been multiplied by a correction factor of 0.70 to adjust for
the fact that our galaxy selection criteria likely select fewer
early-type galaxies than Marzke et al.  See text.}
\end{deluxetable}

\begin{deluxetable}{c  c c   c  c c}
\scriptsize
\tablecolumns{6}
\tablecaption{Parameters of LFs from Method 2}
\tablewidth{0pt}
\tablehead{ 
 \colhead{}   & {$\Omega_{m}=1, \Lambda=0$}  & \colhead{} & 
 {$\Omega_{m}=0.2, \Lambda=0$} & \colhead{} &
 \colhead{$\Omega_{m}=0.3, \Lambda=0.7$} 
  } 
\startdata
\sidehead{Solving for $M^{*}_{B}(0)$}
 $M^{*}_{B}(0)$   & $-20.03 \pm 0.56$  &  & $-20.05 \pm 0.56$ &  &
 $-20.19 \pm 0.57$ \nl 
 $Q$           & $1.43   \pm 0.80$  &  & $1.82  \pm 0.80$  &  &
 $1.89  \pm 0.81$  \nl
 $m$           & $0.46   \pm  0.68$ &  
               & $-0.48 \pm   0.68$ &  
               & $-0.86 \pm 0.68$  \nl
 $\phi^{*}$    & $(1.10 \pm 0.44) \times 10^{-3}$ &   
               & $(1.09 \pm 0.46) \times 10^{-3}$ &   
               & $(0.95 \pm 0.39) \times 10^{-3}$ \nl
\cline{1-6}
\sidehead{Fixed $M^{*}_{B}(0)=-20.14$ from Marzke et al. (1998)}
 $Q$           & $1.29   \pm  0.23$  &
               & $1.70   \pm  0.23$  &
               & $1.97   \pm  0.23$  \nl
 $m$           & $0.64   \pm  0.49$ &  
               & $-0.33 \pm   0.49$ &  
               & $-0.95 \pm   0.48$  \nl
 $\phi^{*}$    & $(1.01 \pm 0.35) \times 10^{-3}$ &   
               & $(1.02 \pm 0.35) \times 10^{-3}$ &   
               & $(0.96 \pm 0.33) \times 10^{-3}$ \nl
\cline{1-6}
\sidehead{Fixed $M^{*}_{B}(0)=-20.54$ from Marinoni et al. (1999)}
 $Q$           & $0.76   \pm  0.23$  &
               & $1.17   \pm  0.23$  &
               & $1.43   \pm  0.23$  \nl
 $m$           & $1.29   \pm  0.52$ &  
               & $0.33   \pm   0.52$ &  
               & $-0.30  \pm   0.51$  \nl
 $\phi^{*}$    & $(0.74 \pm 0.27) \times 10^{-3}$ &   
               & $(0.75 \pm 0.27) \times 10^{-3}$ &   
               & $(0.71 \pm 0.26) \times 10^{-3}$ \nl
\enddata
\end{deluxetable}

\begin{deluxetable}{c  c c   c  c c}
\scriptsize
\tablecolumns{6}
\tablecaption{Parameters of LFs for various $z$ limits}
\tablewidth{0pt}
\tablehead{ 
 \colhead{}   & {$0.2 < z < 1.2$}  & \colhead{$0.05 < z < 0.8$} & 
 {$0.05 < z < 1.0$} & \colhead{$0.2 < z < 0.8$} &
 \colhead{$0.2 < z < 1.0$} 
  } 
\startdata
\sidehead{Solving for $M^{*}_{B}(0)$}
 $M^{*}_{B}(0)$   & $-20.12 \pm 0.57$  &  $-19.92 \pm 0.70$ &
                    $-19.92 \pm 0.57$  &  $-20.01 \pm 0.72$ &
                    $-19.99 \pm 0.58$ \nl 
 $Q$           & $1.73  \pm  0.81$  & $1.99 \pm 1.19$ &
                 $2.05  \pm  0.84$  & $1.85 \pm 1.21$ &
                 $1.96  \pm  0.85$  \nl
 $m$           & $-0.68   \pm  0.73$ & $0.96 \pm 0.96$   
               & $0.16    \pm  0.74$ & $0.83 \pm 1.07$  
               & $0.00    \pm  0.81$  \nl
\enddata
\end{deluxetable}